\documentclass[iop,revtex4,appendixfloats]{emulateapj} 
\usepackage{float,epsfig}
\usepackage{lscape}
\usepackage{natbib}
\usepackage{color}

\def\cxo{{\it Chandra\/}}

\def\rxte{{\it RXTE\/}}

\def\spitzer{{\it Spitzer\/}}

\def\xmm{{\it XMM-Newton\/}}
\def\nustar{{\it NuSTAR\/}}

\def\msun{$M_\odot$}
\def\h2{H$_2$}
\def\ha{H$\alpha$}

\def\ergl{erg~s$^{-1}$}
\def\arcsec{\mbox{$^{\prime\prime}$}}
\def\arcmin{\mbox{$^\prime$}}
\def\ulxa{x299}

\def\aap{A\&A}
\def\apj{ApJ}
\def\apjl{ApJL}
\def\apjs{ApJS}
\def\aj{AJ}
\def\mnras{MNRAS}
\def\araa{ARA\&A}
\def\pasj{PASJ}
\def\nat{Nature}
\def\spose#1{\hbox to 0pt{#1\hss}} 
\def\simlt{\mathrel{\spose{\lower 3pt\hbox{$\sim$}}
        \raise 2.0pt\hbox{$<$}}}
\def\simgt{\mathrel{\spose{\lower 3pt\hbox{$\sim$}}
        \raise 2.0pt\hbox{$>$}}}

\begin{document}

\shorttitle{\nustar\ Observation of M83}

\title{A  Hard X-ray  Study of the Normal Star-Forming Galaxy M83 with {\it NuSTAR}}

\author{
M.~Yukita\altaffilmark{1,2},
A.~E.~Hornschemeier\altaffilmark{2,1},
B.~D.~Lehmer\altaffilmark{1,2,3},
A.~Ptak\altaffilmark{2,1},
D.~R.~Wik\altaffilmark{1,2},
A.~Zezas\altaffilmark{4,5},
V.~Antoniou\altaffilmark{5},
T.~J.~Maccarone\altaffilmark{6},
V.~Replicon\altaffilmark{7},
J.~B.~Tyler\altaffilmark{2,8},
T.~Venters\altaffilmark{2},
M.~K.~Argo\altaffilmark{9}, 
K.~Bechtol\altaffilmark{10}, 
S.~Boggs\altaffilmark{11},
F.~E.~Christensen\altaffilmark{12}, 
W.~W.~Craig\altaffilmark{9,13}, 
C.~Hailey\altaffilmark{14}, 
F.~Harrison\altaffilmark{15},
R.~Krivonos\altaffilmark{11,16},
K.~Kuntz\altaffilmark{1},
D.~Stern\altaffilmark{17}, 
W.~W.~Zhang\altaffilmark{2}
}

\altaffiltext{1}{The Johns Hopkins University, Homewood Campus, Baltimore, MD 21218, USA}
\altaffiltext{2}{NASA Goddard Space Flight Center, Code 662, Greenbelt, MD 20771, USA} 
\altaffiltext{3}{Department of Physics, University of Arkansas, Fayetteville, AR 72701, USA}
\altaffiltext{4}{Physics Department, University of Crete, Heraklion, Greece}
\altaffiltext{5}{Harvard-Smithsonian Center for Astrophysics, 60 Garden Street, Cambridge, MA 02138, USA}
\altaffiltext{6}{Department of Physics, Texas Tech University, Lubbock, TX 79409, USA}
\altaffiltext{7}{Department of Physics, New Mexico Institute of Mining and Technology,  Socorro, NM 87801, USA}
\altaffiltext{8}{Institute for Astrophysics and Computational Sciences, Department of Physics, The Catholic University of America,Washington, DC 20064, USA}
\altaffiltext{9}{Jodrell Bank Centre for Astrophysics, The University of Manchester, Oxford Rd, Manchester M13 9PL, UK}
\altaffiltext{10}{Kavli Institute for Cosmological Physics, Chicago, IL 60637, USA}
\altaffiltext{11}{Space Science Lab, University of California, Berkeley, CA 94720, USA}
\altaffiltext{12}{National Space Institute, Technical University of Denmark, DK-2100 Copenhagen, Denmark}
\altaffiltext{13}{Lawrence Livermore National Laboratory, Livermore, CA, USA}
\altaffiltext{14}{Columbia University, New York, NY, USA}
\altaffiltext{15}{Caltech Division of Physics, Mathematics and Astronomy, Pasadena, CA, USA}
\altaffiltext{16}{Space Research Institute, Russian Academy of Sciences, Profsoyuznaya 84/32, 117997 Moscow, Russia}
\altaffiltext{17}{Jet Propulsion Laboratory, California Institute of Technology, Pasadena, CA 91109, USA}

\begin{abstract}
We present results from sensitive, multi-epoch \nustar\ observations of the late-type star-forming galaxy M83 ($d=4.6$~Mpc),  which is
 the first investigation to spatially resolve the hard ($E>10$~keV)
 X-ray emission of this galaxy.
 The nuclear region and $\sim 20$ off-nuclear point sources, including a previously discovered ultraluminous X-ray (ULX) source, are detected in our \nustar\ observations. 
 The X-ray hardnesses  and luminosities of the majority of the
 point sources are consistent with hard X-ray sources resolved in the
 starburst galaxy NGC~253.  We infer that the hard X-ray
 emission is most likely dominated by intermediate accretion state
 black hole binaries and neutron star low-mass X-ray binaries (Z-sources).    We
 construct the X-ray binary luminosity function (XLF) in the \nustar\
 band for an extragalactic environment for the first time.  The M83
 XLF has a steeper XLF than the X-ray binary XLF in NGC 253,
 consistent with previous measurements by \cxo\ at softer X-ray energies.
 The \nustar\ integrated galaxy spectrum of M83 drops quickly above
 10~keV, which is also seen in the starburst galaxies NGC~253, NGC 3310 and
 NGC 3256.
 The \nustar\ observations constrain any AGN to be either highly
 obscured or to have an extremely low luminosity of
 $\simlt$10$^{38}$~\ergl\  (10--30~keV), implying it is emitting at a
 very low Eddington ratio.
An  X-ray point source consistent with the location of the nuclear star cluster with 
an X-ray luminosity of
a few times 10$^{38}$~\ergl\ may be a low-luminosity AGN but is more
consistent with being an X-ray binary.

\end{abstract}


\begin{deluxetable*}{cccccccc} 
\tabletypesize{\scriptsize}
\tablecolumns{8} 
\tablewidth{0pc} 
\tablecaption{Observation Log\label{t:log}}
\tablehead{     
 \multicolumn{2}{c}{\nustar} &   \colhead{}   & 
\multicolumn{4}{c}{\xmm/\cxo}     \\ 
   \cline{1-3} \cline{5-8} &\colhead{} \\
\colhead{}   & \colhead{}    & \colhead{Net  } & 
\colhead{}    &  \colhead{}& \colhead{}&   \colhead{} &     \colhead{Net } \\
  \colhead{Date}   & \colhead{ObsID}    & \colhead{Exposure} & 
\colhead{}    &  \colhead{Observatory}&  \colhead{Date}&   \colhead{ObsID} &  
 \colhead{Exposure} \\
  \colhead{}   & \colhead{}    & \colhead{(ks)}   & \colhead{}     
 &  \colhead{}    & \colhead{}   & \colhead{}  &  \colhead{(ks)}  }
\startdata 
\multicolumn{8}{c}{{\bf Epoch 1}}\\
  \cline{1-8}
 Aug 7, 2013\phn   & 50002043002 & 42 &  & \xmm\ & Aug 7, 2013 &  0723450101 &  41 \\ 
 Aug 9, 2013\phn   & 50002043004 & 80 &  & &  \nodata& \nodata &   \nodata \\ 
Aug 21, 2013\phn & 50002043006 & 43 & & &  \nodata & \nodata &   \nodata \\ 
             \cline{1-8}   \\ 
\multicolumn{8}{c}{{\bf Epoch 2}}\\          
            \cline{1-8}   \\ 
Jan 19, 2014\phn  & 50002043008 & 81 & &\xmm\ & Jan 11, 2014 & 0723450201  & 20 \\ 
  \cline{1-8}   \\
   \multicolumn{8}{c}{{\bf Epoch 3}}\\    
     \cline{1-8}   \\
Jun 4, 2014\phn  & 50002043010 & 70 & & &  \nodata &\nodata        &  \nodata        \\
Jun 7, 2014\phn  & 50002043012 &110 & &  \cxo & Jun 7, 2014 &   16024    &    30    \\
\enddata
\end{deluxetable*}

\section{Introduction}

The \nustar\ observatory, the first focusing hard X-ray ($E>$10~keV) telescope in orbit \citep{harrison13}, 
is capable of spatially resolving the components of nearby ($d <$ 5~Mpc)
 star-forming galaxies in this energy range for
the first time.  
Hard   X-ray emission in star-forming galaxies lacking an AGN originates primarily
 from X-ray binaries, mainly high-mass X-ray binaries (HMXB) and ultraluminous X-ray sources (ULXs). 
There are other contributions, chiefly low-mass X-ray binaries (LMXBs), but also possibly accretion onto a supermassive 
black hole
(SMBH) and perhaps diffuse non-thermal inverse Compton (IC)
emission \citep{lehmer13,wik14b,ptak15,lehmer15}.  However, the IC component
results from the interaction of cosmic rays with IR and microwave
background photons, and is likely to be negligible at moderately hard X-ray energies (e.g., see Wik et
al. 2014b for the case of NGC 253).

We are conducting a \nustar\ starburst galaxy survey to investigate the hard X-ray properties of a sample of 
six nearby galaxies (NGC 253, Arp 299, M83, M82, NGC 3310, and NGC
3256), spanning star formation rates (SFRs) of 1 -- 100~\msun~yr$^{-1}$.
We are characterizing the 0.5--30~keV spectral energy distribution (SED) of these galaxies \citep[][]{lehmer15,ptak15}, which is 
of great cosmological importance. For example, this SED is used to do important k-corrections for high-z galaxies observed
 in the deepest \cxo\ X-ray surveys \citep{lehmer16}. 
For the closest galaxies ($d<5$~Mpc; NGC~253, M82, and M83),  the very deep \nustar\ exposures 
allows us for detailed studies 
of the spatially-resolved point source population.

The in-depth studies of starburst galaxy NGC~253 \citep{lehmer13,wik14b}
with \nustar\ and \cxo\ showed the utility of coordinated observations with these  two facilities 
and hard X-ray color-color and color-intensity diagnostics to determine
the states of X-ray binaries in external galaxies.  
In NGC~253, the majority of the luminous X-ray binaries 
were found to have properties consistent with (stellar-mass) black hole candidates in the intermediate state, an important measurement of 
the state of the {\it population} of X-ray binaries in a star-forming environment.
We now seek to answer the question of whether or not the preponderance of black holes in intermediate states 
is ubiquitous among star-forming galaxies as well as to further investigate other populations, such as
neutron star (NS) LMXBs, that might have similar colors.  

It also appears that star-forming galaxies are dominated by ULXs and have overall spectra with 
 steep turnovers at  $E>10$~keV (e.g., NGC~253; Wik et al. 2014b; NGC~3310 and NGC~3256; Lehmer et al. 2015).  
This hard energy turn over around 6--10~keV is also seen for bright individual ULXs studied by \xmm\ and \nustar\  \citep{stobbart06,gladstone09,walton13,bachetti13,rana14,walton14} and is
  interpreted as super-Eddington accretion onto a stellar mass black hole instead of an intermediate mass black hole in a  low/hard
  accretion state \citep[e.g.,][]{shakura73,king01,poutanen07,gladstone09,sutton13}.   

Detailed investigation of the spectra of ULXs and other components contributing to the total galaxy spectrum requires observing very nearby galaxies. Even for galaxies only $<$10 Mpc away, X-ray binaries can be unresolved by {\it NuSTAR}'s 58\arcsec\  half-power diameter
point spread function (PSF).

Therefore, we obtained a deep \nustar\ observation of the nearby
star-forming galaxy, M83, a face-on late-type (SABc) spiral galaxy.  
Its proximity  \citep[$d$=4.61$\pm$0.20~Mpc;][]{saha06} and large optical angular extent ($D_{25}$\footnote{The galaxy's size is defined by the $B$-band surface brightness  level of 25 mag arcsec$^2$}=12.9\arcmin$\times$11.5\arcmin, 17$\times$15~kpc),
make it suitable for a point source population study.
We are able to achieve high efficiency in detecting point sources as compared to
 the edge-on galaxy NGC~253,
whose 
resolved sources, owing to a more compact overall distribution, are more spatially blended with each other.
 
M83 has been studied over a variety of wavelengths quite intensively, including the X-ray band.
\cxo\ has observed it for $\sim800$~ks, identifying 398 point sources \citep[][hereafter L14]{Long14}, including
X-ray binaries down to $L_{\rm X} < 10^{36}$~erg~s$^{-1}$ in the 0.35--8.0~keV band.  Notably,
 L14 found the observed X-ray binary luminosity function (LF) does not match with the predicted LF scaled from M83's SFR and stellar mass.  The observed LF suggests that the substantial LMXB population dominates over the HMXBs associated with the current on-going star-forming activity \citep[i.e.,][]{boissier05}.
\citet{ducci13} investigated a number of bright point sources
with \xmm\  and found an overall X-ray binary LF
consistent with the \cxo\ measurements.  There are a couple of bright, individual sources in M83 that are also notable.  
\citet{soria12} discovered a ULX about 1\arcmin\ to the east of the nucleus, that is likely to have a red giant 
 counterpart. 
There is another  ULX candidate at the location of the edge 
of the stellar disk reported by \citet{immler99} and \citet{stobbart06}.

So far, no AGN activity has been reported in M83.  
However, there is an X-ray point source identified in the \cxo\ 
observations as luminous as a few times 10$^{38}$~\ergl, at the location of
 a nuclear star cluster  \citep{soria03,Long14}.  This source is a good candidate 
 for  the center of the galaxy  and
  a low luminosity AGN, but the authors also suggest 
the possibility that it is  an X-ray binary.
  
 The aims of this paper are to identify the point sources detected by \nustar\ at hard energies (4--25~keV), examine 
 the point source population  by investigating luminosities and spectral colors, characterize
 the broadband SED, and constrain the nuclear activity in M83.  
We assume the  distance to M83 is 4.61~Mpc,  for which 1\arcsec~corresponds to 22~pc.    
Unless noted otherwise, quoted uncertainties correspond to 90\% confidence intervals for one interesting parameter.

\section{Data and Data Reduction}\label{S:data}

\subsection{\nustar\ Data}

\begin{figure}
\begin{center}
\includegraphics[angle=-90,width=3.5in]{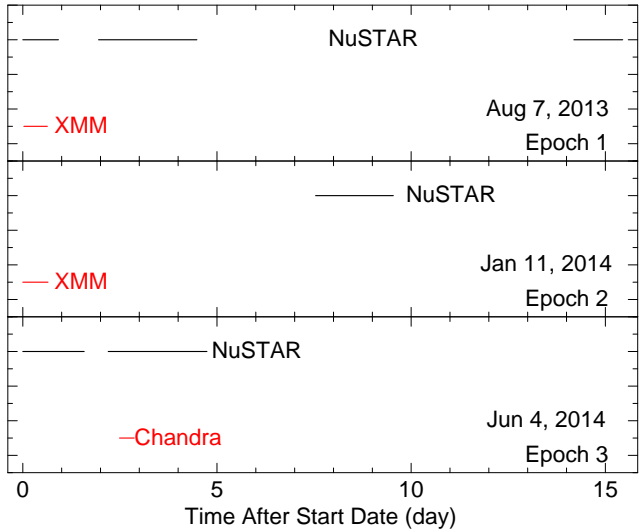}
\end{center}
\caption{The \nustar\ data on M83 were gathered in three separate observing campaigns.   Shown are the relative \nustar, \xmm\, and/or  \cxo\ observation times for each epoch.
The start date for each epoch is indicated.} 
\label{f:coverages}
\end{figure}

{\it NuSTAR} observed M83 in 2013 and 2014 over three epochs as  part of the \nustar\ Starburst Galaxy 
 Survey. 
The three \nustar\ epochs were simultaneous or nearly simultaneous with 
 either \xmm\ or \cxo\ observations,  which constrain the lower energy
 ($E\simlt3$~keV) emission as well as identify sources potentially confused with \nustar 's larger beam.
We show these observational intervals graphically in Figure~\ref{f:coverages} and give the observation log in Table~\ref{t:log}.

We reduced the \nustar\ data using {\tt HEASOFT 6.15} and CALDB v. 20131223.
Specifically,  we processed the data using the {\tt nupipeline} script, which creates calibrated and screened
 level 2 event lists from level 1 data.   
The resultant sum of the good time intervals (GTIs) is also listed in Table~\ref{t:log}.

\begin{figure}
\begin{center}
\includegraphics[width=3.5in]{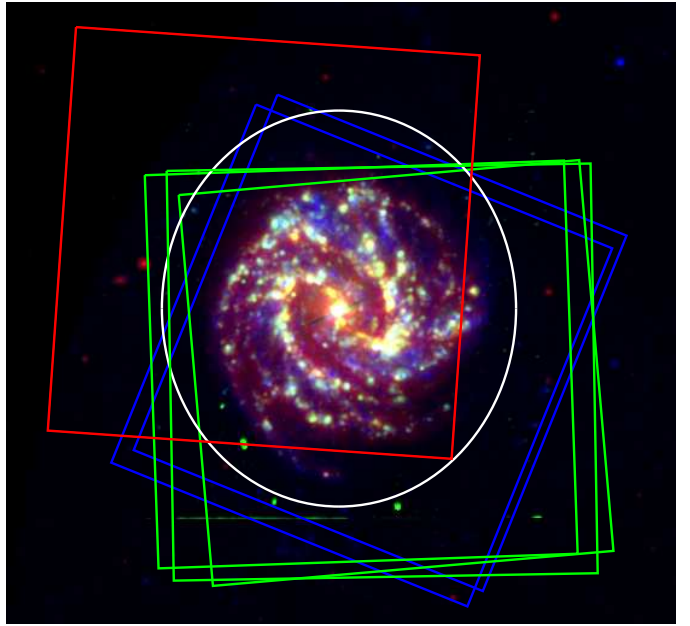}
\end{center}
\caption{{\it GALEX} FUV(blue), continuum-subtracted \ha\ (green) and \spitzer\ 24~$\mu$m (red) images of M83 \citep[from the {\it Spitzer} Local Volume
Survey,][]{dale07}. North is up and East to the left.
White ellipse indicates optical extent ($D_{25}$) of the galaxy.
Boxes represent the \nustar\ FoV (13\arcmin$\times$13\arcmin); green for 
August 2013, red for January 2014 and blue for June 2014 observations.}
\label{f:fov}
\end{figure}

The optical extent of M83 (12.9\arcmin$\times$11.5\arcmin) is comparable to the field-of-view (FoV; Figure~\ref{f:fov}) 
of \nustar\ (13\arcmin $\times$ 13\arcmin); however, 
 the center of the galaxy was placed a few arcminutes offset from the center of the FoV.  
This placement  results in partial
covering of M83 in each observation.  
Since the \nustar\ observations were taken with different roll angles, over the course of the three observations the
entire $D_{25}$ region is covered after merging the data (see Figure~\ref{f:fov}). 

\begin{figure}
\begin{center}
\includegraphics[width=3.5in]{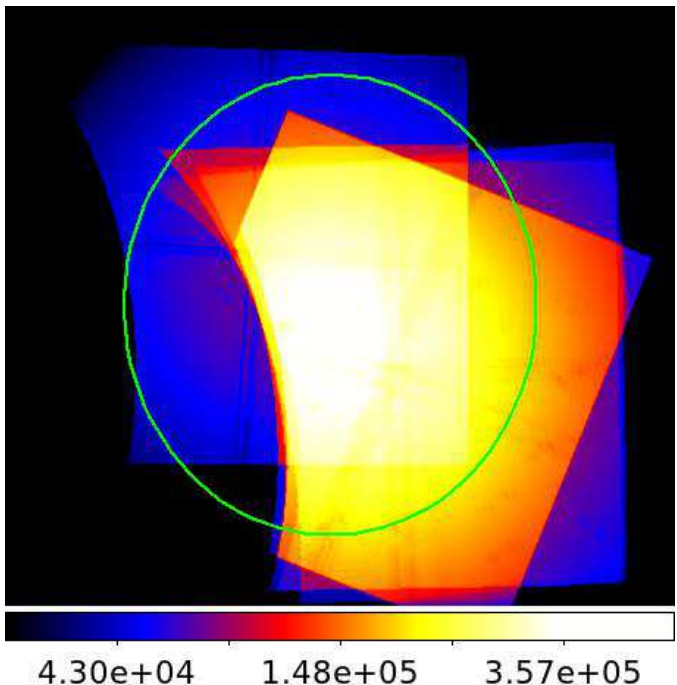}
\end{center}
\caption{The exposure map of the co-added \nustar\ data presents the effective exposure time for different locations of the galaxy.
(North is up and East to the left.)
Green ellipse indicates $D_{25}$. The pixel values are in units of seconds.}
\label{f:exposure}
\end{figure}

We note that there are two  bright objects, the Shapley supercluster and IC 4329A, located within a 1 -- 5 degree annulus centered on M83, which thus result in stray light (unfocused X-rays entering between the optics module and aperture stops) on the \nustar\ detectors.
Unfortunately, the stray light pattern is apparent in all observations.  
However, stray light can be estimated using the location of those bright sources and the observed roll angle.
We used the {\tt nustar\_stray\_light}\footnote{https://github.com/bwgref/nustar\_stray\_light} procedure to estimate the affected regions on each detector for each observation.
IC~4239A contaminated the small region to the east of M83 in both the FPMA and FPMB data.  
The Shapley supercluster only appeared on one of either FPMA or FPMB (i.e., one of the telescopes), but the stray light is extended over a larger area.  
The contamination from both IC~4329A and the Shapley supercluster covered more than half the total area of either the FPMA or FPMB detector.   
We discarded the data which suffered from the Shapley supercluster stray light. 
Specifically, we only utilize FPMA for the Jan 2014 observation and FPMB for the remaining five observations. 
We also eliminated the region affected by IC~4239A in these data.  This resulted in missing partially the southeast region of $D_{25}$ in 
our analysis.   
In Figure~\ref{f:exposure}, we show the co-added \nustar\ exposure map after removing the contaminated 
areas, which presents the area coverage with effective exposures of our M83 observations.  

\begin{figure*}
\begin{center}
\includegraphics[angle=0,height=2.6in]{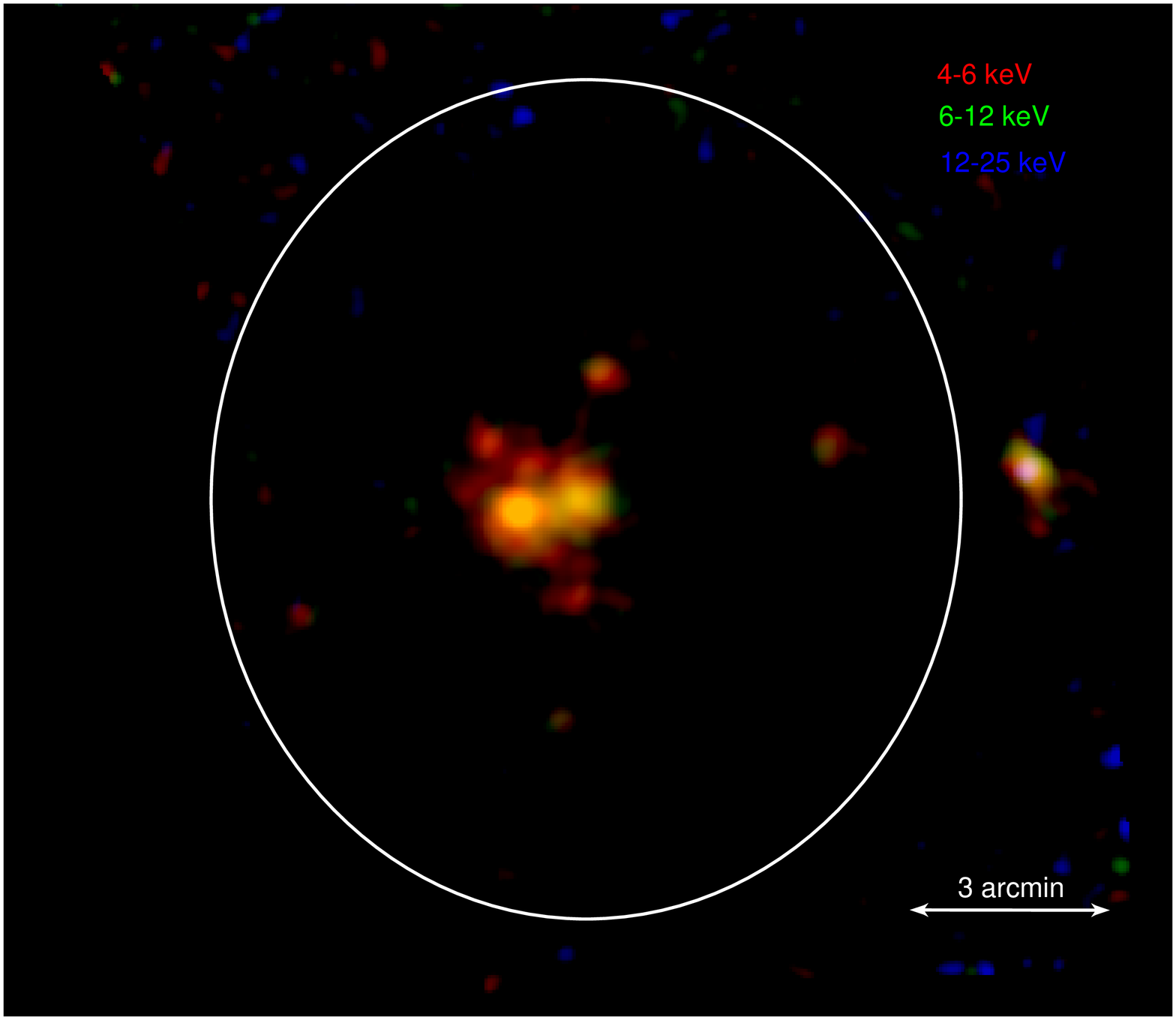}
\includegraphics[angle=0,height=2.6in]{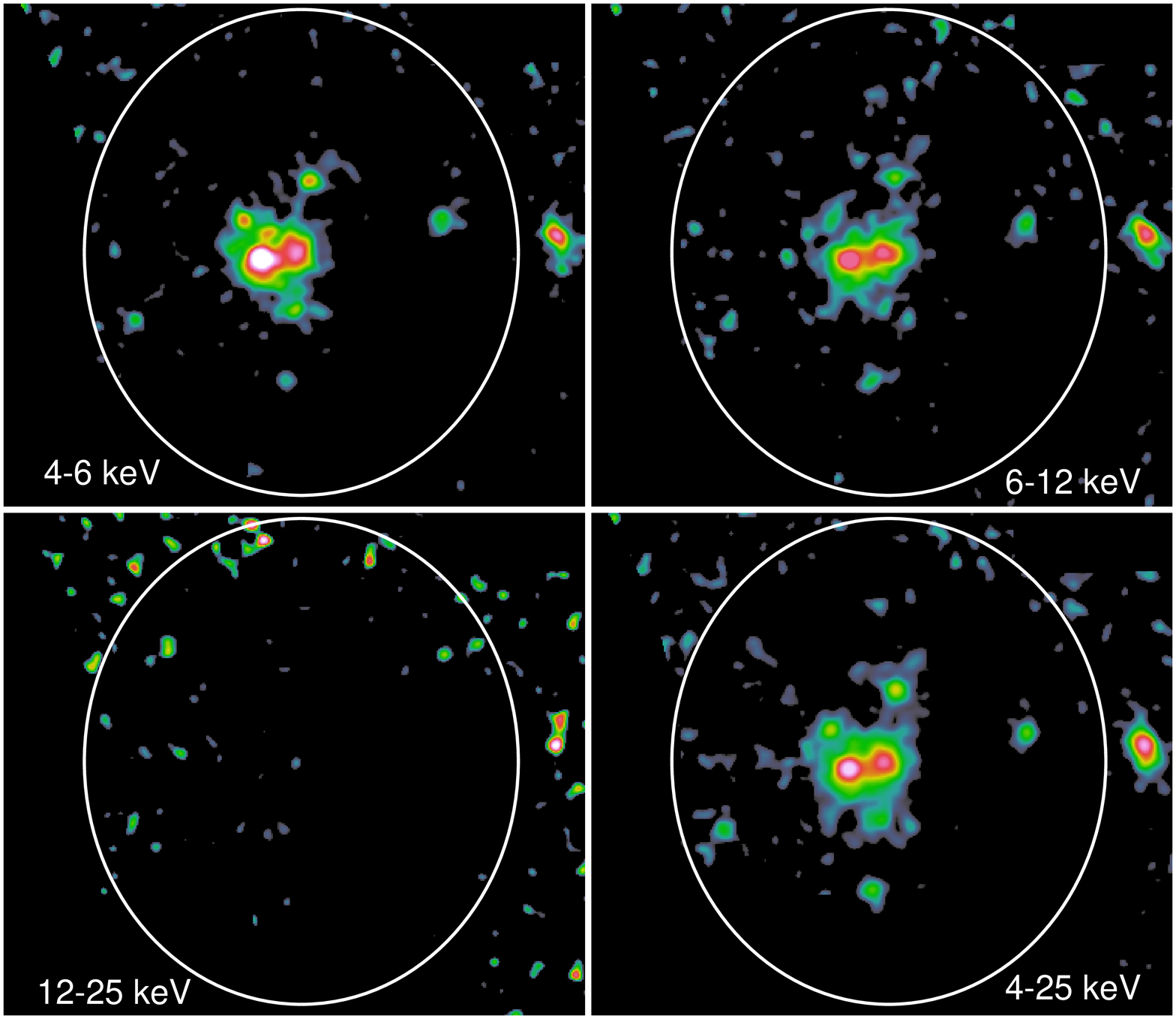}
\end{center}
\caption{Left:  The \nustar\ color image of M83.  Red, green and blue colors correspond to 4--6~keV,
6--12~keV, and 12--25~keV, respectively.   North is up and East to the left. The white ellipse indicates the  $D_{25}$ size of M83.  Each image is
exposure-corrected, background-subtracted (see Section~\ref{S:back}), and smoothed. Right:   The 4--6, 6--12, 4--25, and 12--25~keV (clockwise from the top left) band images
are shown separately.
The 4--6, 6--12, and 4--25~keV images clearly show a number of the resolved point sources in 
M83. }
 \label{f:4bands}
\end{figure*}

\subsection{Simultaneous \xmm\  and \cxo\ Observations}
\xmm\ observations took place on Aug 8, 2013 (epoch 1) and Jan 11, 2014 (epoch 2) as part of
 an AO 12 GO program to monitor the newly discovered ULX, CXO~133705-295207 (P.I. Kuntz).
The detailed \xmm\ analysis of the ULX is presented in \citet{soria15}.
Our primary use of the \xmm\ data is to locate bright point sources and to characterize 
the 0.5--8.0~keV spectral shape of the galaxy and bright sources. 
Because the  \xmm\ PN detector alone achieves a 4--10~keV sensitivity similar to our \nustar\
4--10~keV sensitivity,   we utilized the PN data only to avoid complication
of data analysis between the different \xmm\  detectors.    
The choice of the PN detector over MOS was motivated by the better S/N and larger FoV
  (i.e., due to the loss of two CCD 
chips of MOS 1). 
The \xmm\ data were reduced using SAS v.13.0.0.
The level 1 data were reprocessed using the SAS scripts, {\tt epchain},  and then
{\tt pn-filter} is  applied to eliminate high background periods.
The resulting exposures are tabulated in Table~\ref{t:log}.

\cxo\ observed M83 on June 2, 2014 (epoch 3) using  the Advanced CCD Imaging Spectrometer (ACIS), as a
part of the \cxo\ campaign of the \nustar\ starburst galaxy survey (P.I. Hornschemeier). 
To capture the entire $D_{25}$, we used the ACIS-I array, which has a wide FoV (16\arcmin\ $\times$ 16\arcmin)
that complements well the \nustar\ FoV. 
The \cxo\ data reduction and analysis were performed with {\it CIAO} version 4.6 and CALDB 4.6.3. 
\cxo\ data were reprocessed using the {\it CIAO} tool {\tt chandra\_repro}, which applies time-dependent gain and charge transfer inefficiency (CTI) corrections and
removes bad pixels and bad grade events.  
We removed high background periods applying the sigma-clipping method. 
Namely, we used the {\it CIAO}  {\tt deflare} script, rejecting the intervals that exceeded 3$\sigma$ above the mean. 
The final net exposure was 29.6~ks of the original 30~ks observation. 
Figure~\ref{f:coverages} illustrates the relative \nustar, \xmm, and \cxo\ coverages for each epoch.

Both the \xmm\ and \cxo\ observations were spatially registered to the \cxo\ point source catalog of L14.

\section {Spatial Analysis}\label{S:results}

The \nustar\ M83 observations spatially resolve  multiple point-like X-ray sources including the known ULX source (see Figure~\ref{f:4bands}).  
To maximize the S/N, we co-added observations from all three epochs, resulting in an effective exposure  of $\sim$370~ks
in the central region of M83 (see Figure~\ref{f:exposure}).
The first and third epoch observations were aligned using the position of CXO~133705$-$295207 (\ulxa\ in L14) which 
 was obtained by fitting a circular Gaussian to 4$-$10~keV  images
 of the ULX.   In general using multiple point sources for alignment
 is preferred, but the S/N of the majority of the point sources for
 each observation was low.      
Nevertheless, since \ulxa\ was not bright in the second epoch observation, we used other bright sources to
 align the data to the remaining observations.
All the aligned event lists were merged using the FTOOLS {\tt ftmerge} utility to create a combined, deep exposure event list. 
The final event list was registered using the position of the ULX to the published \cxo\ coordinates of \ulxa\ by L14. 
In our astrometric solution,  we assume no rotation since there is not a sufficient number of sources to constrain it.

The left panel of Figure~\ref{f:4bands}~shows the \nustar\ false color image of M83 for the merged data. 
The individual 4--6, 6--12, 12--25, and 4--25~keV images are displayed in the
right panel of Figure~\ref{f:4bands}.  All the images are exposure-corrected, background-subtracted  and smoothed.
The 4--6, 6--12, and 4--25~keV images clearly show that a number of  point sources are resolved within the optical extent of
M83. 
These resolved sources are not obviously seen in the 12--25~keV band image. 
In the next subsection, we identify the resolved point sources and measure their fluxes  in the \nustar\ images.

\subsection{Point Source Identification}
\nustar\ fluxes for  point  sources were obtained via point spread function
(PSF) fitting as opposed to aperture photometry, because some sources contain flux from broad PSF wings underneath them.
We used  {\it Sherpa} (included as a part of {\it CIAO}) for spatial
fitting of the \nustar\ images. 
In this section, we describe the method to identify and measure the \nustar\ fluxes of 21 resolved point sources.
The  outline procedure is to (1) create the \xmm/\cxo\ point source catalog, (2) measure \nustar\ fluxes based
on the \xmm/\cxo\ source positions, and (3) reject faint sources whose fluxes  are below 90\% confidence levels.

\subsubsection{\nustar\ Background Modeling}\label{S:back}

For photometry, the \nustar\ background must also be taken into account. 
The \nustar\ background consists of the focused cosmic X-ray background,    
stray light from the cosmic X-ray background (referred as the 
 aperture background), instrument background, and reflected solar X-rays \citep{wik14}.
The spatial variations of the background surface brightness are significant,
therefore we 
use the {\tt nuskybgd} script  \citep{wik14} to create model background images for each observation.
The  {\tt nuskybgd} script characterizes the spectral and spatial parameters 
of the observed background in the source-free region of the observations and then extrapolates 
across the FoV  \citep[see][for details]{wik14}. 
In {\it Sherpa}, the simulated background images are fitted to a 2D polynomial
function to analytically determine  the spatial distribution of the background model. 
We also note that an on-axis PSF model \citep{an14,madsen15} is applied even to off-axis sources.
This approach may therefore introduce systematic uncertainties due to possibly oversimplified background and PSF models.

\begin{deluxetable*}{llcrrrrrr}
\tabletypesize{\scriptsize}
\tablecolumns{9} 
\tablewidth{0pc} 
\tablecaption{M83 NuSTAR Point Sources\label{t:nustarpt}}
\tablehead{ 
\colhead{}   & \colhead{}   & \colhead{L14-{\it NuSTAR}$^b$}  & \colhead{Effective} & 
\colhead{}    & \multicolumn{4}{c}{Count Rates}   \\
\colhead{R.A.$^a$}   & \colhead{Dec.$^a$}   & \colhead{Offset}  & \colhead{Exp} & 
\colhead{ID$^c$}    & \colhead{(4--6~keV)}   & \colhead{(6--12~keV)}    & \colhead{(12--25~keV)}
& \colhead{(4--25~keV)}\\
\multicolumn{2}{c}{(J2000) } & \colhead{(arcsec)}   & \colhead{(ks)} & 
\colhead{L14}    &\multicolumn{4}{c}{(10$^{-3}$ ct~s$^{-1}$) }}
\startdata 
13:37:04 & $-$29:49:27  &3.1 & 322.1 & x281  &$  0.41_{-0.11}^{+0.12}$ & $  0.54_{-0.14}^{+0.15}$ & $ < 0.11$ & $ 1.07_{-0.22}^{+0.22}$\\
13:36:43 & $-$29:51:01   &  6.3 & 221.1& x029  & $  1.09_{-0.17}^{+0.18}$ & $  1.10_{-0.19}^{+0.20}$ & $  <0.11$ & $ 2.22_{-0.29}^{+0.30}$\\
13:36:49 & $-$29:52:30  &0.3  & 309.9& x048  &$  0.15_{-0.09}^{+0.10}$ & $  0.21_{-0.10}^{+0.11}$ & $  <0.13$ & $ 0.47_{-0.17}^{+0.19}$\\
13:36:52 & $-$29:53:34  &0.5 & 305.2& x073  &$  0.17_{-0.09}^{+0.10}$ & $  0.22_{-0.11}^{+0.12}$ & $  <0.08$ & $ 0.45_{-0.17}^{+0.18}$\\
13:36:57 & $-$29:49:11  &1.3  & 318.2 & x145  &$  0.46_{-0.12}^{+0.13}$ & $  0.57_{-0.14}^{+0.15}$ & $  < 0.20$ & $ 1.21_{-0.22}^{+0.23}$\\
13:36:57 & $-$29:50:58  &6.7  & 361.5& x165  &$  0.34_{-0.11}^{+0.12}$ & $  0.51_{-0.13}^{+0.14}$ & $  < 0.15$ & $ 0.96_{-0.21}^{+0.22}$\\
13:36:57 & $-$29:53:38  &1.8  & 345.1 & x152  &$  0.55_{-0.12}^{+0.12}$ & $  0.58_{-0.13}^{+0.14}$ & $ <0.05$ & $ 1.13_{-0.20}^{+0.21}$\\
13:36:58 & $-$29:47:25 &3.9  & 255.1& x155  &$  0.17_{-0.12}^{+0.13}$ & $  0.36_{-0.16}^{+0.17}$ & $   <0.22$ & $ 0.73_{-0.25}^{+0.27}$\\
13:36:59 & $-$29:50:01  &1.5  & 342.9 & x185  &$  1.43_{-0.16}^{+0.16}$ & $  1.18_{-0.16}^{+0.17}$ & $  0.17_{-0.11}^{+0.12}$ & $ 2.86_{-0.26}^{+0.26}$\\
13:37:00 & $-$29:52:02  &3.3  & 349.8& x193  &$  1.08_{-0.25}^{+0.26}$ & $  1.20_{-0.27}^{+0.27}$ & $ <0.13$ & $ 2.25_{-0.40}^{+0.41}$\\
13:37:01 & $-$29:47:46  &3.1 & 264.0& x252  &$  0.32_{-0.12}^{+0.13}$ & $  0.52_{-0.16}^{+0.18}$ & $  <0.15 $ & $ 1.11_{-0.25}^{+0.27}$\\
13:37:01 & $-$29:51:31  &10.9 & 305.2 & x246  &$  0.93_{-0.26}^{+0.27}$ & $  0.46_{-0.25}^{+0.27}$ & $  < 0.17$ & $ 1.48_{-0.40}^{+0.41}$\\
13:37:01 & $-$29:51:56 &0.3& 343.9 & x233  &$  1.93_{-0.34}^{+0.36}$ & $  2.25_{-0.35}^{+0.37}$ & $  0.33_{-0.17}^{+0.19}$ & $ 4.50_{-0.53}^{+0.54}$\\
13:37:01 & $-$29:53:25  &2.0  & 349.3 & x248  &$  1.29_{-0.16}^{+0.16}$ & $  0.66_{-0.14}^{+0.15}$ & $  0.23_{-0.11}^{+0.12}$ & $ 2.28_{-0.25}^{+0.25}$\\
13:37:02 & $-$29:55:19  &1.5& 292.3 & x258  &$  0.83_{-0.14}^{+0.15}$ & $  1.27_{-0.18}^{+0.19}$ & $  0.27_{-0.13}^{+0.14}$ & $ 2.46_{-0.27}^{+0.28}$\\
13:37:04 & $-$29:51:19 &2.7  & 335.7& x286  &$  0.45_{-0.18}^{+0.18}$ & $  0.10_{-0.16}^{+0.17}$ & $  < 0.16$ & $ 0.63_{-0.27}^{+0.28}$\\
13:37:05 & $-$29:52:07 &0.0$^*$  & 353.0 & x299 &  $4.85_{-0.26}^{+0.27}$ & $3.59_{-0.24}^{+0.25}$ & $ 0.20_{-0.11}^{+0.12}$ & $8.71_{-0.38}^{+0.38}$\\
13:37:05 & $-$29:53:59 &4.8  & 345.5& x284  &$  0.43_{-0.11}^{+0.12}$ & $  0.31_{-0.12}^{+0.13}$ & $   0.14_{-0.11}^{+0.12}$ & $ 0.92_{-0.20}^{+0.21}$\\
13:37:07 & $-$29:51:01 &0.5  & 346.2 & x321  &$  1.32_{-0.16}^{+0.17}$ & $  0.90_{-0.16}^{+0.17}$ & $  <0.17$ & $ 2.38_{-0.25}^{+0.26}$\\
13:37:13& $-$29:52:01 & 6.2& 56.5& x366  &$  0.45_{-0.23}^{+0.28}$ & $  0.56_{-0.27}^{+0.32}$ & $  <0.15$ & $ 0.73_{-0.42}^{+0.48}$\\
13:37:20& $-$29:53:43   &6.1 & 49.7& x403  &$  0.89_{-0.28}^{+0.33}$ & $  0.46_{-0.29}^{+0.34}$ & $  0.30_{-0.26}^{+0.32}$ & $ 1.03_{-0.46}^{+0.51}$\\
\enddata
\tablenotetext{a}{{\it NuSTAR} source position.}
\tablenotetext{b}{Offset between \nustar\ source position and the \cxo\ catalog by L14.}
\tablenotetext{c}{Source ID by L14.}
\tablenotetext{*}{Astrometry is registered to this source.}
\end{deluxetable*}

\clearpage
\begin{turnpage}
\begin{deluxetable}{crrrrrrrrrrrrrr} 
\tabletypesize{\scriptsize}
\tablecolumns{15} 
\tablewidth{0pc} 
\tablecaption{M83 NuSTAR Point Sources\label{t:nupt3ep}}
\tablehead{ 
 \colhead{}   & \multicolumn{5}{c}{Epoch 1} & \multicolumn{5}{c}{Epoch 2} &  \multicolumn{4}{c}{Epoch 3} \\
 \cline{2-5}  \cline{7-10}  \cline{12-15}\\ 
\colhead{ID$^a$}     & \multicolumn{4}{c}{Count Rates} & 
& \multicolumn{4}{c}{Count Rates} 
&  & \multicolumn{4}{c}{Count Rates} \\
\colhead{L14}    & \colhead{(4--6~keV)}   & \colhead{(6--12~keV)}    & \colhead{(12--25~keV)} 
& \colhead{(4--25~keV)} & & \colhead{(4--6~keV)}   & \colhead{(6--12~keV)}    & \colhead{(12--25~keV)}
& \colhead{(4--25~keV)} & & \colhead{(4--6~keV)}   & \colhead{(6--12~keV)}    & \colhead{(12--25~keV)}
& \colhead{(4--25~keV)} }
\startdata 
x281 & $  0.27_{-0.18}^{+0.21}$ & $  0.62_{-0.23}^{+0.25}$ & $  0.23_{-0.19}^{+0.21}$ & $  1.40_{-0.37}^{+0.39}$ &
          &  $  0.57_{-0.24}^{+0.27}$ & $  0.70_{-0.28}^{+0.32}$ & $  0.51_{-0.27}^{+0.31}$ & $  1.39_{-0.45}^{+0.49}$  &
         & $  0.52_{-0.22}^{+0.22}$ & $  0.45_{-0.21}^{+0.23}$ & $  <0.15$ & $  0.75_{-0.32}^{+0.34}$ \\  
x029 & $  1.18_{-0.23}^{+0.25}$ & $  1.03_{-0.25}^{+0.27}$ & $ <0.16$ & $  2.35_{-0.39}^{+0.42}$ &
&\nodata    & \nodata     &\nodata    & \nodata     &
& $  0.95_{-0.26}^{+0.28}$ & $  1.14_{-0.27}^{+0.29}$ & $  <0.32$ & $  2.08_{-0.41}^{+0.43}$ \\
x048 & $  <0.16$ & $  <0.17$ & $  <0.26$ & $  <0.55$ &
          &  $  0.29_{-0.24}^{+0.29}$ & $  0.33_{-0.25}^{+0.30}$ & $  <0.43$ & $  0.93_{-0.46}^{+0.51}$ &
          & $  0.21_{-0.13}^{+0.19}$ & $  0.36_{-0.16}^{+0.18}$ & $ <0.15$ & $  0.53_{-0.25}^{+0.28}$ \\
x073 & $  0.18_{-0.13}^{+0.16}$ & $  0.22_{-0.16}^{+0.18}$ & $ <0.16$ & $  0.43_{-0.26}^{+0.29}$ &
          &  $  <0.36$ & $  0.48_{-0.30}^{+0.35}$ & $  <0.59$ & $  1.03_{-0.47}^{+0.53}$ &
          & $  0.19_{-0.15}^{+0.18}$ & $  <0.32$ & $  <0.20$ & $  0.35_{-0.26}^{+0.28}$ \\
x145 & $  0.58_{-0.21}^{+0.23}$ & $  0.66_{-0.24}^{+0.26}$ & $  <0.35$ & $  1.48_{-0.38}^{+0.40}$ &
          &  $  0.49_{-0.24}^{+0.28}$ & $  0.57_{-0.27}^{+0.31}$ & $  0.41_{-0.25}^{+0.29}$ & $  1.11_{-0.44}^{+0.48}$  &
          & $  0.45_{-0.23}^{+0.23}$ & $  0.53_{-0.22}^{+0.24}$ & $  <0.25$ & $  0.99_{-0.34}^{+0.37}$ \\
x165 & $  0.22_{-0.17}^{+0.20}$ & $  0.65_{-0.22}^{+0.25}$ & $  <0.22$ & $  1.10_{-0.35}^{+0.37}$ &
          &  $  0.27_{-0.20}^{+0.24}$ & $  0.44_{-0.27}^{+0.32}$ & $  0.29_{-0.25}^{+0.30}$ & $  0.66_{-0.39}^{+0.44}$  &
          & $  0.53_{-0.24}^{+0.22}$ & $  0.48_{-0.20}^{+0.22}$ & $  <0.31$ & $  1.09_{-0.33}^{+0.35}$ \\
x152 & $  0.46_{-0.18}^{+0.20}$ & $  0.53_{-0.20}^{+0.22}$ & $  <0.15$ & $  1.11_{-0.32}^{+0.34}$ &
          &  $  0.43_{-0.24}^{+0.29}$ & $  0.49_{-0.28}^{+0.34}$ & $  0.34_{-0.27}^{+0.32}$ & $  1.03_{-0.46}^{+0.51}$ &
          & $  0.68_{-0.23}^{+0.19}$ & $  0.68_{-0.20}^{+0.22}$ & $  <0.15$ & $  1.30_{-0.30}^{+0.32}$ \\
x155 & $  0.62_{-0.25}^{+0.28}$ & $  0.54_{-0.27}^{+0.31}$ & $  0.33_{-0.26}^{+0.29}$ & $  1.45_{-0.46}^{+0.49}$ &
          &  $  <0.39$ & $  0.60_{-0.30}^{+0.34}$ & $  0.43_{-0.28}^{+0.32}$ & $  0.59_{-0.43}^{+0.48}$ &
          & $ <0.18$ & $  <0.36$ & $ <0.24$ & $ <0.39$ \\
x185 & $  1.46_{-0.27}^{+0.26}$ & $  1.13_{-0.26}^{+0.28}$ & $  <0.39$ & $  2.94_{-0.43}^{+0.45}$ &
          &  $  1.22_{-0.29}^{+0.33}$ & $  0.95_{-0.31}^{+0.35}$ & $  0.82_{-0.30}^{+0.34}$ & $  2.28_{-0.49}^{+0.52}$ &
          & $  1.56_{-0.27}^{+0.32}$ & $  1.41_{-0.26}^{+0.28}$ & $  0.24_{-0.17}^{+0.19}$ & $  3.17_{-0.41}^{+0.43}$ \\
x193 & $  1.23_{-0.42}^{+0.41}$ & $  1.46_{-0.43}^{+0.45}$ & $ <0.20$ & $  2.60_{-0.64}^{+0.66}$ &
          &  $  0.61_{-0.48}^{+0.52}$ & $  1.02_{-0.54}^{+0.59}$ & $  0.87_{-0.52}^{+0.57}$ & $  1.54_{-0.79}^{+0.84}$  &
          & $  1.16_{-0.45}^{+0.52}$ & $  1.01_{-0.42}^{+0.44}$ & $  <0.49$ & $  2.35_{-0.64}^{+0.66}$ \\
x252 & $  0.25_{-0.20}^{+0.23}$ & $  0.62_{-0.28}^{+0.31}$ & $  <0.47$ & $  1.12_{-0.43}^{+0.47}$ &
          &$  0.39_{-0.21}^{+0.25}$ & $  0.83_{-0.32}^{+0.36}$ & $  0.62_{-0.30}^{+0.34}$ & $  1.46_{-0.48}^{+0.52}$ &
          & $  0.48_{-0.24}^{+0.26}$ & $  0.43_{-0.26}^{+0.29}$ & $  <0.36$ & $  1.00_{-0.40}^{+0.43}$ \\
x246 & $  1.13_{-0.49}^{+0.47}$ & $  0.82_{-0.48}^{+0.51}$ & $  <0.40$ & $  2.24_{-0.75}^{+0.78}$ &
          &  $  1.14_{-0.59}^{+0.65}$ & $ <0.61$ & $  <0.61$ & $  1.15_{-0.87}^{+0.94}$  &
          & $  0.81_{-0.47}^{+0.36}$ & $  0.38_{-0.33}^{+0.36}$ & $ <0.23$ & $  1.09_{-0.53}^{+0.56}$ \\
x233 & $  1.63_{-0.49}^{+0.61}$ & $  2.18_{-0.56}^{+0.60}$ & $  0.37_{-0.25}^{+0.28}$ & $  4.24_{-0.83}^{+0.86}$ &
         &  $  1.75_{-0.66}^{+0.73}$ & $  2.49_{-0.70}^{+0.76}$ & $  2.49_{-0.70}^{+0.75}$ & $  4.56_{-1.05}^{+1.11}$ &
         & $  2.26_{-0.63}^{+0.71}$ & $  2.25_{-0.56}^{+0.59}$ & $  <0.59$ & $  4.77_{-0.86}^{+0.89}$ \\
x248 & $  1.20_{-0.25}^{+0.26}$ & $  0.62_{-0.22}^{+0.24}$ & $  <0.32$ & $  2.28_{-0.39}^{+0.41}$ &
          &  $  1.10_{-0.33}^{+0.37}$ & $  0.90_{-0.35}^{+0.40}$ & $  0.74_{-0.33}^{+0.38}$ & $  2.48_{-0.58}^{+0.63}$ &
          & $  1.43_{-0.30}^{+0.27}$ & $  0.58_{-0.21}^{+0.23}$ & $  0.22_{-0.16}^{+0.18}$ & $  2.21_{-0.37}^{+0.39}$ \\          
x258 & $  0.59_{-0.19}^{+0.20}$ & $  1.03_{-0.24}^{+0.26}$ & $  <0.23$ & $  2.25_{-0.37}^{+0.39}$ &
          &  $  0.92_{-0.45}^{+0.55}$ & $  1.32_{-0.58}^{+0.69}$ & $  0.98_{-0.54}^{+0.65}$ & $  3.27_{-0.98}^{+1.08}$ &
          & $  0.90_{-0.24}^{+0.19}$ & $  1.34_{-0.23}^{+0.25}$ & $  0.24_{-0.16}^{+0.19}$ & $  2.44_{-0.35}^{+0.37}$ \\          
x286 & $  0.65_{-0.29}^{+0.32}$ & $  <0.35$ & $  <0.42$ & $  1.03_{-0.46}^{+0.48}$ &
          &  $  0.60_{-0.44}^{+0.50}$ & $  0.54_{-0.39}^{+0.46}$ & $  0.48_{-0.38}^{+0.45}$ & $  1.13_{-0.66}^{+0.72}$ &
          & $  <0.45$ & $ <0.24$ & $ <0.18$ & $<0.46$ \\
x299 & $  6.96_{-0.49}^{+0.49}$ & $  5.74_{-0.46}^{+0.48}$ & $  <0.38$ & $ 13.39_{-0.70}^{+0.72}$ &
          &  $  1.77_{-0.40}^{+0.43}$ & $  0.50_{-0.32}^{+0.36}$ & $  0.34_{-0.30}^{+0.35}$ & $  2.38_{-0.57}^{+0.61}$  &
          & $  4.25_{-0.35}^{+0.53}$ & $  2.88_{-0.34}^{+0.35}$ & $  0.20_{-0.17}^{+0.19}$ & $  7.35_{-0.54}^{+0.56}$ \\
x284 & $  0.22_{-0.16}^{+0.20}$ & $  <0.38$ & $  0.20_{-0.19}^{+0.22}$ & $  0.99_{-0.34}^{+0.36}$ & 
          &  $  0.17_{-0.16}^{+0.21}$ & $  <0.22$ & $  <0.19$ & $  0.61_{-0.36}^{+0.41}$  &
          & $  0.61_{-0.20}^{+0.27}$ & $  0.57_{-0.21}^{+0.23}$ & $  <0.20$ & $  1.19_{-0.34}^{+0.36}$ \\
 x321 & $  1.31_{-0.28}^{+0.28}$ & $  1.05_{-0.28}^{+0.30}$ & $  <0.35$ & $  2.97_{-0.45}^{+0.48}$ &
           &  $  1.51_{-0.33}^{+0.36}$ & $  0.69_{-0.27}^{+0.30}$ & $  0.55_{-0.26}^{+0.29}$ & $  2.27_{-0.49}^{+0.53}$ &
           & $  1.44_{-0.33}^{+0.24}$ & $  0.87_{-0.23}^{+0.25}$ & $  <0.19$ & $  2.24_{-0.37}^{+0.39}$\\  
x366 &\nodata    & \nodata     &\nodata    & \nodata  &
&  $  0.77_{-0.26}^{+0.30}$ & $  0.54_{-0.26}^{+0.30}$ & $  0.37_{-0.24}^{+0.28}$ & $  1.19_{-0.45}^{+0.49}$  &
&\nodata    & \nodata     &\nodata    & \nodata \\
x403 &\nodata    & \nodata     &\nodata    & \nodata  &
   &$0.70_{-0.26}^{+0.31}$ & $  0.37_{-0.27}^{+0.32}$ & $  <0.47$ & $  1.57_{-0.50}^{+0.55}$ &
    &\nodata    & \nodata     &\nodata    & \nodata \\
\enddata
\tablenotetext{a}{Source ID by L14}
\tablenotetext{.}{Notes -- The count rates are given in units of $10^{-3}$~ct~s$^{-1}$.}
\end{deluxetable}
\clearpage
\end{turnpage}
\clearpage

To estimate systematic errors, we simulate a background image by adding Poisson noise onto
a {\tt nuskybgd} background model image and then added simulated point sources at  arbitrary locations.
We compared the input and measured simulated point source fluxes
and concluded that  using an analytic background model 
and an on-axis PSF model introduces $<$10\% systematic uncertainties  
for sources within 5\arcmin\ from the optical-axis.  All 
of our point sources are detected within 5\arcmin\  from the optical axis, and statistical errors are 
typically much larger than 10\%.  Therefore, the systematic error is negligible.   

We also validate our method by comparing it to the tool developed by \citet{wik14b},
which uses both the simulated background images and customized PSF models for individual sources.
 We applied the \citet{wik14b} method to one of our three epochs and confirmed that flux measurements done by both methods
agree with each other within 90\% confidence.

\begin{figure*}
\begin{center}
\includegraphics[angle=0,width=7in]{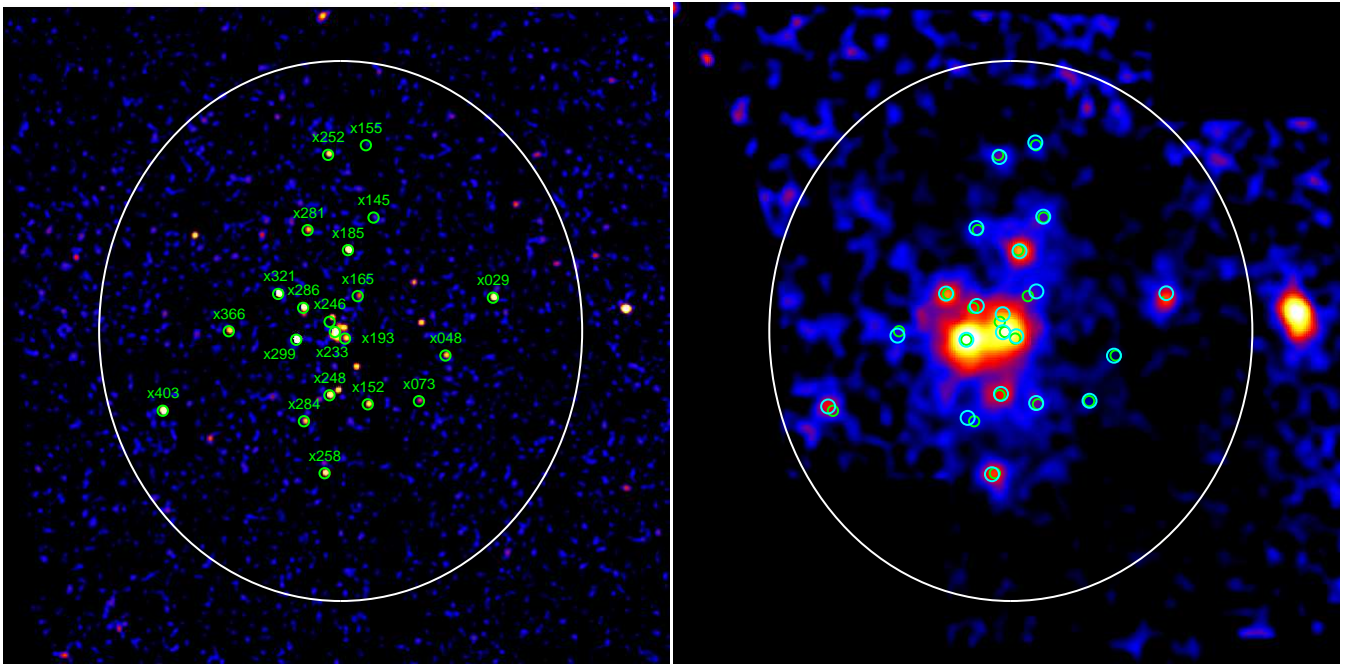}
\end{center}
\caption{Left: The \cxo\ 4--8~keV image taken in June 2014. North is up and East to the left.
The white ellipse indicates the optical extent ($D_{25}$) of the galaxy.  Green circles indicate the \cxo\  positions of the point sources with
\nustar\ flux measurements.  The labels for each point source are the source IDs from L14.  
The \cxo\ image is smoothed with a Gaussian of width 1.5\arcsec.
Right:
The  \nustar\ co-added 4--25~keV image.  Cyan circles are the \nustar\ positions of the point sources. Green circles are the \cxo\ source
positions. We note  that  only 21 out of  39  \cxo\ sources have \nustar\ flux measurements.  Since bright sources are white in the \cxo\ image, it is clear that sources without 
\nustar\ fluxes are fainter.   The \nustar\ image is  smoothed with a Gaussian of 7.5\arcsec width. 
The bright source located outside the galaxy $D_{25}$ in the West has  $log (L_{\rm X}/L_{\rm opt})=$ $-1$
\citep{ducci13}, which is a typical value for background AGNs \citep[see][]{tzanavaris06}. }
\label{f:sourceposition}
\end{figure*}

\subsubsection{Determining the NuSTAR Point Source List}

Since our \nustar\ observations are taken simultaneously or nearly-simultaneously with either \xmm\ or \cxo,
candidate	 point source detection is performed using these higher spatial resolution images  
 to  mitigate source confusion.  
We point out that the \cxo\ and \xmm\ exposure times were sufficiently long to ensure
that the \cxo\ and \xmm\ observation sensitivities in the  4--8~keV band 
exceed the sensitivity of 
our \nustar\ observations in the same band. 
We ran the SAS source detection tool {\tt detect\_chain} with 16 spline nodes and 
an upper limit likelihood of 15  on
the  \xmm\ PN images in the 4--10~keV band, chosen to maximize the energy range
which overlaps with \nustar. 
For the \cxo\ data, 
we  ran {\tt wavdetect} with scale parameters  of 1.0, 1.4, 2.0, 2.25, 4.0, and 5.56  on the 4--8~keV image,  
 since the \cxo\
effective area drops off rapidly above 7 keV.  
Then, we merged the three point source lists from each epoch to create a master point source catalog.

We note that the nuclear region is detected as one source in the \xmm\
observations, whereas it is clearly resolved into at least seven sources in the \cxo\ 4--8~keV
image. 
Our final \cxo/\xmm\ source list contains 39 sources, 38 of which were reported in L14.
One source that was not included in L14 is located outside of the FoV of their \cxo\ observations.

We characterized the \nustar\ point sources  that were detected at or above the 90\% confidence level
in the co-added \nustar\ 4--10~keV image via PSF-fitting.  
In our  PSF-fitting procedure,
 \nustar\ point sources were modeled with 2D Gaussians with widths of $\sim$5\arcsec; these Gaussians were
 convolved with the PSFs to model the sources.  
We chose to use a 2D Gaussian instead of a delta function to take  into account the uncertainties in the PSF shape. 
We note that the  model background parameters constrained from the {\tt nuskybgd} images were fixed during the fitting.  

Using the \xmm/\cxo\ point source position in the master catalog, 
we initially fit the brightest point source to the \nustar\ 4--10~keV image. 
The source positions were derived parameters, so offsets, distortions, and rotations of the images were allowed to fit. 
Once we obtained the best-fit parameters for a source, we added the
 next brightest source in the \xmm/\cxo\ catalog to the model and continued thereafter source by source, moving down in count rate.  
When the  resulting source flux was negative and/or the best-fit position was more than 12\arcsec\ (5 pixel) away from
 the original point source position in the master catalog, the source was considered to be undetected.
 We also discarded sources whose fluxes were not significant at the 90\%
 confidence level. 
Having identified which \cxo\ or \xmm\ point sources are detected
significantly in the 4--10~keV \nustar\ image, we used their positions (fixed during the PSF fitting)
to measure
 their  4--6, 6--12, 12--25, and 4--25~keV fluxes while fixing the source
 positions during PSF-fitting. 
 We note that in our method we should be able to identify and to measure sources 
with fluxes above $\sim$2.3$\times$10$^{-4}$~ct~s$^{-1}$, $\sim$0.80$\times$10$^{-4}$~ct~s$^{-1}$, $\sim$1.1$\times$10$^{-4}$~ct~s$^{-1}$, $\sim$1.1$\times$10$^{-4}$~ct~s$^{-1}$and $\sim$1.5$\times$10$^{-4}$~ct~s$^{-1}$ in the 
4--10, 4--6, 6--12, 12--25, and 4--25~keV, respectively
for a 350~ks exposure.
However, these values may fluctuate depending on source positions due to background
variations and neighboring sources.

\begin{figure*}
\begin{center}
\includegraphics[angle=0,width=2.1in]{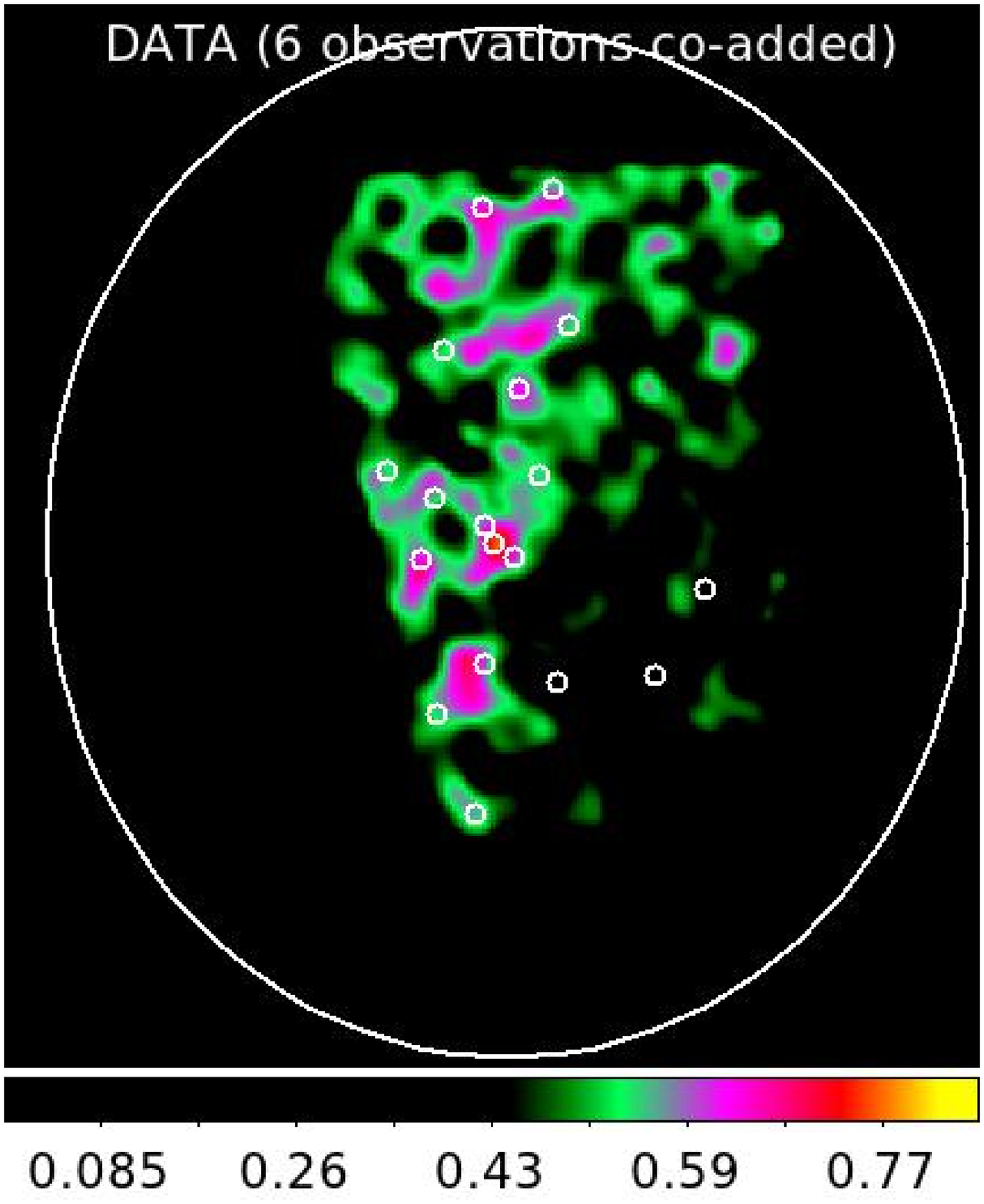}
\includegraphics[angle=0,width=2.1in]{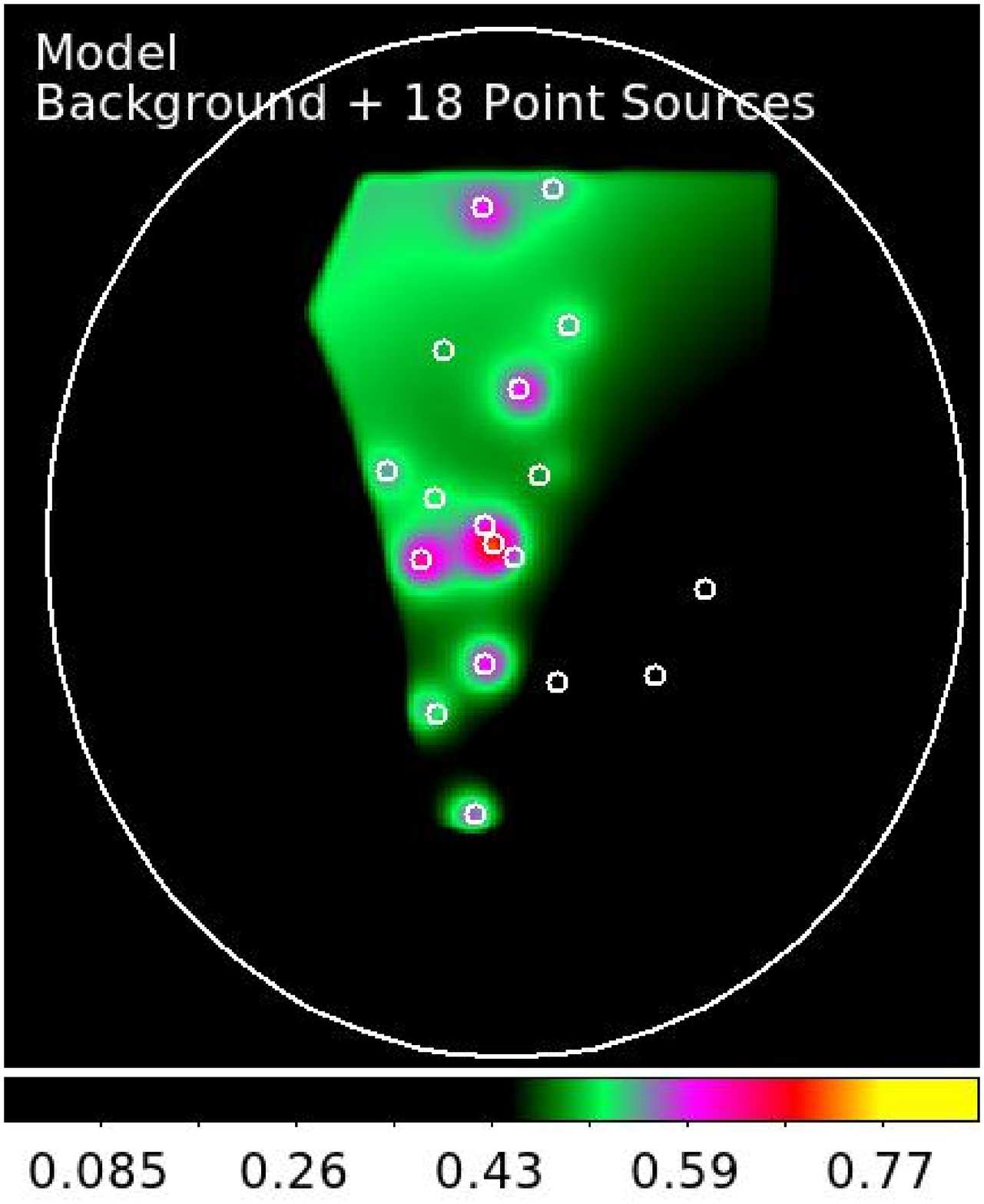}
\includegraphics[angle=0,width=2.1in]{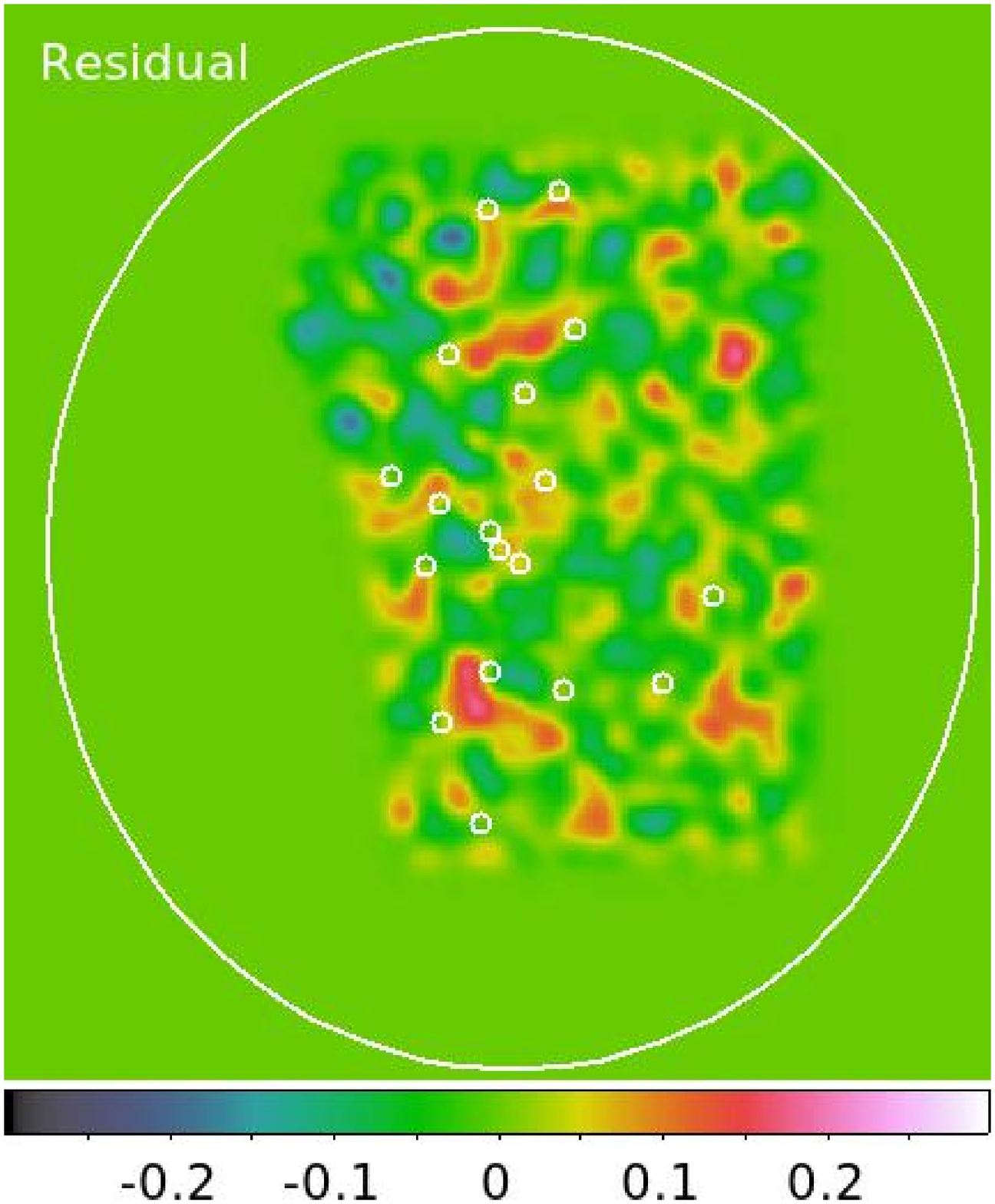}
\end{center}
\caption{From left to right: The \nustar\ 12--25~keV data, the PSF-fitting model, and fit residual images. We note that only the overlapped area from
all six observations is shown. The images are smoothed using a Gaussian of 15\arcsec\ width for display purposes.  
The the hard X-ray emission of the nuclear region shows a peak. The residual image shows no strong excess emission.  This indicates that
no hard sources have been missed in our model. The white ellipse depicts the size of $D_{25}$.  The small circles indicate the \nustar\ sources detected in the 4--10~keV image.
We note that three sources (x403, x29, and x366) are excluded from the images as these sources were missed in one of the three epochs.} 
\label{f:hard_fitting}
\end{figure*}

We point out that  we may have missed very obscured sources that do not appear in the \xmm\ and/or \cxo\ images.
However, we note that  no very bright sources appeared  in the {\it NuSTAR} 12--25~keV image
that were not also detected in the \cxo/\xmm\  observations (see Figure~\ref{f:hard_fitting}).

There were a total of 21 \nustar\ sources with flux measurements.  The results are listed in Table~\ref{t:nustarpt} along with
the source ID of L14.  We also calculate and list position differences between the \nustar\ 
and \cxo\ L14 positions.  
We note that some of the offsets listed in the table are large (6\arcsec\--10\arcsec).  This may be
due to rotation and PSF uncertainties that could not  completely be taken into account  by using a 2D Gaussian model.  
We have compared the \nustar\ 4--10~keV images with the concurrent \xmm\ and \cxo\ images in the same band.
It would be less likely that sources detected in the \xmm\ or \cxo\ images
decreased in flux in the \nustar\ observations at the same energy band
and different \xmm/\cxo\ undetected sources increased in flux to be above the
detection threshold within a few
arcseconds. 
 However,  we cannot exclude the possibility that a measured \nustar\ flux may not be
 truly associated with the identified \cxo\ or \xmm\ source.

Figure~\ref{f:sourceposition} illustrates the location of the \nustar\ detected point sources with \cxo\ source IDs (from L14).
Overall, about half of the \cxo/\xmm\ sources were detected in the
\nustar\ observations, while often sources were not detected due to source confusion.
For example, two \nustar\ sources are identified in the nuclear region with our
PSF-fitting approach, whereas \cxo\  clearly resolved the region into seven point sources (see Figure~\ref{f:m83_nuc}). 
Similarly, there are several sources detected in the \xmm\ and \cxo\
images near source x248 that  are detected as a single source by \nustar.

\begin{figure}
\begin{center}
\includegraphics[angle=0,width=3.5in]{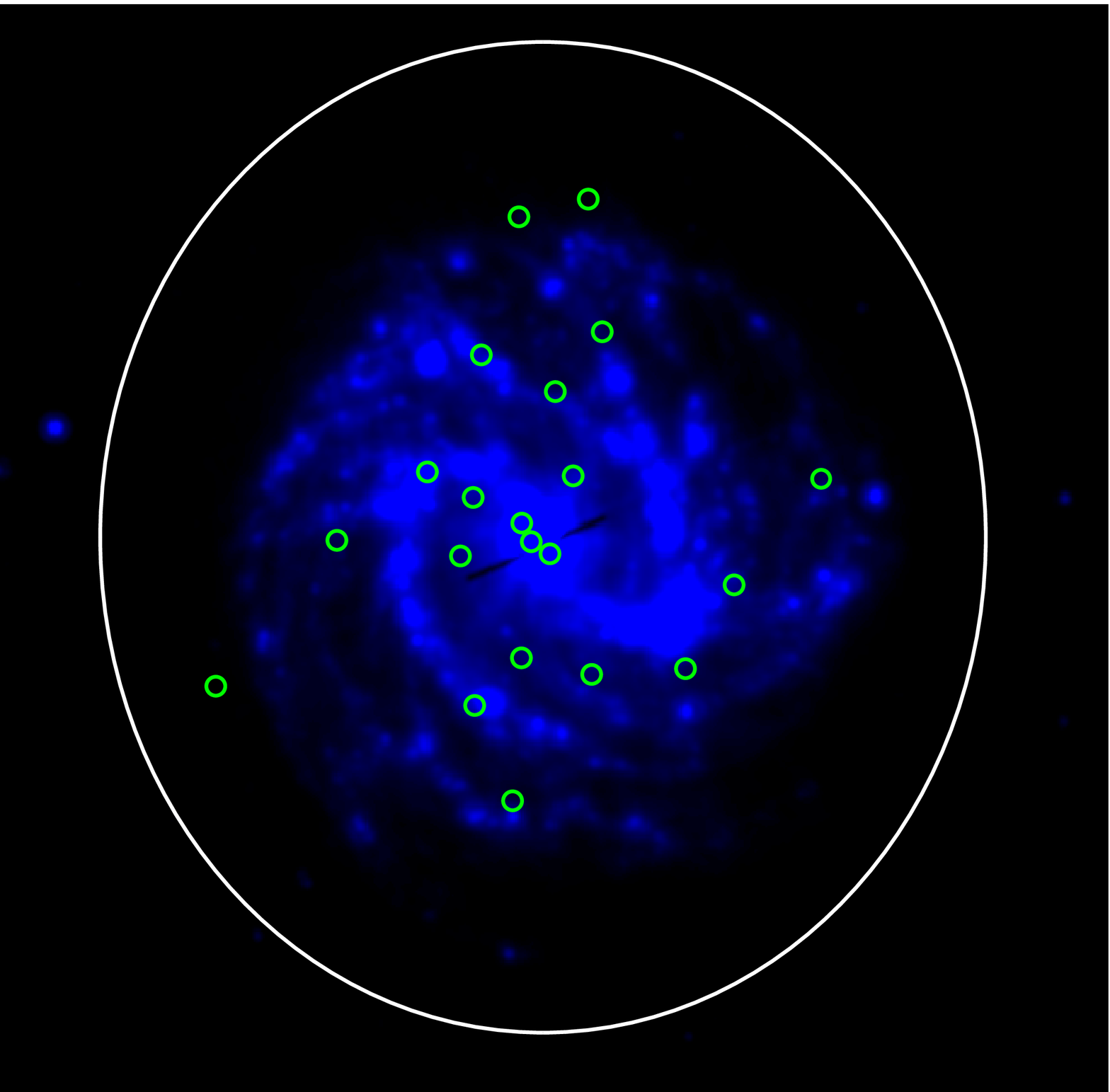}
\end{center}
\caption{The \nustar\ point sources overlaid on the {\it GALEX} FUV image.  About 25\% of the \nustar\ sources are located in  
active star-forming regions.  North is up and East to the left.  The white ellipse depicts the $D_{25}$ size.}
\label{f:FUV_Xpos}
\end{figure}

Figure~\ref{f:FUV_Xpos} shows the correlation between the active star-forming regions and the location of the \nustar\
point sources. About 25\% of the sources are located in intense star-forming regions; however, 
there is no strong relation between the X-ray source locations and the spiral arms.

\begin{figure}
\begin{center}
\includegraphics[angle=0,width=3.5in]{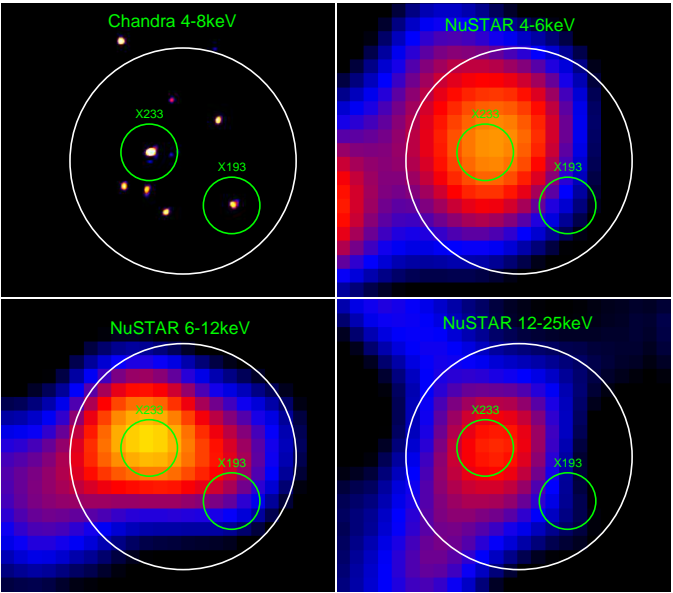}
\end{center}
\caption{The \nustar\ images of the central region of M83. North is up and East to the left. Source x233 is located at the position of
the optical nucleus (nuclear star cluster).  There are $\sim$ 7  X-ray sources resolved in the \cxo\ image.
The white circle indicates a 20\arcsec\ radius aperture used for extracting the nuclear spectrum in Section~\ref{s:nuc}.
Green circles indicate the sources with \nustar\ flux measurements.}
\label{f:m83_nuc}
\end{figure}

\begin{figure}
\begin{center}
\includegraphics[angle=-90,width=3.5in]{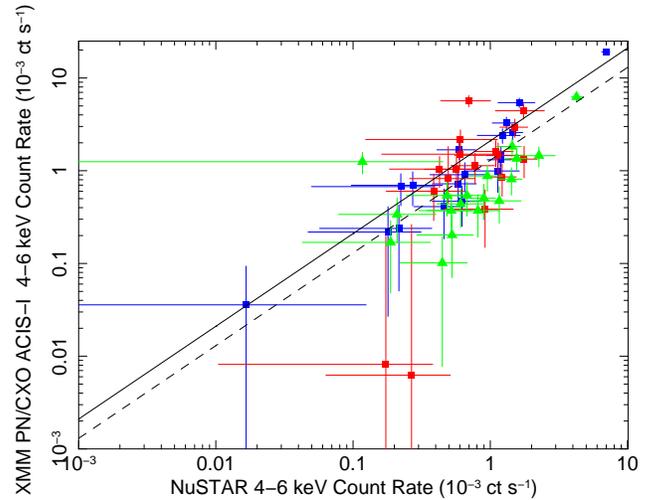}
\end{center}
\caption{Count rates in the 4--6~keV band for the same sources in the \cxo/\xmm\ and \nustar\
observations.  Red and blue data points  depict \nustar~\xmm\ (epochs 1 and 2, respectively) PN observations.  Green points indicate
\nustar~\cxo\ observations. The solid line corresponds to the expected relation on count rates between  \xmm\ and \nustar\ for a spectral shape of a power law
with index $\Gamma$ of 2.  The dash line indicates the relation between \cxo\ and \nustar\ for the same spectral model.}
\label{f:countrates}
\end{figure}

Instead of using simply the net exposure, the source count rates were obtained using the ``effective exposures", listed in Table~\ref{t:nustarpt},  which include corrections based on effective area at the
detector locations,
vignetting,  and removal of detector area affected by stray light. 
To study the long-term X-ray variability in the 4--25~keV band probed by  \nustar\ (see Section~\ref{s:pointv}),
we also measured the \nustar\ fluxes at each epoch.  The measured fluxes are listed in Table~\ref{t:nupt3ep}.
To validate our \nustar\ flux measurements, we compared the 
measured \nustar\ count rates to the \cxo/\xmm\ count
rates in the 4--6~keV energy band for the same sources (see
Figure~\ref{f:countrates}) and found reasonable agreement.
 
\begin{figure*}
\begin{center}
\includegraphics[angle=0,width=7in]{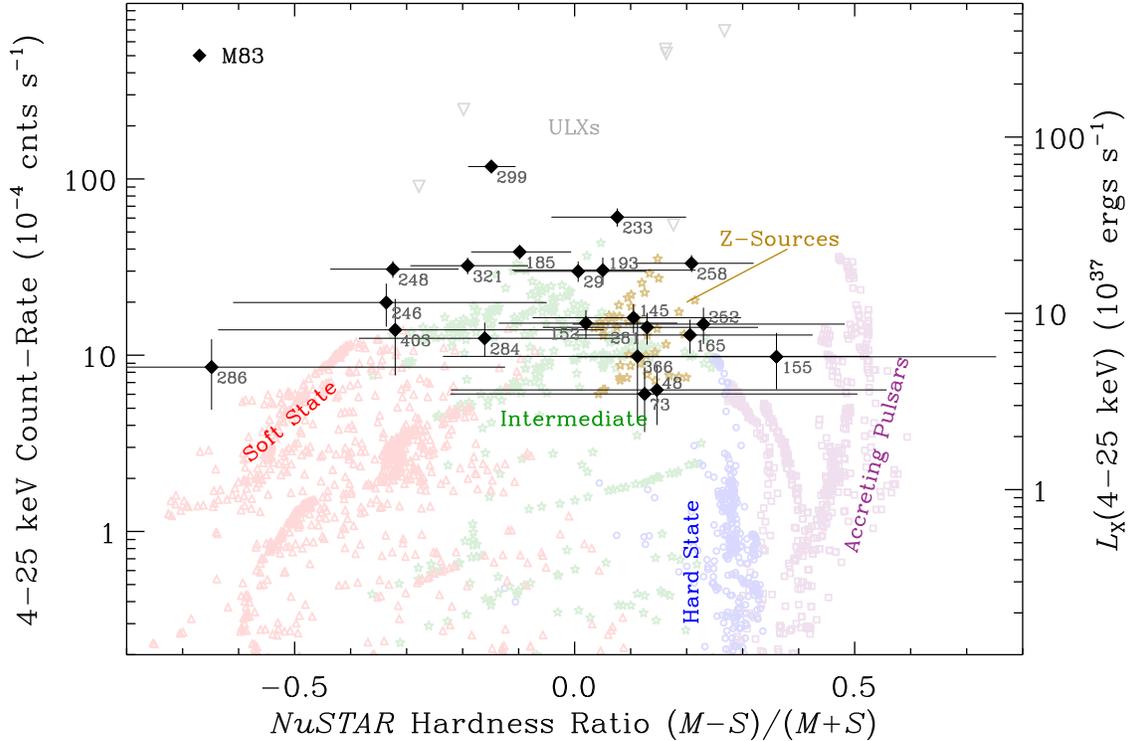}
\end{center}
\caption{Hardness-intensity diagram for M83 point sources plotted with black diamonds (marked by their L14 IDs). The colors and rates are
averaged over the effective exposure times.  Note that the plotted count rates are
scaled to the distance of 4~Mpc.  The count rate of 1$\times$10$^{-4}$~ct~s$^{-1}$ corresponds to  $L_{\rm X}$ $\sim$ 5$\times$10$^{36}$~\ergl. The \nustar\ soft
hardness ratio is defined as (M--S)/(M+S), where the medium band, M, is 6--12~keV, and the soft band, S, is 4--6~keV.
Red, green, blue and magenta symbols depict simulated \nustar\ colors and count rates of seven Galactic LMXBs in the soft, intermediate and hard accretion state and 8 accreting Be pulsar binaries, respectively, based on
{\it RXTE} spectral fits.    Gray triangles and orange stars  indicate the \nustar\ results for ULX sources 
and Galactic NS LMXBs, respectively (see text for details).}
\label{f:colorrate_ave_m83}
\end{figure*}

\begin{figure}
\begin{center}
\includegraphics[angle=0,width=3.5in]{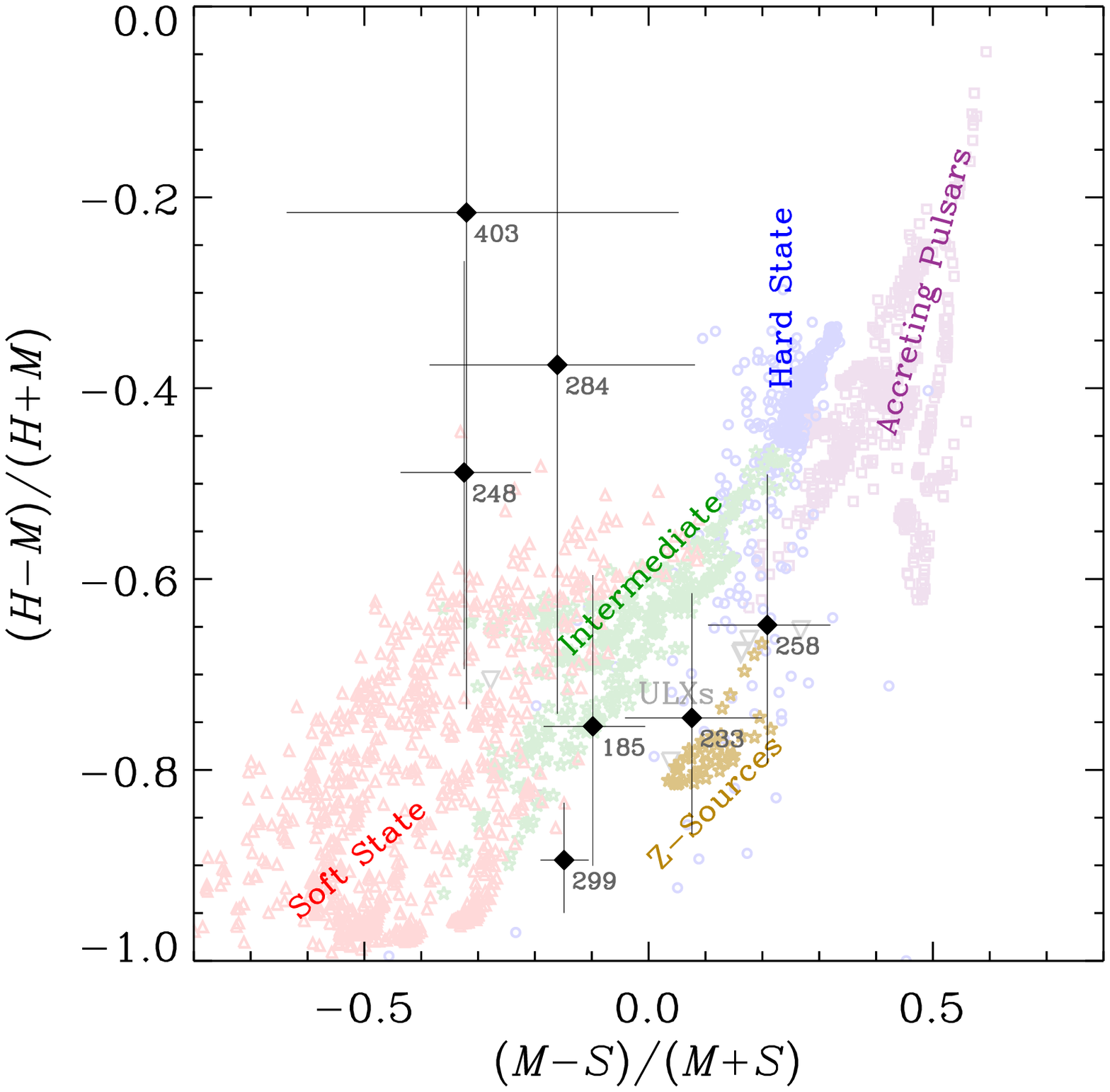}
\end{center}
\caption{The \nustar\ color-color diagram for the M83 point sources (black diamonds marked by their L14 IDs).  The \nustar\ soft color is the same as Figure~\ref{f:colorrate_ave_m83}.
The \nustar\ hard color is defined as (H-M)/(H+M), where the hard band, H, is 12--25~keV.  The sources with an upper limit only
in the hard band are omitted. The symbols are the same in Figure~\ref{f:colorrate_ave_m83}.}
\label{f:colors_ave_m83}
\end{figure}

\subsection{\nustar\ Color Diagnostics for Compact Binaries}
Differentiating BH binaries from NS LMXBs can be done by investigating 
curvature in their spectra, disk temperature and inner radius as we have shown
 for the bright M31 sources with very high S/N ratio spectra when extending up to 20~keV
 \citep{maccarone16}.
Similarly, as black hole binaries move through different accretion states, 
their X-ray spectra change dramatically at $E>$~10~keV, 
much more so than at lower energies  \citep[e.g.,][]{tananbaum72,done07}. 
\nustar's great increase in sensitivity at $>$ 10~keV over past missions has provided the opportunity
to constrain the nature of the resolved sources and to investigate  accretion states of black hole binary populations in nearby 
 galaxies for the first time (see also the NGC~253 study of Wik et al. 2014b).
 The color-intensity and color-color 
diagrams are powerful tools for diagnose point sources which are too faint for spectral analysis.
 In this section, we construct \nustar\ color-intensity and color-color 
diagrams to examine both the identities and accretion states of the X-ray
point sources in M83.
  
Similar to 
\citet{wik14b}, 
we compute the hardness ratios of the point sources
 using the \nustar\ count rates we obtained in the previous subsection. 
 The hardness ratios are defined as follows:  The \nustar\ hard color 
is (H--M)/(H+M), where H is the hard band
 (12--25~keV) count rate, and M is the medium band count rate (6--12~keV).  The \nustar\ soft 
color is (M--S)/(M+S), where S is the soft band (4--6~keV).  
These colors and intensities can be compared with well-studied 
Galactic BH binaries in different accretion states.  Using 
high S/N hard X-ray spectra of Galactic BH binaries from
{\it RXTE} observations, \nustar\ colors and intensities were calculated for seven Galactic BH binaries, over the full range of accretion 
states,  using the \nustar\ response.  Intensities were
computed assuming a distance  of 4~Mpc.  Similarly, a sample of
8 pulsar binaries were included in the same fashion (see A. Zezas et al., in prep for more details).
 
The simulated colors and intensities for the Galactic BH binaries are shown 
for different accretion states. 
Different color symbols  indicate different BH accretion states  in
Figures~\ref{f:colorrate_ave_m83} and \ref{f:colors_ave_m83}.    Specifically, soft, intermediate, and hard accretion states 
correspond to red, green, and blue points, respectively.
Galactic accreting pulsars are plotted with magenta open squares.
We also plot Galactic NS LMXBs using spectral fits from \citet{church12}. Although high S/N ratio spectra analysis distinguish BH binaries from NS, NS 
LMXBs possess similar 
\nustar\ colors and intensities to the Galactic BH binaries in our color diagnostic diagrams.
The NS LMXBs in the plots are all "Z-sources"; another subgroup of NS LMXBs, atoll sources, are
relatively faint in general and located below our M83 observation detection limit in the diagrams.  
We note, however, that the errors on our X-ray colors are large near the detection limit, and higher quality data would be required to clearly differentiate between a high-luminosity atoll source and a Z-source.

In addition to the Galactic BH binaries, accreting pulsars and "Z-sources", several ULXs observed by \nustar\ 
\citep{walton13,bachetti13,rana14,walton14} are included as gray upside-down triangles.
We note that the ULXs are well-separated from the Galactic binary tracks in the color-intensity plot (Figure~\ref{f:colorrate_ave_m83}), as expected given their high luminosities. 
The Galactic NS LMXBs overlap with Galactic BH binaries in the intermediate accretion state in the 
color-intensity plot. However, the two types may be differentiated in the color-color diagram, since
NS binaries are likely to be softer in the \nustar\ hard  band (12--25~keV).  
We plot the M83 \nustar\ sources with filled black diamonds and have annotated their L14 IDs.   
Fourteen out of  the 21 \nustar\ sources have upper limits on the hard band flux.
Therefore, we plot in Figure~\ref{f:colors_ave_m83} only the seven sources with hard \nustar\  detections.

The brightest source in Figure~\ref{f:colorrate_ave_m83} is x299, which appears to have colors similar to the other \nustar\ ULXs. 
The second brightest is x233, which is in the
nuclear region,  and may be confused with neighboring faint sources, making interpretation of this source difficult (see Figure~\ref{f:m83_nuc}).

One source, x286, has very soft X-ray colors placing it in the locus of the
soft-state black holes.
The remaining sources are consistent with intermediate accretion state BH  binaries or NS
LMXBs. 
We note 
x403 has been reported as an ULX candidate before \citep{immler99,stobbart06, ducci13}; the current \nustar\ observed colors and intensities suggest that this source is more consistent with  a soft or intermediate accretion state BH binary. 
Unfortunately, we only have reliable \nustar\ fluxes for x403 from epoch 2 due to the stray light contamination from IC~4239A.
We measure an $L_{\rm X}$ of 2 $\times$ 10$^{39}$~\ergl\ 
in the 0.5--8.0~keV band with \xmm, which would qualify this source as a ULX based on the observed soft X-ray luminosity (see Appendix~\ref{S:app}). However,  the \nustar\ properties are not consistent with other ULXs studied with \nustar.

\subsection{\nustar\ Source Variability}\label{s:pointv}

Accreting binary populations are highly variable and \nustar\ has
detected statistically significant variations in the X-ray binary
populations in the NGC~253 monitoring
campaign on timescales of several weeks to several months \citep{lehmer13,wik14b}.  
 Here we extend this 4--25~keV variability analysis 
to M83 on 4--6~month timescales.

Source x299 changed in brightness significantly over the three epochs of the \nustar\ observations 
(90\% error confidence; see Table~\ref{t:nupt3ep}).
However, the majority of sources do not show statistically significant variability.
This may be due to their large 
statistical errors. 
 In one of the three epochs we have only an upper limit on the luminosity of three sources: x155, x048 and x286.

\begin{figure*}
\begin{center}
\includegraphics[angle=0,width=7in]{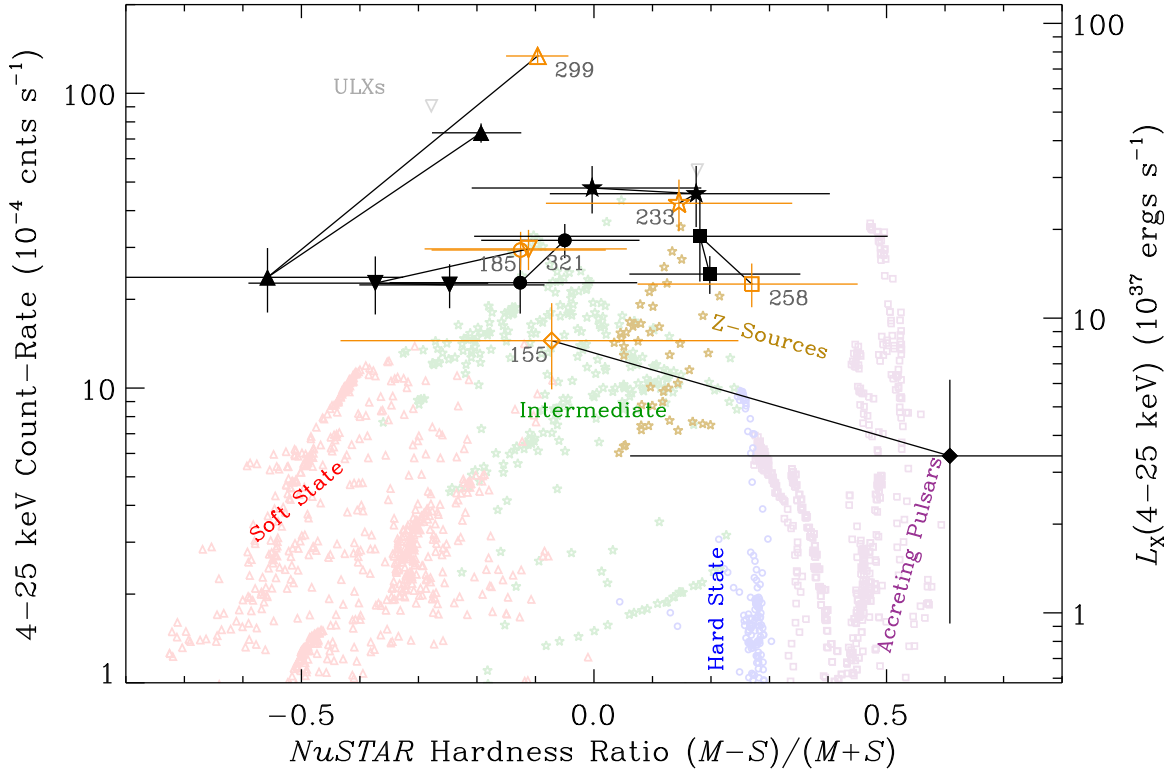}
\end{center}
\caption{Hardness-intensity diagram for the five  brightest  sources and x155.
Colors and symbols for the Galactic sources, and ULXs are the same as in Figure~\ref{f:colorrate_ave_m83}.
M83 sources are indicated in orange or black (epoch 1 or epochs 2 \& 3, respectively) with the L14 source ID.}
\label{f:m83_color_rate_v}
\end{figure*}

We also checked the variability of the \nustar\ hardness ratio (only the soft color).
Figure~\ref{f:m83_color_rate_v} shows the variability of the five brightest sources and x155 
(which shows some color variability) in the color-intensity plot.
Due to the large errors in the colors of the remaining sources, 
they are not shown in the figure. 
The intensities and colors of the sources x233, x185, x321 and x258 did not change over the three epochs.
\ulxa\ exhibits a possible state transition from ULX to soft/intermediate, and back to ULX.
The  \xmm\ spectral analysis (the same datasets we used and an additional
observation taken in July 2014) also suggests a similar state transition for this source \citep{soria15}.
We present the detailed \nustar\ spectral analysis of this sources in the next section. 

The source x155 has the same color and typical intensity as a BH-XRB in the intermediate state in epoch 1; however,
this source was observed to be harder and dimmer 5 months later in epoch 2.  Then, we only obtain an upper limit on its intensity
another 5 months later  in epoch 3.
The source may be changing its accretion state from intermediate, soft, and to hard, with the soft state occurring sometime between
epoch 1 and 2. 
However, within the formal statistical errors, we cannot
rule out that the source stays in the intermediate accretion state while changing in brightness. 
We note that the color and intensity in epoch 2 alone is consistent with an accreting pulsar, yet
this is not likely due to the observation during epoch 1.
We also note that it is possible that the source is a NS binary.  
 
We compare this \nustar\ intensity and color variability to the soft X-ray variabilities obtained from \xmm\ and \cxo\ observations.
Tables~\ref{t:softxct} and \ref{t:softxspec} 
list count rate, luminosity, and spectral parameters from
the \cxo/\xmm\ observations for spectra with at least 200~net counts.
Some sources show variability in luminosity (see Table~\ref{t:softxct}). However, there are no
sources showing a spectral model change from disk blackbody to power law or  vice versa,
 besides x185. For x185, the preferred spectral model
changed from  a power law to disk blackbody model statistically, but the overall spectral 
shape stayed similar having either a power law of 
$\Gamma \sim$1.9 or a disk blackbody with $kT~\sim$ 1.4~keV 
as well as no significant intensity change. 
This is consistent with the observed lack of variability for x185 in
the \nustar\  hardness-intensity plot.

\section{Spectral Analysis} 
One of the \nustar\ starburst galaxies survey goals is to characterize the galaxy-wide  0.5--30~keV
spectral properties of nearby star-forming galaxies, which ultimately is used to calibrate k-corrections
for high-z galaxies (e.g., Lehmer et al. 2016).
In this section, we perform joint spectral analysis using \nustar\ and \xmm/\cxo\
to investigate the spectral properties of the
brightest point sources along with the integrated galaxy spectra.

For the \nustar\ data, the source spectra and spectral response files (i.e., ARFs and RMFs) were created by the script {\tt nuproducts}.
We used the {\it CIAO} {\tt specextract} script to create the corresponding files for the \cxo\ data.
We utilized SAS {\tt arfgen} and {\tt rmfgen} to create the \xmm\
response files.

We note that the background spectra for the \cxo\ data  were created
using a local background extracted from the vicinity of the source regions in the observation.
Local backgrounds were also used for the \xmm\ point source analysis; however, we used the ESAS analysis methods
 for the integrated galaxy analysis.\footnote{http://heasarc.gsfc.nasa.gov/docs/xmm/xmmhp\_xmmesas.html}
   Specifically, the instrument
 background was subtracted, but the soft proton, as well as
  foreground and cosmic background, were modeled during the fitting process.  
  In order to characterize the soft proton, foreground, and cosmic backgrounds, we include 
  a source-free spectrum using a 7\arcmin\--10\arcmin\ annulus around the
  galaxy center during the fitting procedure.  
 As discussed in \citet{wik14}, the \nustar\ background strongly depends on detector position. 
Therefore, a local background is often not adequate for \nustar\
spectral analysis.   
We therefore obtained  \nustar\  background spectra using
{\tt nuskybgd}.

The \nustar\ and \xmm\ source spectra were grouped
to have at least 
 one count per spectral bin \citep[see][]{wik14}, whereas the \cxo\ spectra were not grouped.
The joint fitting was performed in {\it XSPEC} using the C-statistic.
The fitted energy is 0.5--7.0~keV for \cxo\ and \xmm\, (although the 0.5--12.0~keV bandpass was used for the \xmm\
ESAS analysis), and
 3.0--20.0~keV for \nustar.   
 The galaxy is not a strong hard X-ray emitter, and \nustar\ has strong instrumental line emission  between
 20--30~keV; therefore, we exclude data above 20~keV instead of attempting to model the complicated
  line emission. 
Except for the integrated galaxy spectra for epoch 1 (see next section for detail), we co-added spectra of  obsids   
50002043002 and 50002043004 for epoch 1 and obsids 
50002043010 and 50002043012 for epoch 3 using the FTOOL {\tt addascaspec}
 in order to increase the S/N ratios.

\subsection{ULX: \ulxa}
This source has the highest  average luminosity over three epochs in the \nustar\ band (see Table~\ref{t:nustarpt}).
The source  was discovered in 2010 with \cxo\ \citep{soria12}  as a ULX reaching 
$L_{\rm X}\sim$ 5$\times10^{39}$~\ergl\ in 2011 in the 0.3--10~keV band.  It 
has been bright for close to four years showing
some color and intensity variability \citep{soria15}. 
The detailed \xmm\ spectral analysis  applying  various models is published in \citet{soria15}.
We note that the third  observation in  \citet{soria15} was done with \xmm\ in July 2014,
which is about a month after our epoch 3 observation (with \cxo), although the first two \xmm\
observations in their paper are also used in this work.
Within a month, the source showed color variability, becoming harder in the {\it Swift} color band as shown in the Figure~2 of 
 \citet{soria15}.
In  summary, 
the high signal-to-noise \xmm\ spectra taken in 2013 and 2014 show a curvature,
although the \cxo\ spectra in  2010--2011 were dominated by 
a power law component. 
\citet{soria15}  concluded that the \xmm\ spectra were consistent with a slim disk model and 
 that the source is varying between the ultraluminous and soft/high states.
 
In this paper, we jointly fit \nustar\ and \cxo/\xmm\ spectra over the 0.5--20~keV band,
applying four models to this source:
an absorbed power law (PL), absorbed disk blackbody (DB), absorbed power law plus disk blackbody (PL+DB) or
an absorbed broken power law (BPL).
The resulting fit parameters are tabulated in Table~\ref{t:speculx}. 
Overall, an absorbed power-law  model is not preferred for epochs~1 and 3. However, an absorbed power law model and 
an absorbed disk blackbody model give comparable results (Cstat/dof  of 525.1/611 and 524.6/611, respectively)
during epoch 2, when the source is at its faintest level ($L_{\rm X}$ $\sim$ (5--10) $\times$10$^{38}$~\ergl).
The power law model was not preferred during epoch 2 when fitting only \xmm\ data as shown in Table~\ref{t:softxspec}.
  \citet{soria15} also reported that a power law model was rejected during the 2013--2014 \xmm\ observations.
 \nustar\ could be detecting a power law-like spectrum at  10--20~keV,
where  \xmm\ is not sensitive. 

We also tried an absorbed disk blackbody plus power law model.
During epoch 1, the second component (power law) provides statistically significant improvement to the fit ($\Delta C > $100 for
2 additional dof).
However, during epochs 2 and 3, the additional component only improves $\Delta C\sim$15.
The obtained fitted parameter values for epochs 1 and 2 are consistent with \citet{soria15}. 
Although the power-law component for epoch 3 is not well constrained with our \cxo\ spectrum,
our result for epoch 3 agrees with \citet{soria15} within the errors. 

We performed a statistical test using {\it XSPEC} {\tt simftest} with 1000 trials to see
whether or not a power law component is required  during epochs 1 and 2.
The \cxo\ spectrum does not have enough S/N to confirm this, so we do not  perform the test for
the lower S/N epoch 3.
During epoch 1, both \xmm\ only and joint \nustar\ analyses require a power-law component 
in addition to a disk blackbody component.
The power-law component was required at the 1$\sigma$ level during epoch 2 with the \xmm\ only analysis.
However, this power-law component was required at more than 3$\sigma$ 
 when including \nustar\ data.  
We note that below 5~keV, the disk blackbody component dominates. 
Overall, the ULX shows  very similar spectral properties to BH binaries in the intermediate accretion state.

\begin{figure}
\begin{center}
\includegraphics[angle=-90,width=3.5in]{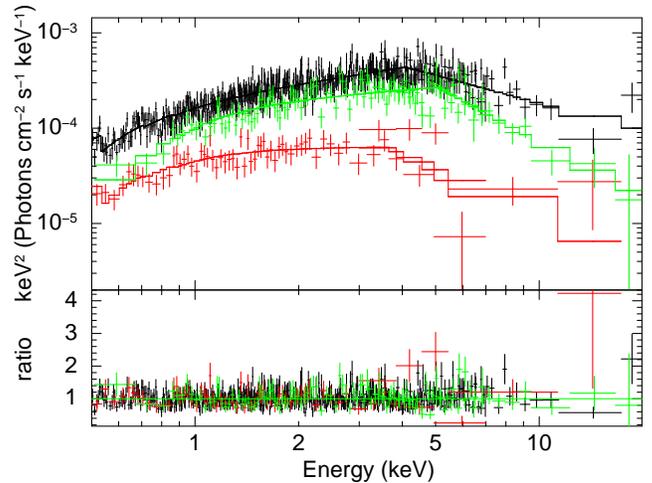}
\end{center}
\caption{The unfolded 0.5--20~keV joint \cxo/\xmm\ and \nustar\  spectra and fitted broken power-law model  in $E^2f(E)$ and data to model ratio of  \ulxa\ (ULX) for epochs 1 (black), 2 (red), and 3 (green).
All spectra show a break at 3--5~keV.  This is consistent with other
ULXs studied with \nustar.}\label{f:ulxjoint}
\end{figure}

In order to compare the \ulxa\ spectral features to the integrated galaxy  in  Section 4.3, we also fitted  with an
absorbed broken power law model, which  also fits  the data well.  
Figure~\ref{f:ulxjoint} shows the unfolded broken power-law
joint \nustar\ and \cxo/\xmm\ spectra of \ulxa\ for each epoch,
exhibiting how spectral shapes are changing over three epochs.

\begin{deluxetable*}{ccrrrrrcc} 
\tabletypesize{\scriptsize}
\tablecolumns{8}
\tablewidth{0pc} 
\tablecaption{ULX (x299) 0.5--20~keV Joint Fit Results\label{t:speculx}} 
\tablehead{ 
 \colhead{}  &  \colhead{}   & \colhead{}  & \colhead{}    & \colhead{}
& \colhead{}  & \colhead{}  & \colhead{log$L_{\rm X}^c$}  
  \\   
     \colhead{}      & \colhead{$N_{\rm H}$}   & \colhead{}     & \colhead{Norm (PL)/$E_{br}^{b}$}  & \colhead{$kT_{in}/\Gamma^b_2$}
& \colhead{Norm} & \colhead{}  & \colhead{(0.5--30~keV)}\\
   \colhead{Model$^a$}   & \colhead{(10$^{21}$~cm$^{-2}$)}   & \colhead{$\Gamma/\Gamma_1^b$}     & \colhead{(10$^{-4}$)/(keV)}  & \colhead{(keV)/ --}
& \colhead{(10$^{-4}$)} & \colhead{cstat/dof} & \colhead{(erg~s$^{-1}$)} }
\startdata 
\multicolumn{8}{c}{{\bf Epoch 1}}\\
 PL& $1.9_{-0.1}^{+0.1}$ & $1.94_{-0.04}^{+0.04}$ &$2.91_{-0.13}^{+0.14}$ &\nodata & \nodata &  1481.0/1342  & 39.7 \\
 DB & $0.4_{-0.4}^{+0.0}$ & \nodata & \nodata &$1.52_{-0.03}^{+0.04}$ & $114.67_{-8.8}^{+9.5}$ & 1341.9/1342  & 39.5 \\
 PL+DB& $0.7_{-0.7}^{+0.5}$ &  $2.01_{-0.26}^{+0.40}$& $0.74_{-0.29}^{+0.40}$& $1.63_{-0.11}^{+0.13}$ & $65.10_{-17.51}^{+18.32}$ & 1238.3/1340 & 39.6 \\    
 BPL &  $0.9_{-0.2}^{+0.2}$ & $1.41_{-0.10}^{+0.10}$& $4.1_{-0.4}^{+0.5}$ &$3.05_{-0.23}^{+0.34}$&$2.04_{-0.14}^{+0.17}$&1237.8/1340& 39.6\\  
 \multicolumn{8}{c}{}\\
\multicolumn{8}{c}{{\bf Epoch 2}}\\
 PL & $2.3_{-0.4}^{+0.5}$ & $2.50_{-0.15}^{+0.16}$ &$0.82_{-0.12}^{+0.14}$ &\nodata & \nodata & 525.1/611  & 39.0\\
DB & $0.4_{-0.4}^{+0.2}$ & \nodata  & \nodata &$0.92_{-0.06}^{+0.06}$ & $148.21_{-38.97}^{+69.33}$ & 524.6/611 &  38.7\\
PL+DB& $0.4_{-0.4}^{+0.2}$ &$1.85_{-1.10}^{+0.88}$&  $0.12_{-0.12}^{+0.41}$&$0.88_{-0.11}^{+0.20}$ & $132.22_{-32.41}^{+40.29}$ &   510.3/609&  38.8\\
BPL &     $2.2_{-0.7}^{+0.7}$ & $2.03_{-0.61}^{+0.27}$&$3.5_{-1.4}^{+0.8}$ &$3.70_{-0.99}^{+1.37}$&$0.60_{-0.13}^{+0.15}$&512.3/609&38.8\\  
 \multicolumn{8}{c}{}\\
\multicolumn{8}{c}{{\bf Epoch 3}}\\
PL & $5.1_{-0.7}^{+0.7}$ & $2.41_{-0.08}^{+0.08}$ &$3.67_{-0.42}^{+0.48}$  &\nodata&\nodata& 788.0/670   & 39.4\\
DB& $0.4_{-0.4}^{+0.3}$ & \nodata &\nodata &$1.53_{-0.06}^{+0.06}$ & $74.32_{-9.90}^{+12.87}$ & 662.2/670 & 39.3\\
PL+DB& $0.4_{-0.4}^{+0.4}$ &$-0.64_{-2.47}^{+2.70}$ & $0.0002_{-0.0002}^{+0.3000}$  &$1.50_{-0.07}^{+0.07}$ & $79.30_{-11.69}^{+22.30}$ & 654.4/668 & 39.6\\
BPL &  $2.2_{-0.7}^{+0.7}$   & $1.74_{-0.14}^{+0.13}$& $5.0_{-0.4}^{+0.4}$&$3.98_{-0.38}^{+0.46}$ &$1.83_{-0.26}^{+0.31}$&665.7/668&39.4\\  
\enddata
\tablenotetext{a}{PL: power law; DB: disk blackbody; BPW: broken power law}
\tablenotetext{b}{Broken power-law model }
\tablenotetext{c}{Intrinsic  luminosity}
\end{deluxetable*}

\subsection{Nucleus}\label{s:nuc}
The nuclear region of M83 also appears as a bright \nustar\ source. 
The high spatial resolution \cxo\ images already reveal that there are several
bright X-ray point sources near  the center of M83 (see Figure~\ref{f:m83_nuc}).  
Yet, these sources are not resolved by the \nustar\ PSF.
At most, we were able to identify via PSF-fitting two \nustar\  (x233 and x193) sources, separated by 15 arcsecond.
Yet, these two sources cannot be separated when extracting spectra.
Therefore, we utilize a 20\arcsec\ radius aperture (shown in Figure~\ref{f:m83_nuc}) to capture both sources
as much flux as possible for spectral analysis while avoiding contamination from \ulxa.  
In order to be consistent with the \nustar\ source spectra,
we extract the  \xmm\ and \cxo\ spectra using the same  20\arcsec\ radius aperture as the \nustar\ aperture.
Aperture corrections were  applied to the \nustar\ and \xmm\ data, since the 
sizes of the PSFs for these telescopes 
are larger than 20\arcsec.  
For the \cxo\ data, we used  area-weighted ARFs, since \cxo's PSF
 size is much smaller than 20\arcsec.

We applied three (PL, DB, BPL) models for the nuclear region but 
added a thermal gas component using a two temperature model ({\it XSPEC} {\tt apec} + {\tt apec}) to account for the strong diffuse soft
X-ray emission seen in the \cxo\ data.
The fitting  results and luminosities for the nuclear region are tabulated in Tables~\ref{t:specnuc} and \ref{t:specnucl}.

For the two temperature model, 
the fitting suffered from a degeneracy between the column density and
 thermal gas temperatures, resulting in a high $N_{\rm H}$ and a lower temperature.  
We, therefore, fixed the column density  $N_{\rm H}$  to the Galactic value \citep[4$\times$10$^{20}$~cm$^{-2}$;][]{dickey90}.
The obtained temperatures are 0.3--0.6~keV and 0.9--1.5~keV (see Table~\ref{t:specnuc}).  
The lower temperature   is  consistent with what \citet{soria02} 
found ($kT$ = 0.6 keV) in the nuclear  16\arcsec\  starburst  region  using  \cxo.
 We note that the second  temperature is  less constrained  for the \cxo\ spectrum due to the poor S/N compared to the \xmm\ data.
The higher temperature ($\sim$1~keV) is higher than those typically measured in the galaxy-wide spectra of nearby star-forming galaxies with \cxo\ \citep{mineo12b}.
 However, we limit our study to the nuclear region in M83, where intense star formation is on-going. 
It is possible that we detect a very hot gas component which is
directly associated with supernova ejecta \citep{strickland07} as seen for example in the central region of M82.

\begin{figure}
\begin{center}
\includegraphics[angle=-90,width=3.5in]{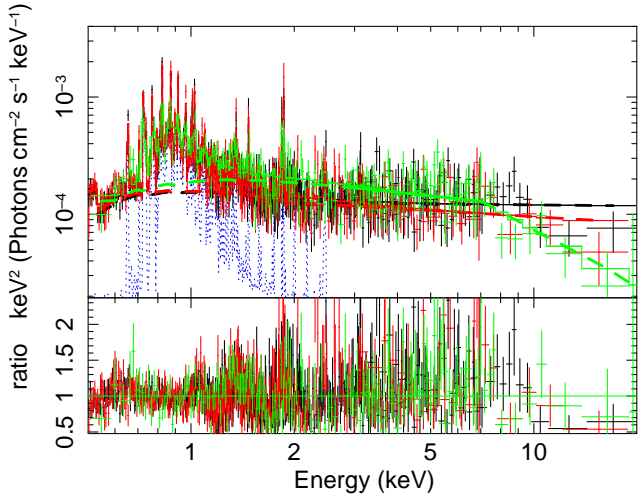}
\end{center}
\caption{The unfolded data  and fitted absorbed broken power law plus thermal gas model in $E^2f(E)$ and ratio for the
nuclear 20\arcsec\ radius region of M83 for epochs 1 (black), 2 (red) and 3 (green).   The thick dashed lines indicate the 
broken power-law component, and the blue dotted line represent the thermal gas emission.
The spectra of the point sources in the nuclear region became steeper at the hard energies, especially  in epoch 3.  However, the statistical uncertainties are large above 10~keV.}\label{f:nucjoint}
\end{figure}

Contrary to \ulxa, a power law model with
a steeper slope of $\Gamma \sim$2.0--2.5 is preferred to a disk blackbody model.  This is similar  to the results obtained for the resolved nuclear sources  x233 and x193
in Table~B using \cxo\ and \xmm\ alone.  
The luminosity in the 0.5--30~keV band does not
change over the three epochs. 

An absorbed broken power law fits well too, but it does not significantly improve the fit statistics.
For epochs 1 and 2,
residuals above 10~keV from a single power law indicate a break. 
When an absorbed broken power law is applied, the  break energy is not constrained, and
 even after fixing it to  the best-fit values (9--10~keV), the parameters of the two photon
 indices  are almost identical, indicating that the break is not supported by the data.
However, for epoch 3 we find strong support for a broken power-law model.
The measured break  energy is $E_{\rm br}$ =$6.91_{-1.96}^{+0.91}$~keV 
as well as a steep power-law
slope at harder energies (Table~\ref{t:specnuc}). 
These  spectral characteristics are similar to what we expect with ULXs, intermediate accretion state BH binaries and
NS binaries (Z-sources) \citep{church12,bachetti13,lehmer13,walton13,walton14}.

Since the nuclear region spectra are a collection of  multiple sources, we do not interpret the fitted parameter values
physically.  
Instead, we are curious how the spectral shape of the nuclear region varies over three epochs.
The unfolded \cxo/\xmm\ and \nustar\ spectra
 of the nuclear region   in $E^2f(E)$ for each epoch are shown in Figure~\ref{f:nucjoint}.
The nuclear spectra  are more or less consistent over the three epochs.  
The spectrum for epoch~3 becomes steeper at harder energies; however, the statistical
uncertainties are very large above 10~keV for all three epochs.

One of \nustar's strengths is to constrain the presence of  an obscured AGN, which may not be
visible in the soft X-ray bandpass with \cxo\ and \xmm.
To test this, we also applied the MyTorus model \citep{murphy09} to the nuclear region spectra.
MyTorus is a model that  self-consistently computes X-ray reprocessing for a power-law
component in a toroidal geometry.   
For the fitting, the spectral model included a power-law component
absorbed by a line-of-sight component (MyTorusZ) along with reflection
components (MyTorusL and MyTorusS) which would likely be due to
reflection from the opposite side of the toroidal obscuration as well as scattering from  gas clouds 
in the vicinity of the nucleus.
We assumed that some scattered emission (i.e., from highly ionized, optically thin
gas near the AGN) is likely present and modeled this as being
the same power law  as used with MyTorus but with the normalization
fixed at 30\% of the main power-law (i.e., 30\% of the AGN flux is
scattered around the torus).  Finally, we included components for
thermal ({\it XSPEC} {\tt apec}) and a second absorbed power-law to account for the
X-ray binaries.  All components included optically thin ``global'' absorption (i.e., toward the M83 system and
within the extranuclear region).  
 
To obtain a better constraint, we jointly fit three epoch spectra.  
We fixed  global $N_{\rm H}$ to  $5 \times 10^{20} \rm\
cm^{-2}$ as we used for the previous fits. The fitted thermal component temperatures were $\sim$ 0.62~keV and $\sim$1.0~keV with 
$L_{\rm X} \sim$  $1.0 \times 10^{39}$ erg~s$^{-1}$.  The best-fit AGN power-law normalization was 0, showing that no AGN
component is required by the data.  To establish the upper-limit on
any intrinsic AGN luminosity, we fixed the AGN power-law slope to 1.8
and set MyTorus $N_{\rm H}$ values at 1,
2 and 4 $\times 10^{24} \rm \ cm^{-2}$.  We then increased the AGN power-law
normalization until there was a change in C-statistic of 4.6.  We then
computed the intrinsic AGN luminosity by setting $N_{\rm H}$ to 0, which
then gave 90\% $L_{10-30\ \rm keV}$ upper-limits of 7.8 $\times
10^{37}$, 8.3 $\times 10^{37}$
and 1.1 $\times 10^{38}$ erg s$^{-1}$ for the three $N_{\rm H}$ values from the spectral fits.  The slope and  luminosities of
the X-ray binary power-law component were $\Gamma\sim$2.2 and $L_{\rm X} \sim$ 4.4 $\times 10^{38}$ \ergl, and 2.1 $\times$ 10$^{39}$~\ergl\ for 10--30~keV and 0.5--30~keV, respectively.

\begin{deluxetable*}{rrrrrrrrrl}
\tabletypesize{\scriptsize}
\tablecolumns{10}
\tablewidth{0pc}
\tablecaption{Nuclear Region 0.5--20~keV Joint Fit Results\label{t:specnuc}}
\tablehead{ \colhead{} & \colhead{} & \colhead{} &
\colhead{} & \colhead{} & \colhead{} &
\colhead{}  &  \colhead{Norm} & \colhead{}  &  \colhead{Norm}\\ 
\colhead{Model$^a$} & \colhead{$N_{\rm H}$} & \colhead{} &
\colhead{$E_{br}^b$} & \colhead{$kT_{in}/\Gamma_2^b$} & \colhead{Norm} & \colhead{$kT_{e1}$} & \colhead{(gas$_{1}$)}
 & \colhead{$kT_{e2}$} &  \colhead{(gas$_{2}$)} \\
\colhead{+2 apec} & \colhead{(10$^{21}$~cm$^{-2}$)} & \colhead{$\Gamma/\Gamma_1^b$} &
\colhead{(keV)} & \colhead{(keV)/--} & \colhead{(10$^{-4}$)} 
& \colhead{(keV)} & \colhead{(10$^{-4}$)}  & \colhead{(keV)} & \colhead{(10$^{-4}$)}}
\startdata
\multicolumn{10}{c}{{\bf Epoch 1}}\\
PL & $0.5_{-0.2}^{+0.2}$ & $2.11_{-0.07}^{+0.07}$ & \nodata
&\nodata &$1.44_{-0.12}^{+0.13}$  &$0.59_{-0.14}^{+0.09}$ &
$0.69_{-0.20}^{+0.31}$&  $0.97_{-0.07}^{+0.16}$ &$0.84_{-0.32}^{+0.21}$  \\
DB & $0.4$ & \nodata  & \nodata &$1.58_{-0.10}^{+0.11}$ &
$37.8_{-8.7}^{+11.2}$ & $0.32_{-0.01}^{+0.01}$ & $1.13_{-0.08}^{+0.08}$&  $0.96_{-0.02}^{+0.02}$ &
 $1.14_{-0.07}^{+0.07}$\\ 
BPL & 0.4 & $2.10_{-0.06}^{+0.05}$ & 9.05
& $2.11_{-0.76}^{+0.94}$ & $1.43_{-0.08}^{+0.08}$ & 
$0.57_{-0.10}^{+0.09}$ & $0.07_{-0.17}^{+0.33}$ & $0.96_{-0.06}^{+0.14}$ & $0.09_{-0.33}^{+0.16}$\\
\multicolumn{10}{c}{{}}\\
\multicolumn{10}{c}{{\bf Epoch 2}}\\
PL & 0.4 & $2.21_{-0.06}^{+0.06}$ & \nodata
& \nodata &$1.53_{-0.11}^{+0.11}$ &$0.62_{-0.14}^{+0.09}$ &
$0.79_{-0.33}^{+0.28}$& $0.98_{-0.10}^{+0.18}$&$0.73_{-0.36}^{+0.30}$ \\
DB & 0.4 & \nodata & \nodata
&$1.44_{-0.12}^{+0.13}$ & $51.6_{-15.1}^{+20.8}$ & $0.31_{-0.02}^{+0.02}$ &
$1.20_{-0.11}^{+0.11}$&   $0.94_{-0.03}^{+0.03}$ & $1.41_{-0.10}^{+0.10}$\\
BPL & 0.4 & $2.20_{-0.07}^{+0.07}$ & 10.3& $3.84_{-1.70}^{+2.76}$ 
& $1.51_{-0.12}^{+1.13}$ & $0.61_{-0.14}^{+0.09}$ & $0.76_{-0.30}^{+0.31}$ 
& $0.96_{-0.09}^{+0.18}$&$0.96_{-0.36}^{+0.28}$  \\
\multicolumn{10}{c}{{}}\\
\multicolumn{10}{c}{{\bf Epoch 3}}\\
PL & 0.4 & $2.37_{-0.08}^{+0.08}$ &\nodata  & \nodata & $2.37_{-0.26}^{+0.28}$
& $0.54_{-0.17}^{+0.48}$ & $0.37_{-0.33}^{+0.54}$
& $0.96_{-0.64}^{+0.38}$ &$0.77_{-0.34}^{+0.21}$ \\
DB & 0.4 & \nodata & \nodata
&$1.91_{-0.19}^{+0.23}$ & $16.0_{-6.8}^{+1.5}$  & $0.62_{-0.06}^{+0.15}$ & 
$1.30_{-0.14}^{+0.15}$& $1.57_{-0.32}^{+0.25}$& $2.14_{-0.49}^{+0.27}$\\
BPL & 0.4 & $2.21_{-0.14}^{+0.10}$ & $6.91_{-1.96}^{+0.91}$
& $3.64_{-0.80}^{+0.92}$ & $2.06_{-0.30}^{+0.29}$  & $0.54_{-0.19}^{+0.25}$ 
& $0.53_{-0.32}^{+0.35}$ & $1.00_{-0.12}^{+0.17}$  &$0.83_{-0.61}^{+0.24}$ \\
\enddata
\tablenotetext{a}{PL: power law; DB: disk blackbody; BPL: broken power law; A solar abundance \citep{anders89} is assumed for the {\tt apec} component.} 
\tablenotetext{b}{Broken power-law model}
\end{deluxetable*}

\begin{deluxetable}{rrrrr}
\tabletypesize{\scriptsize}
\tablecolumns{5}
\tablewidth{0pc}
\tablecaption{Nuclear Region Fit Statistics and 0.5--30 keV  Luminosities\label{t:specnucl}}
\tablehead{ \colhead{Model$^a$} &  \colhead{} & \colhead{log $L_{\rm X}^{obs}$} & \colhead{log $L_{\rm X}^{int}$} &
\colhead{log $L_{\rm X~gas}^{int}$}\\ 
\colhead{$+$2 apec} & \colhead{C-stat/dof}  & \colhead{(\ergl)} & \colhead{(\ergl)} &
\colhead{(\ergl)}}
\startdata
\multicolumn{5}{c}{{\bf Epoch 1}}\\
PL & 1017.6/1112 & 39.5 & 39.5 & 39.0 \\
DB & 1266.9/1113 & 39.4 & 39.5 & 39.2 \\
BPL & 1014.8/1112 & 39.4 & 39.5 & 39.0 \\
\multicolumn{5}{c}{{}}\\
\multicolumn{5}{c}{{\bf Epoch 2}}\\
PL & 793.4/905 & 39.4 &39.5 &39.0 \\
DB &  921.6/905 & 39.4 &39.4 &39.2 \\
BPL &790.9/904 & 39.4 & 39.5 &39.0\\
\multicolumn{5}{c}{{}}\\
\multicolumn{5}{c}{{\bf Epoch 3}}\\
PL & 721.6/674  & 39.5&  39.5 &38.9\\
DB & 775.1/674 & 39.4& 39.5 & 39.2 \\
BPL & 703.5/672&39.5 & 39.5 &  39.0 \\
\enddata
\tablenotetext{a}{PL: power law; DB: disk blackbody; BPL: broken power law}
\end{deluxetable}

\subsection{The 0.5--20~keV Integrated Galaxy Spectrum}

The integrated galaxy spectra were extracted using the  $D_{25}$ area covered by
the observations, excluding data within regions of significant stray light (c.f., Figure~\ref{f:exposure}). 
Due to the different roll angles between  the observations, the size of the extraction area
(see Figure~\ref{f:fov}) for the
galaxy spectra were different among the six spectra (FPMA and FPMB for
the three epochs).  
We note that all the  \nustar\ sources are included  in the \nustar\ spectra for all three epochs.
In our NGC~253 work, we found that a small number of brightest sources dominate 
 the integrated galaxy spectrum above 10~keV, which is characterized well as a broken power law showing
 rapid decrease at harder energies \citep{wik14b}. 
In this section, we examine which sources contribute to the total galaxy emission in the
\nustar\ band and characterize the shape of the broadband X-ray spectrum of M83.

\begin{figure}
\begin{center}
\includegraphics[angle=-90,width=3.5in]{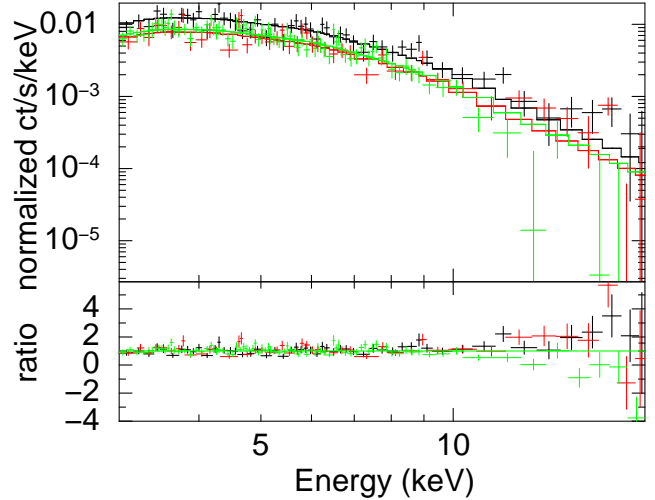}
\end{center}
\caption{The \nustar\ spectra of the integrated M83 galaxy. Black, red, and green indicate the spectrum for epochs 1, 2, and 
3, respectively.  A broken power-law model was fitted simultaneously tying the power law indices.
The obtained parameters are $\Gamma_1 \sim$ 2.3 (for the lower energy) and $\Gamma_2 \sim$ 
for 3.0 (for the higher energy) with the break at
$\sim$6~keV. These values are consistent with the starburst galaxy, NGC~253.
The spectra have been rebinned to achieve at least 4$\sigma$ for  display purposes.}\label{f:nustargalaxy}
\end{figure}

First, we performed the
 \nustar-only spectral analysis, which  we show in Figure~\ref{f:nustargalaxy}.
In general, all three epochs can be characterized well with a power law of $\Gamma \sim$ 2--3, 
although the integrated M83 \nustar\  spectra hint at a rapid decrease at harder
energies, meaning  that a broken power law would result in a good fit  but with poorly
constrained parameters  in epochs 2 and 3.
We also jointly fitted all three epoch spectra tying the power law indices  but
varying the normalization component.
This would average the spectral shape but increase the S/N ratio.    
A simple power-law model gives a photon  index of 2.6$\pm$0.1 with 1359/1268 (C-stat/dof). 
A broken power-law model gives a better fit with 1342/1266 (C-stat/dof), but the fit statistic is not
significantly improved. 
 The obtained fit  parameters are
 $\Gamma_1 =$ 2.3$\pm$0.2  and $\Gamma_2$= 3.0$^{+0.4}_{-0.2}$ with a break at
6.1$^{+1.6}_{-0.9}$~keV.  These values are consistent with the integrated \nustar\ spectrum of NGC~253 \citep{wik14b}.
Figure~\ref{f:nustargalaxy} shows the \nustar\ spectra for the integrated galaxy with the broken power-law model and data over model ratio.

For the joint \nustar\ and \xmm/\cxo\ spectral analysis, we fit two models, either a power law or a broken power law,   to the
integrated galaxy spectra. 
Since the galaxy contains a significant amount of  hot gaseous emission due to
star forming  activity, a thermal plasma component  is also necessary.  
The resulting  fit parameters  and luminosities are listed in Tables~\ref{t:specgal} and \ref{t:specgall}.

\begin{figure}
\begin{center}
\includegraphics[angle=-90,width=3.5in]{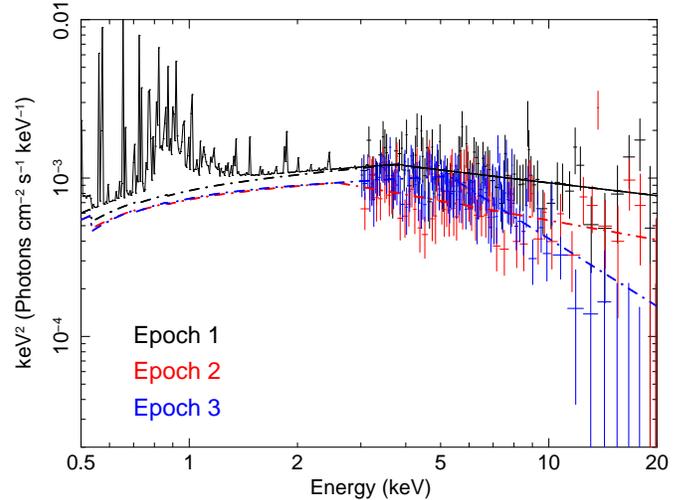}
\end{center}
\caption{Black line indicates an unfolded model  in $E^2f(E)$ for the
whole galaxy for epoch 1, when an absorbed broken power law and two temperature
model is applied.  The dash-dot lines present the broken power law
component for epochs 1 (black), 2 (red) and 3 (green).
The \nustar\  data points are also plotted in this figure to present large statistical uncertainties 
at harder (especially above 10~keV) energies.}\label{f:xspecgalaxy}
\end{figure}

The measured diffuse gas temperatures are 0.65--0.75~keV and 0.2--0.3~keV for both models.  The
0.7~keV gas is consistent with the nuclear region; however, the second temperature is
much lower compared to what we found in the spectral-fit result for the nuclear region ($\sim$ 1~keV).
These two temperatures are consistent with what  \citet{mineo12b} found in the nearby star-forming
galaxies.  
Unsurprisingly, the thermal gas component  contribution is unchanged over three epochs. 
For all three epochs, the absorbed broken power-law model fits better than a single power-law model.
The fit parameters for the broken power-law model are  $\Gamma_1$ of $\sim$1.8 and 
$\Gamma_2$ of 2--3.5 with a break energy of 2--5.5~keV.
We note that the power-law index for the harder energies and the break energy are somewhat difficult
to be constrained due to low number counts in the \nustar\ spectra. 
Since the background/foreground components in the \xmm\ spectra are not subtracted but modeled (see the 
ESAS threads and Section 4 above), we only show the  models (with a broken power law) for the
soft X-ray band but with the unfolded \nustar\ data points to show  the statistical uncertainties  in the hard X-ray energies in
Figure~\ref{f:xspecgalaxy}.

\begin{figure}
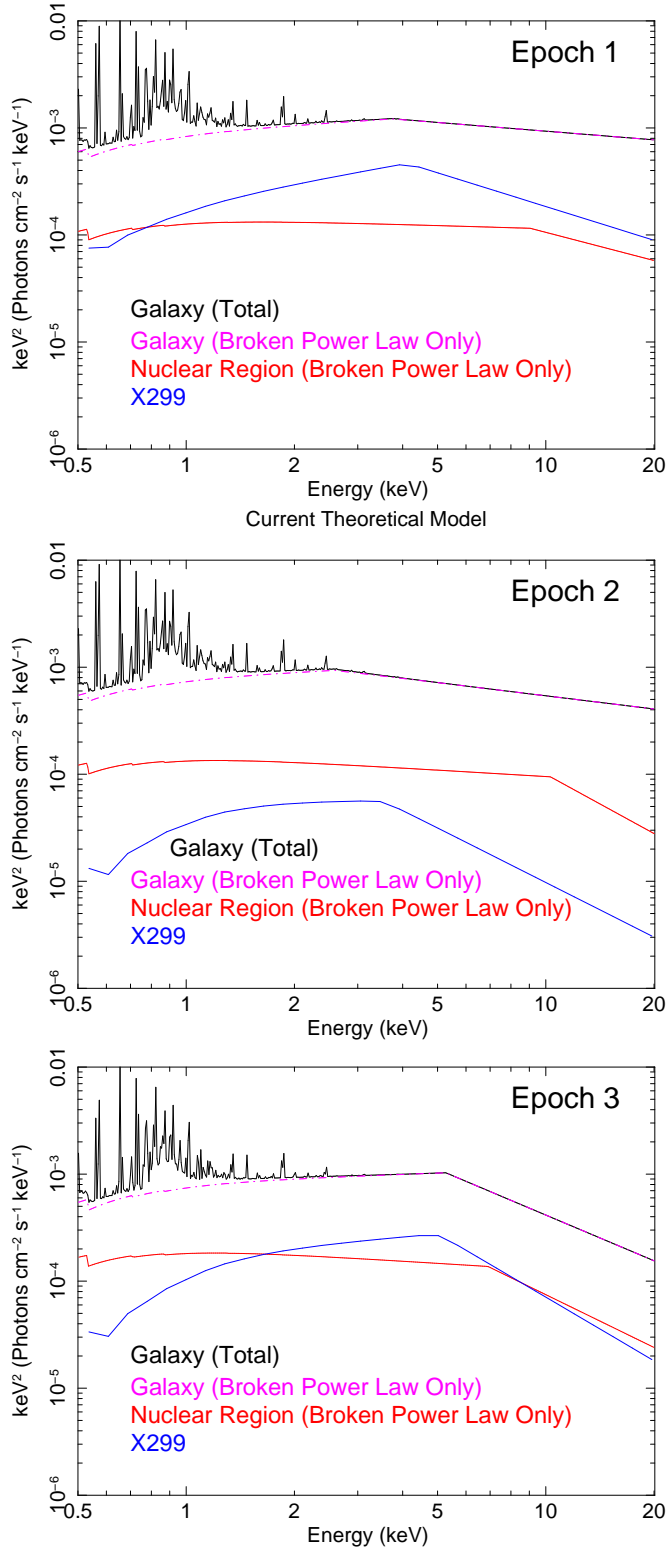

\begin{center}
\includegraphics[angle=-90,width=3.5in]{f17a.eps}
\includegraphics[angle=-90,width=3.5in]{f17b.eps}
\includegraphics[angle=-90,width=3.5in]{f17c.eps}
\end{center}
\caption{From top to bottom: the unfolded  models in $E^2f(E)$ for the integrated galaxy including the thermal component (black), nuclear region without thermal component (red), and \ulxa\ (blue) for
epochs 1, 2, and 3.  The integrated galaxy spectra show very similar shapes to the nuclear spectra.}\label{f:galaxy}
\end{figure}

Figure~\ref{f:galaxy} presents the  model for the integrated galaxy spectrum for each epoch,
along with the  models of \ulxa\ (blue) and the nucleus (red, without thermal gas)
to show how these components contribute to the integrated galaxy emission.  
In all cases, we show the absorbed broken power-law models.
Source \ulxa\ shows about a factor of six change in its 0.5--30~keV
luminosity, and this source effectively determines the shape of  the integrated galaxy spectra, especially
the location of the break energy.
Specifically, the integrated galaxy spectra break around the energies where \ulxa\ shows breaks
for epochs 1 and 3  when
 the source is bright.  During epoch 2,
 \ulxa\ became  dim, and the integrated galaxy spectrum  shows a break  at $\sim$2.5~keV, whereas the nuclear region and \ulxa\ show breaks at higher energies.

Table~\ref{t:specLX} lists the observed X-ray luminosities for 
the point source contribution 
to the galaxy, nuclear region, and x299 in the 0.5--30~keV and 10--30~keV.
The nuclear region and x299 contribute about 15--30\% of the total galaxy luminosity at 10--30~keV.
The remaining flux is likely to be contributed by fainter point sources. 

\begin{deluxetable*}{crrrrrrrrrrl}
\tabletypesize{\scriptsize}
\tablecolumns{10}
\tablewidth{0pc}
\tablecaption{Integrated Galaxy 0.5--20~keV Joint Fit Results\label{t:specgal}}
\tablehead{ \colhead{} & \colhead{} & \colhead{} &
\colhead{} & \colhead{} & \colhead{} &
\colhead{}  &  \colhead{Norm} & \colhead{}  &  \colhead{Norm}\\ 
\colhead{Model$^a$} & \colhead{$N_{\rm H}$} & \colhead{} &
\colhead{} & \colhead{} & \colhead{Norm} & \colhead{$kT_{e1}$} & \colhead{(gas$_{1}$)}
 & \colhead{$kT_{e2}$} &  \colhead{(gas$_{2}$)} \\
\colhead{+2 apec} & \colhead{(10$^{21}$~cm$^{-2}$)} & \colhead{$\Gamma/\Gamma_1^b$} &
\colhead{$E_{br}^b$} & \colhead{$\Gamma_2^b$} & \colhead{(10$^{-4}$)} 
& \colhead{(keV)} & \colhead{(10$^{-4}$)}  & \colhead{(keV)} & \colhead{(10$^{-4}$)}}
\startdata
\multicolumn{10}{c}{{\bf Epoch 1}}\\
PL & $0.5_{-0.1}^{+0.2}$ & $1.93_{-0.03}^{+0.03}$ &\nodata
&\nodata & $9.46_{-0.46}^{+0.43}$  &$0.74_{-0.02}^{+0.02}$ &
$3.72_{-0.17}^{+0.44}$&  $0.20_{-0.01}^{+0.01}$ &$4.62_{-1.40}^{+2.07}$  \\
BPL & 0.4 & $1.76_{-0.08}^{+0.04}$ & $ 3.74_{-0.51}^{+0.42}$
& $2.27_{-0.12}^{+0.07}$ & $9.22_{-0.41}^{+0.34}$
  & $0.75_{-0.01}^{+0.01}$ & $3.79_{-0.40}^{+0.19}$  &  $0.20_{-0.01}^{+0.01}$&
 $4.86_{-0.36}^{+0.71}$  \\
\multicolumn{10}{c}{}\\
\multicolumn{10}{c}{{\bf Epoch 2}}\\
PL & $1.1_{-0.4}^{+0.3}$ & $2.23_{-0.05}^{+0.04}$ & \nodata & \nodata &$10.12_{-0.63}^{+0.51}$  & $0.72_{-0.02}^{+0.02}$ & $4.53_{-0.75}^{+0.66}$ & $0.19_{-0.01}^{+0.01}$ & $10.3_{-4.5}^{+4.6}$\\
BPL & $0.4_{-0.0}^{+0.3}$ & $1.85_{-0.11}^{+0.11}$ & $2.58_{-0.25}^{+0.44}$
& $2.33_{-0.08}^{+0.12}$ & $8.12_{-0.46}^{+0.53}$
 & $0.75_{-0.02}^{+0.02}$ & $3.75_{-0.19}^{+0.53}$ &  $0.20_{-0.01}^{+0.01}$&
 $4.98_{-0.41}^{+2.41}$  \\
\multicolumn{10}{c}{}\\
\multicolumn{10}{c}{{\bf Epoch 3}}\\
PL & 0.4 & $2.20_{-0.05}^{+0.05}$ & \nodata & \nodata & $10.63_{-0.50}^{+0.25}$  &  $0.68_{-0.07}^{+0.08}$ & $1.19_{-1.24}^{+1.58}$ &$0.27_{-0.10}^{+0.08}$&
 $4.98_{-0.41}^{+2.41}$  \\
BPL & 0.4 & $1.88_{-0.07}^{+0.07}$ & $5.27_{-0.49}^{+0.62}$
& $3.42_{-0.28}^{+0.40}$ & $8.44_{-0.57}^{+0.59}$
& $0.71_{-0.06}^{+0.08}$ & $3.16_{-0.99}^{+0.75}$ & $0.23_{-0.09}^{+0.10}$&
 $3.26_{-1.65}^{+4.10}$ \\
\enddata
\tablenotetext{a}{PL: power law; BPL: broken power law; A solar abundance \citep{anders89} is assumed for the {\tt apec} component.} 
\tablenotetext{b}{Broken power law model}
\end{deluxetable*}

\begin{deluxetable}{rrrrr}
\tabletypesize{\scriptsize}
\tablecolumns{5}
\tablewidth{0pc}
\tablecaption{Integrated Galaxy Fit Statistics and 0.5--30 keV  Luminosities\label{t:specgall}}
\tablehead{ \colhead{Model$^a$} &  \colhead{} & \colhead{log $L_{\rm X}^{obs}$} & \colhead{log $L_{\rm X}^{int}$} &
\colhead{log $L_{\rm X~gas}^{int}$}\\ 
\colhead{$+$2 apec} & \colhead{C-stat/dof}  & \colhead{(\ergl)} & \colhead{(\ergl)} &
\colhead{(\ergl)}}
\startdata
\multicolumn{5}{c}{{\bf Epoch 1}}\\
PL & 4391.3/4987 & 40.3 & 40.3 & 39.6 \\
BPL & 4343.5/4985 & 40.3 & 40.3 & 39.7 \\
\multicolumn{5}{c}{{}}\\
\multicolumn{5}{c}{{\bf Epoch 2}}\\
PL & 3677.0/4970 & 40.2 &40.3 &39.8 \\
BPL &3631.3/4937& 40.2 & 40.2 &39.6\\
\multicolumn{5}{c}{{}}\\
\multicolumn{5}{c}{{\bf Epoch 3}}\\
PL &1023.2/862  & 40.2&  40.2 &39.4\\
BPL & 862.5/860 &40.1  & 40.2 &  39.6 \\
\enddata
\tablenotetext{a}{PL: power law; BPL: broken power law}
\end{deluxetable}

\begin{deluxetable*}{crrrrrr}
\tabletypesize{\scriptsize}
\tablecolumns{7}
\tablewidth{0pc}
\tablecaption{Observed Broken Power Law  X-ray Luminosities\label{t:specLX}}
\tablehead{ \colhead{Epoch} & \multicolumn{2}{c}{Galaxy} & \multicolumn{2}{c}{Nucleus} &
\multicolumn{2}{c}{x299}\\
\colhead{}  & \colhead{0.5--30~keV} & \colhead{10--30~keV}  &  \colhead{0.5--30~keV}  & \colhead{10--30~keV} &  \colhead{0.5--30~keV} & \colhead{10--30~keV}\\
\colhead{}  & \colhead{(10$^{39}$~\ergl)} & \colhead{(10$^{39}$~\ergl)}  &  \colhead{(10$^{39}$~\ergl)}  & \colhead{(10$^{39}$~\ergl)} &  \colhead{(10$^{39}$~\ergl)} & \colhead{(10$^{39}$~\ergl)}} 
\startdata
1 &  15.2 & 3.5   & 1.7 & 0.30 & 4.0 & 0.45\\
2 &  10.7 & 1.9    & 1.6 & 0.19 & 0.52 & 0.02\\
3 &  10.4 & 0.93 & 2.1 & 0.15 & 2.0 & 0.12\\
\enddata
\end{deluxetable*}

\section{Discussion}\label{S:discussion}

\subsection{Nature of the NuSTAR Point Sources in M83}
We have identified a total of 21 point sources in M83 with \nustar\ and all
of these sources are listed in the \cxo\ point source catalog compiled by L14.
There are no bright hard X-ray sources that appear  in the \nustar\ observations but do not have
\cxo\ counterparts.
One (x165) of these sources is classified as a background AGN based on
 its association with an optical counterpart (L14). 
It is expected that background AGN would have \nustar\ colors similar to intermediate-state
BH binaries and indeed this is the case for x165.
However, for x165, the \nustar\ source has a large positional offset (6.7\arcsec)  from the nearest
 \cxo\ counterpart, and may be a false match.
We note that the number
of expected background sources above a 4--25~keV count rate of $\sim 8 \times 10^{-4}$ cts s$^{-1}$
is $\sim 3.5$ for the entire $D_{25}$ based on the \nustar\ $\log N$-$\log S$ 
measurements of \citet{harrison15}.   This is far
fewer than the total 21 point sources and should not affect our main results.  We further note
 that  the \cxo\ sources have been thoroughly studied using multi-wavelength data  and since any moderate-redshift
AGN would have optical/IR counterparts already identified,   the remaining point
sources identified in our \nustar\ observations are unlikely to be background AGN.

The remaining 20 sources are  thus likely to be associated with M83. 
Our diagnostics suggest that the majority of the \nustar\ point sources in M83 are consistent with 
BH binaries in the intermediate accretion state and NS LMXBs.  
We note that this could be due to a selection bias. The intermediate state in our diagrams by selection traces the Very High State \citep[VHS;][]{done07} also referred to as the Steep Power Law \citep[SPL; see review in][]{terarenko16} and thus likely our more X-ray luminous sources are in this higher state.  The commonality with the intermediate state for lower X-ray luminosity black holes is that the disk and non-thermal components are of comparable magnitude.
 We also point out that NS LMXBs with high X-ray luminosity on the plot are likely to be Z-sources, accreting  very close to their Eddington limit.
 
Fourteen out of the 21 \nustar\ sources have been studied  in detail via
  \cxo\ spectral analysis in L14, who have classified  
five of them as soft state accreting BH binaries.
This includes source x286, which is also 
classified as a BH-XRB in the soft accretion state from our \nustar\ color diagnostics. 
For the remaining four  BH binaries (x321, x248, x284, x403) that were identified as soft state sources by
\cxo,
we measured \nustar\ colors that are more consistent with the intermediate state; however, 
they are still consistent with the soft state within the (large) errors.  
We also point out that the boundaries of the different states in our diagrams are not sharp. 

Differentiating between intermediate and soft accretion states is difficult using only the \cxo\ band
as the spectra are more subtly different in this band \citep[e.g., Figure~9 of][]{done07}.  
\nustar\ has a larger lever arm, akin to the hard X-ray capabilities of \rxte, for detecting the harder
components and more sensitively discriminating the intermediate accretion
state for these sources. It is possible that detecting objects in this state is thus simply due to a detection limit
bias.  

The other previously identified ULX, x403 \citep{immler99, ducci13},  reached luminosity of (2--3)$\times$10$^{39}$~\ergl\ in 
the 0.5--8.0~keV in Jan. 2014 (see Table~\ref{t:speculx}).  This source is not very bright in the \nustar\ data, indicating
a steeper spectrum. In our spectral fitting of x403 (see Appendix~\ref{S:app}), an absorbed power law is preferred to
an absorbed blackbody disk model.
Applying an absorbed disk-blackbody plus
power-law model to the \xmm\ spectra (both epochs 1 and 2) resulted 
in a power-law photon index $\Gamma \sim$3--3.5, and this power-law component dominates even in the softer energy band.
The disk-blackbody component is still required above 2$\sigma$ significance.
We find that the inner disk temperature $kT_{in}$ is $\sim$0.9--1.1~keV, which is
consistent with the  $kT_{in}$ measured from the \xmm\ data taken in 2013 and 2014 as well as the \cxo\ observations
taken in 2012 (L14), and also is a typical value for BH binaries.

L14 classified five of the 21 \nustar\ sources (x29, x185, x152, x281, x193)   as 
NSs in XRBs with accretion rates near or above the Eddington limit.
Their classification was based on the X-ray luminosity and disk temperature.
We note that NS binaries likely have similar \nustar\ soft colors to those of BH binaries, and it is difficult to
distinguish between the two populations in the color-intensity diagnostic alone as shown in the previous section.  The \nustar\ intensity and (soft) color of 
the four sources (x29, x152, x281, x193) are consistent with Galactic NS LMXBs as well as
BH binaries with intermediate accretion state (Figure~\ref{f:colorrate_ave_m83}).
The \nustar\ color-color diagnostic may separate two types; however, the hard colors for the
three sources are not constrained well. 
The colors and intensity for x185 is more consistent with BH binaries with intermediate accretion state.
Hard X-ray luminosity is a useful 
discriminant  \citep[see e.g., Figure~13 of ][]{barret00}, and it is rather unlikely to find NS binaries
with $L_{\rm X} \simgt 10^{38}$erg~s$^{-1}$ (4--25~keV).   It is thus worth exploring if the high-luminosity sources in M83
may indeed be NSs.

From the measured  \nustar\ count rates,  we estimate that two of the NS binary candidates  (x152 and x281)  
have $L_{\rm X}$  just below 10$^{38}$~erg~s$^{-1}$ in the 4--25~keV band, suggesting that they could be NS binaries.
Three other sources have $L_{\rm X}$  $\sim$ a few times 10$^{38}$~erg~s$^{-1}$ (4--25~keV), placing them in the BH 
luminosity range.
However, L14  mention that x185 is a NS X-ray binary radiating above 2--3 times its Eddington luminosity with 
an inner accretion disk temperature  $kT_{in}=$2.2~keV, which corresponds to $r_{in} cos \theta \sim$10~km.
When we apply a power-law plus disk-blackbody model to the \xmm\ epoch 1 spectrum of x185, we also obtain a high
disk temperature value, $kT_{in}$  $\sim$2.6~keV, although our temperature is not well-constrained.  
As L14 argue, this temperature would be too high for a BH binary.
We also applied an absorbed blackbody plus disk-blackbody model, which is a typical spectral
model for a NS compact binary.  
This model fit is comparable to the power-law plus disk-blackbody model fit, resulting in  $kT_{in}$ = 2.2~keV
and a blackbody temperature of $\sim$0.3~keV.   Since it is expected that NSs have $T_{BB}\sim 1$--2~keV \citep[e.g.][]{church14}, 
it is likely that we are unable to decouple the NS blackbody from the disk component.

Sometimes, the optical counterpart of X-ray sources help to distinguish the nature of the sources.
Z-sources in Figure~\ref{f:colorrate_ave_m83} are considered to be LXMBs; therefore, we do not expect these
systems to contain young stars as their companions.   
We note that all five sources are located in/near star-forming regions (L14).
However, L14 did not find unique optical counterparts for three sources (x152, x185, x281), suggesting that
at least these sources could be LMXBs and possibly Z-sources.

The ULX, \ulxa, is the brightest object in the 4--6, 6--12, and 4--25~keV bands except for 
the Jan 2014 observation when the source was faint  ($<10^{39}$~\ergl).
 \nustar\ colors and intensity among three epochs confirm that 
the source moves between the ultraluminous state and high/soft or intermediate accretion states, as  \citet{soria15} suggested from the \xmm\ spectral analysis.
\citet{soria15}  estimated the mass of the BH to be 10--20~\msun, resulting in $L_{\rm X} \sim 0.3-2~L_{\rm Edd}$.
The broadband spectral analysis indicates that the bulk of the emission is radiated below 10~keV.

\subsection{Comparison to the NGC 253 NuSTAR source population}

In this section we compare the \nustar\ sources detected in  NGC~253 and M83, which are star-forming
galaxies at comparable distances, with deep \nustar\ coverage, but slightly different
star formation properties. 

 Based on the {\it GALEX} FUV and {\it Spitzer} 24~$\mu$m luminosities by \citet{lee11} and \citet{dale09} and 
using the SFR equations from \citet{calzetti13}, we find that M83 and NGC~253
have SFRs of 3.2~\msun~yr$^{-1}$ and 6.1~\msun~yr$^{-1}$, respectively. 
The stellar mass estimated for M83 and NGC 253 from the 2MASS $K$-magnitudes \citep{jarrett03} and the conversion by \citet{bell03} are 3.3~$\times$10$^{10}$~\msun\ and
 5.7~$\times$10$^{10}$~\msun, respectively. 
There are 21 point sources total detected in M83 (see \S~5.1) and 22 point sources total
in NGC~253 \citep{wik14b}, with M83 exhibiting fewer luminous sources (one ULX) than
 NGC~253 (at least four ULXs).
Based on their SFRs and the XLF of nearby
galaxies, we expect 1.5 ULXs for M83 and 3 ULXs for NGC~253 \citep{swartz11,mineo12}.

\begin{figure*}
\begin{center}
\includegraphics[angle=0,width=7in]{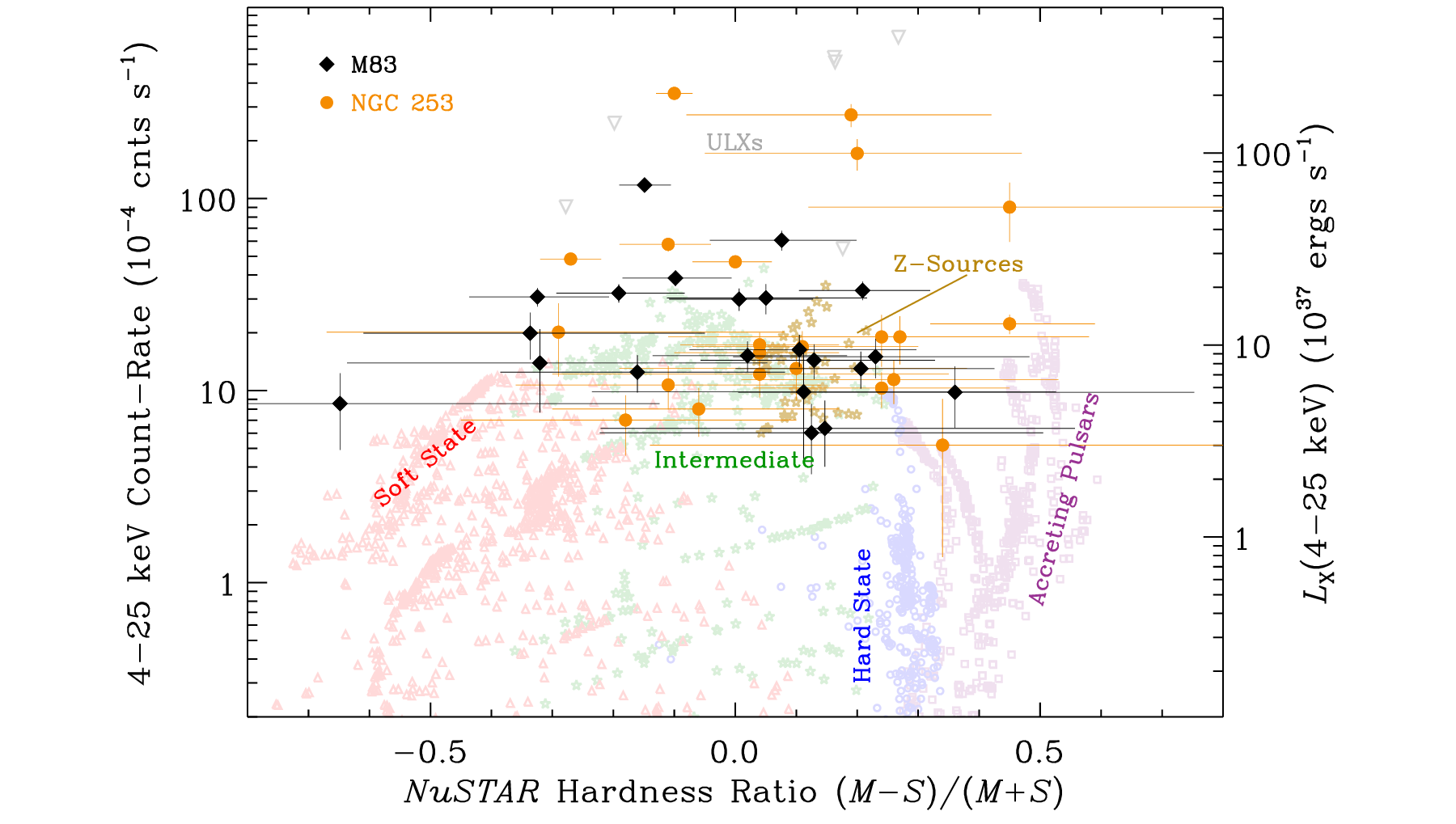}
\end{center}
\caption{Hardness ratio-intensity plot diagram for M83 (black diamonds) and NGC~253 (orange circles) sources.  Other colors and symbols are the same as Figure~\ref{f:colorrate_ave_m83}.
M83 and NGC~253 have a similar point source population, such that the majority of sources are   BH binaries in the
intermediate accretion state. Note that NGC~253 has more ULXs.}
\label{f:ngc253_m83}
\end{figure*}

Figure~\ref{f:ngc253_m83} displays both the  M83 and  NGC~253 \nustar\ detected sources in the
intensity-hardness plot which is useful for discriminating between BH  accretion states  (Zezas et al., in prep.). 
Most of the sources in NGC 253 and M83 have similar colors and intensities to 
Milky Way BH binaries in the intermediate and/or very high state.
However, we note that, albeit at slightly lower X-ray luminosity, NS LMXBs occupy a similar region of the hardness-intensity diagram as
BH-XRBs \citep[Figure~\ref{f:colorrate_ave_m83}; also][]{done07}.  Therefore,  most of these sources are also
consistent with Z-sources.  The color-color diagram shown in Figure~\ref{f:colors_ave_m83} constrains the nature
of the sources better; however,  not all the XRBs have sufficiently high S/N ratios in the 12--25~keV band to allow the use of this more discriminatory color-color diagnostic.
We note that there has been heightened interest in luminous NS XRB populations recently given the recent discovery by \nustar\ of an extremely luminous pulsar in M82 \citep[e.g.,][]{bachetti14}.   To address the question of the nature of higher $L_{\rm X}$ non-magnetized NS XRBs, a system like M83 is ideal.

We found one accreting pulsar candidate (no pulsation found) in NGC 253, whereas no such candidates were found in M83.  
Our detection limit of $\sim$10$^{38}$~\ergl\  is near the upper end of the luminosity for Galactic accreting pulsars. 
An accreting pulsar with such a luminosity could be either 
 a Type II burst accreting pulsar (Be-XRB, the most numerous subclass of HMXBs in the Milky Way and the Magellanic Clouds; Reig et al. 2008)
 \nocite{reig08}  or a Roche lobe overflow pulsar with a supergiant companion like SMC X-1.
 Be XRBs are generally found to be associated with
  young bursts of star formation \citep[10--40~Myr; ][]{antoniou10,antoniou16}.
However, Type II X-ray bursts are rare and short-lived (generally lasting only a few orbital timescales);
therefore, detecting such a system points to a host galaxy with a very young stellar population.
Contrary to the transient behavior of Be XBRs,  Roche lobe overflow systems with a supergiant, which are also very young systems, are persistent,
but they are a small fraction of the total population. 
We note that 
 the persistent nature of the pulsar candidate in NGC~253 may argue against a BeXRB nature but it could be a supergiant pulsar.
We did not found any pulsar binaries in M83, because we assume that the most intense star-forming activity is the nuclear region of the galaxy which may be confused  
at  the resolution of NuSTAR.  In the disk, the XRB populations are much older,
and the number of BeXRBs we would expect to observe in a single exposure is very small. The same holds for Roche lobe overflow supergiants.

It has been more than two decades since the first measurement of the strong correlation of X-ray 
emission from star-forming galaxies with host galaxy
properties such as SFR \citep[e.g.,][]{david92} and it is now relatively
well established that much of this strong correlation, particularly at
$E>2$~keV, is due to X-ray binary populations 
\citep[e.g.,][and references therein]{fabbiano06}.  
One very useful tool
for studying binary populations is the XLF.
Specifically, it is now well established that the normalization of the binary XLF 
 for HMXBs and ULXs scales with SFR \citep{grimm03,mineo12,swartz11}
whereas the LMXB XLF scales with stellar mass \citep{gilfanov04,kim10}.
The XLF for resolved point sources is a 
powerful tool for characterizing point source populations in nearby 
galaxies. 
The slope of the binary XLF in the 0.5--10~keV bands gets steeper as
the underlying stellar population evolves and the dominant population
shifts from HMXBs to LMXBs \citep[e.g.,][]{wu01}.
The high-$L_{\rm X}$ slope of the \cxo\ XLF for M83 is found to be steeper ($\gamma$=1.38) than that of NGC~253 ($\gamma$=0.81)
\citep{kilgard02}, implying that NGC~253 is likely to be HMXB dominated whereas M83 (especially in the disk)
is LMXB dominated (see also L14).
The XLF slopes also differ within M83, showing a flatter slope for the nuclear starburst region and a
steeper slope for the disk region (L14).

\begin{figure}
\begin{center}
\includegraphics[angle=-90,width=3.5in]{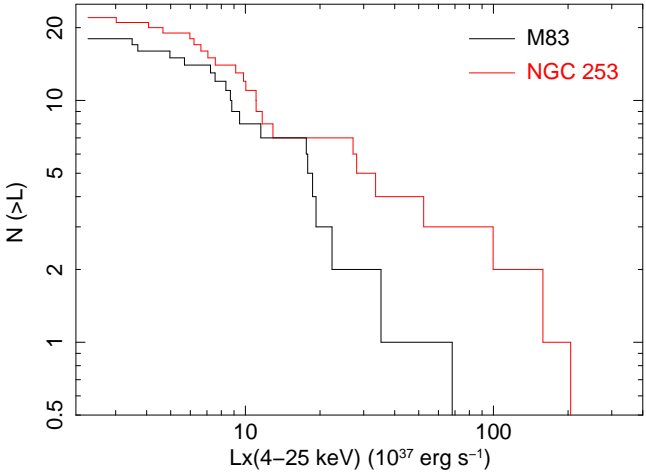}
\end{center}
\caption{The cumulative  \nustar\ point source XLFs for M83 (black) and NGC~253 (red).   The LFs become 
flatter below 5$\times$ 10$^{37}$~\ergl\ (0.001~cts~s$^{-1}$), indicating that completeness corrections may need to be taken into account.
The slope of the M83 LF is steeper than that of NGC253.}
\label{f:lf}
\end{figure}

We constructed \nustar\ XLFs using the 4--25~keV band for both M83 and NGC~253, which are shown in Figure~\ref{f:lf}.
We excluded three (x403, x029, x366) sources in M83, because they were not in the \nustar\ FoV for all three
epochs. 
The remaining sources have roughly similar detection sensitivities in the different epochs  
(a 30\% difference in the effective exposures). 
All 22  \nustar\ sources in NGC~253 were in the \nustar\ FoV for all epochs, thus no sources were excluded.

The flattening
below 5$\times$ 10$^{37}$~\ergl\ (0.001~cts~s$^{-1}$) suggests that incompleteness may be an issue below this limit.
We therefore confine our  comparisons to above this limit.
We clearly see that the high-$L_{\rm X}$ slope of the M83 \nustar\ LF is 
steeper than that of NGC~253, which is similar to the soft X-ray LFs for these galaxies with \cxo.   
A more detailed analysis of the \nustar\ XLF will be presented in a forthcoming paper.

\subsection{Lack of  an Active Nucleus?}

The position of the galactic center of M83 has been debated \citep[see][]{knapen10}.  
M83 possesses a nuclear star cluster, which is referred to as the optical nucleus \citep{gallais91}. 
The kinematic center and photometric center are coincident with each other; however, 
  the nuclear star cluster is $\sim$3\arcsec\ offset from this location
(see Figure~1 of Knapen et al. 2010 as well as Sakamoto et~al. 2004, Muraoka et al. 2009).
\nocite{sakamoto04, muraoka09, knapen10}
This offset suggests there may be a hidden nuclear mass \citep[i.e.,][]{thatte00}.
However, recent studies \citep{houghoton08,knapen10} prefer to adopt the nuclear star cluster 
as the nucleus.

Our \nustar\ observations detect hard ($>$10~keV) X-ray emission from the nuclear region; however,
our analysis suggests that 
the emission is likely to be the integrated emission from X-ray binaries instead of
the X-ray emission from an obscured AGN based on the steep spectra and the resolution
of this emission into several sources by \cxo.
If M83 contained a highly obscured AGN such as the Compton-thick AGN in Arp 299 \citep{ptak15}, then we would have expected a rising power law to 20--40~keV, which we do not 
observe.   

\citet{soria03} and L14 have shown that there is an X-ray counterpart (x233) coincident with the position of the star cluster nucleus.
The averaged spectral slope during 2010 of this X-ray source  in the soft X-ray band is a power law with 
photon index of $\Gamma\sim$1.4 and $L_{\rm X}$  of 3 $\times$10$^{38}$~\ergl\  in the 0.5--8.0~keV band
(L14).  
They suggest that this could be a SMBH with a mass of 1.3 $\times$10$^7$~\msun\  \citep{thatte00} accreting 
with 10$^{-7}$~$L_{\rm Edd}$, but they cannot reject the possibility that it is a compact binary at the location of  the nuclear star cluster
(e.g., \citeauthor{laparola03} 2003 for M33, and \citeauthor{yukita07} 2007 for NGC~2403).

X233  is also detected in our observations taken in June 2014. 
 Taking a 2\arcsec\ radius extraction region (331 net counts in the 0.5--8.0~keV band) in the \cxo\ data in order to reduce the contamination of
  underlying thermal diffuse emission (note that Tables~\ref{t:softxct} and \ref{t:softxspec} use a 5\arcsec\  radius),  an absorbed power-law model gives $\Gamma$=2.3$^{+0.7}_{-0.4}$  and $N_{\rm H}$  = 1.3$^{+1.8}_{-1.0} \times$10$^{21}$~cm$^{-2}$ with  $L_{\rm X}^{obs} \sim$  2.8 $\times$ 10$^{38}$~\ergl\
 in the 0.5--8.0~keV band.  The intrinsic luminosity is about 1.3 times higher than the
 observed luminosity.
The spectral slope reported here is steeper (in June 2014) than that observed in 2010 reported by L14, but the X-ray luminosity in the \cxo\ band is comparable.

If we extrapolate the spectral shape using a power law with photon
index of $\Gamma=$2.3,  obtained from the \cxo\ spectrum,  into the \nustar\ band, we expect x233 has \nustar\ count rates of 
4$\times$10$^{-4}$~ct~s$^{-1}$, 5$\times$10$^{-4}$~ct~s$^{-1}$, and 1$\times$10$^{-4}$~ct~s$^{-1}$ for the
4--6, 6--12, 12--25~keV bands during epoch 3, respectively.
These expected 4--6~keV and 6--12~keV count rates are about a factor of 5 lower than the count rates 
measured by \nustar\ for the nuclear region, implying x233 makes only a small contribution to the nuclear region emission (see Table~\ref{t:nupt3ep}).

We only obtained an upper limit for 
the observed 12--25~keV count rate, but the expected power-law  model count rate in the hard band is below
the upper limit of 5.9 $\times$ 10$^{-4}$~ct~s$^{-1}$.
The faint sources near x233, that were resolved in the \cxo\ image, were not resolved in the \nustar\ images, and 
these sources may contribute 
to the measured x233 count rate in the 4--12~keV band.
Therefore, the \nustar\ data do not reject that x233
is   M83's SMBH  accreting at very low level, whose
0.5--25~keV spectrum is characterized by a single power law with an index of $\Gamma\sim$2.3.
We measure a very low sub-Eddington value for the total X-ray luminosity of the central source.
AGNs with these very low levels of ${L}_{\rm 2-10keV}/{{L}_{\rm Edd}}$ typically have spectral
 slopes of $\Gamma= 1.4 \pm 0.4$ \citep[e.g.,][]{shemmer06,younes11}, 
  shallower than the
constraints we have for the nuclear source.  
This makes it very unlikely that x233 is powered by an AGN.
We note that the measured X-ray luminosity and spectral values for the nuclear source are also consistent with an X-ray binary.

\section{Summary}\label{S:summary}
We present the first-ever spatially-resolved study of point sources in M83 above 10 keV, which was made possible with new \nustar\ observations.  
NuSTAR's harder energy bandpass constrains and classifies the nature of the resolved sources and
accretion states for both BH and NS systems  (Wik et al. 2014b, Maccarone et al. 2016).
This is only the second deep \nustar\ survey of ``resolved'' compact object populations in a star-forming galaxy outside of the MW following our first study of the nearby starburst NGC~253.
Our main results are listed below:

\begin{itemize}
\item Twenty  point sources, which are likely associated with the galaxy and are thus XRBs, were  found in the \nustar\ images.  
We found that all the \nustar-detected sources are detected in the extremely deep archival \cxo\ data available for M83, albeit with a broader range of hard X-ray colors than is seen in the soft X-ray data alone.    

Overall, the \nustar\ M83 XRB population differs from that of NGC~253 studied by \citet{wik14b} mainly in that M83 possesses fewer  luminous sources (fewer ULXs) than NGC~253  
in the \nustar\ band, which is expected based on the higher SFR (6~\msun~yr$^{-1}$) of NGC~253.

\item
We have constructed the \nustar\ X-ray binary LF in the 4--25~keV for an extragalactic environment for the first time.
The slope of the \nustar\ XRB XLF for M83 is steeper than that of NGC~253, consistent with previous \cxo\ results studied at 
soft X-ray energies.  Note one might expect major differences in the XRB LF at $E>10$~keV as we probe to lower $L_{\rm X}$ ($\sim 10^{36}$~erg s$^{-1}$)  typical of the obscured HMXB population observed in the Milky Way with INTEGRAL \citep{lutovinov13}. 

\item  We classified XRBs using the NuSTAR point source color-intensity and color-color diagrams.
Based on their \nustar\ colors and intensities, the majority of the sources are likely to be intermediate accretion state BH binaries and NS LMXBs (Z-sources). 
 One source is likely to be  a soft/high-state BH binary.  The known ULX,
 \ulxa, has a luminosity and color (turning over at higher energies) that is similar to other ULXs studied with \nustar.
\ulxa\ shows long term variability and possible spectral state transitions, from ULX to soft/intermediate and then back to the ULX state.
This state-transition behavior of \ulxa\ is similar to what was observed by \citet{soria15}  using \xmm\ data.
Other sources do not show long term variability or  state transitions.

\item The 0.5--20~keV  spectra of the nuclear region indicate  no strong reprocessed X-ray emission from the nucleus of M83, 
suggesting that there is no buried/highly obscured actively accreting SMBH present in this galaxy.  We place an upper limit of $\sim10^{38}$~\ergl\ in the 10--30~keV band
for an obscured AGN.
The X-ray point source, x233,  found at the location of
the nuclear star cluster can be described  as  a single power law with an index of $\Gamma$$~\sim$2.3 up to 25~keV, a spectral shape that is more consistent with a luminous X-ray binary.  However, our data do not rule out the possibility that the
source is a low luminosity AGN.

\item  The galaxy-wide spectrum of M83 becomes steeper (with power law index $>$ 2) at harder energies,  also seen
in other starburst galaxies observed by \nustar.  Either ULX-like sources, intermediate accretion state BH binaries, or NS LMXBs (Z-sources)  dominate the 10--30~keV
emission in these galaxies.   This revised 0.5--30~keV SED is of great utility in doing proper k-corrections of high-$z$ X-ray constraints on galaxies in the deep \cxo\ surveys (see, e.g., Lehmer et al. 2016).
\end{itemize}

\vspace{1cm}

We thank the referee for useful comments and suggests that improved this paper. 
This work was supported under NASA Contract No. NNG08FD60C, and made use of data from the 
\nustar\ mission, a project led by the California Institute of Technology, managed by the Jet Propulsion Laboratory, 
and funded by the National Aeronautics and Space Administration. 
We sincerely thank Karl Forster and Brian Grefenstette for their work on the complicated 
stray light pattern and for assistance in designing the \nustar\ 
 observational setup for minimizing this contamination.  
 We also thank \nustar\ and \cxo\ mission planners for making the
 \cxo\ and \nustar\ observations simultaneous.  
We also thank P. Tzanavaris and D. Swartz for helpful discussion. 
Support for this work was provided by the National Aeronautics and Space Administration through Chandra Award Number GO4-15086Z issued by the Chandra X-ray Observatory Center, which is operated by the Smithsonian Astrophysical Observatory for and on behalf of the National Aeronautics Space Administration under contract NAS8-03060.
RK acknowledges support from Russian Science Foundation (grant 14-12-01315).

Facilities: \cxo, \nustar, \xmm

\appendix
\section{\xmm/\cxo\ Point Source Properties}\label{S:app}
We briefly summarize the \xmm\ and \cxo\ properties for the \nustar\ identified point sources. 
Table~\ref{t:softxct}  lists the count rates and luminosities in the 0.5--8.0~keV band using
 the 15\arcsec\ and 5\arcsec\ radius aperture for  the \xmm\ and \cxo\  data, respectively.  
Aperture correction has been applied.
The background is obtained utilizing nearby source-free regions.
The luminosities are estimated using a power law with a photon index $\Gamma$=2.
The variability is flagged if the source flux change more than 3$\sigma$ from the averaged value among the three epochs.

For the spectral analysis, we use the 0.5--8~keV energy band and applied two models, 
an absorbed power-law or an absorbed disk-blackbody model \citep[{\it XSPEC}  model {\tt diskbb};][]{mitsuda84}.  
The nuclear sources, x233 and x199, exhibit an excess of soft emission that could be due to
hot gaseous emission from the nuclear starburst.  
Hence, we add an optically thin thermal plasma
component \citep[{\it XSPEC} {\tt apec};][]{smith01} to obtain better fit statistics.  
The fitting results are tabulated in Table~\ref{t:softxspec}.

\begin{deluxetable}{lrrrrrrrrc} 
\tabletypesize{\scriptsize}
\tablecolumns{10}
\tablewidth{0pc} 
\tablecaption{M83 Point Sources: \xmm/\cxo\ Count Rates and Luminosities\label{t:softxct}} 
\tablehead{ 
 \colhead{}   & \multicolumn{3}{c}{Epoch 1} & \multicolumn{3}{c}{Epoch 2} &  \multicolumn{2}{c}{Epoch 3} \\
 \cline{2-3}  \cline{5-6}  \cline{8-9}  &  \colhead{} \\ 
\colhead{}     & \colhead{Rate}   &  \colhead{log$L_{\rm X}$} & & 
\colhead{Rate}   & \colhead{log$L_{\rm X}$}  & & 
\colhead{Rate}   & \colhead{log$L_{\rm X}$} & \\
\colhead{ID}    & \multicolumn{2}{c}{(0.5--8.0~keV)}  &  
& \multicolumn{2}{c}{(0.5--8.0~keV)}    &  
& \multicolumn{2}{c}{(0.5--8.0~keV)} & \\ 
\colhead{L14}    & \colhead{($10^{-3}$~ct$^{-1}$~s$^{-1}$)  }   & \colhead{(10$^{37}$~\ergl)}    & 
& \colhead{($10^{-3}$~ct$^{-1}$~s$^{-1}$) }   & \colhead{(10$^{37}$~\ergl)}    & 
& \colhead{($10^{-3}$~ct$^{-1}$~s$^{-1}$)  }   & \colhead{(10$^{37}$~\ergl)} & \colhead{variability$^a$}  }
\startdata 
x281&8.8$\pm$0.8 & 8.0$\pm$0.7 && 10.7$\pm$1.0 & 9.7$\pm$0.9 && 3.0$\pm$0.3 & 10.0$\pm$1.1 & \\
x029&13.7$\pm$0.7 & 12.4$\pm$0.6 && 9.3$\pm$0.7 & 8.5$\pm$0.6 && 6.2$\pm$0.5 & 20.6$\pm$1.5& v\\
x048&\nodata & \nodata &&  1.7$\pm$0.7 & 1.5$\pm$0.6 && 1.2$\pm$0.2 & 4.1$\pm$0.7& \\
x073&4.3$\pm$0.7 & 4.0$\pm$0.7 && \nodata & \nodata && 1.4$\pm$0.2 & 4.7$\pm$0.8 &\\
x145&6.5$\pm$0.7 & 5.9$\pm$0.7  && 7.3$\pm$0.9 & 6.7$\pm$0.8 && 2.3$\pm$0.3 & 7.7$\pm$0.9& \\
x165&6.1$\pm$0.7 & 5.6$\pm$0.7 && \nodata & \nodata &&  \nodata & \nodata& \\
x152&8.7$\pm$0.9 & 7.9$\pm$0.8 && 6.6$\pm$1.1 & 6.0$\pm$1.0 && 3.9$\pm$0.4 & 12.9$\pm$1.2& v\\
x155&3.7$\pm$0.4 & 3.4$\pm$0.4 && 1.0$\pm$0.3 & 0.9$\pm$0.3 && \nodata & \nodata &\\
x185&28.9$\pm$1.1 & 26.3$\pm$1.0 && \nodata & \nodata && 11.1$\pm$0.6 & 36.6$\pm$2.0 & \\
x193&205.6$\pm$2.4 & 187.1$\pm$2.2 &&   219.3$\pm$3.1 & 199.6$\pm$2.8 &&  2.5$\pm$0.5 & 8.1$\pm$1.6& \\
x252&\nodata & \nodata &&  4.5$\pm$0.5 & 4.1$\pm$0.5 && 3.3$\pm$0.3 & 10.9$\pm$1.1&\\
x246&21.4$\pm$1.3 & 19.5$\pm$1.1 && 41.0$\pm$1.7 & 37.3$\pm$1.6 && 0.9$\pm$0.2 & 3.0$\pm$0.7&\\
x233&321.0$\pm$2.9 & 292.1$\pm$2.6 && 312.2$\pm$3.8 & 284.1$\pm$3.4 && 12.4$\pm$1.0 & 40.8$\pm$3.4&\\
x248&29.7$\pm$1.0 & 27.0$\pm$1.0 && 24.6$\pm$1.4 & 22.4$\pm$1.2 && 8.9$\pm$0.6 & 29.1$\pm$1.8&\\
x258&8.2$\pm$0.7 & 7.5$\pm$0.6 && \nodata & \nodata && 2.8$\pm$0.3 & 9.3$\pm$1.0&\\
x286&30.3$\pm$1.1 & 27.6$\pm$1.0  && 48.8$\pm$1.7 & 44.4$\pm$1.5 && 16.2$\pm$0.8 & 53.2$\pm$2.5& v\\
x299&252.3$\pm$2.6 & 229.6$\pm$2.3  && 57.0$\pm$1.8 & 51.8$\pm$1.6  && 57.5$\pm$1.4 & 189.4$\pm$4.6 &v\\
x284&25.2$\pm$1.0 & 23.0$\pm$0.9 && \nodata & \nodata && 8.2$\pm$0.5 & 27.1$\pm$1.8&\\
x321&28.5$\pm$1.1 & 25.9$\pm$1.0 && 35.1$\pm$1.5 & 32.0$\pm$1.3 && 10.2$\pm$0.6 & 33.7$\pm$2.0&\\
x366&\nodata & \nodata && 4.6$\pm$0.8 & 4.2$\pm$0.7 && 2.0$\pm$0.3 & 6.6$\pm$0.9&\\
x403&132.7$\pm$1.8 & 120.8$\pm$1.7 &&  171.6$\pm$2.7 & 156.2$\pm$2.4 && 25.9$\pm$0.9 & 85.3$\pm$3.1&v\\
\enddata
\tablenotetext{.}{Notes -- The 15\arcsec\ or 5\arcsec\ radius aperture is used for \xmm\ or \cxo\ data, respectively. 
Luminosity is calculated using a power law index of $\Gamma$=2 and Galactic column density, and
aperture correction is applied.  When a source fell in a chip gap and/or bad pixels, no measurement is listed.} 
\tablenotetext{a}{Marked if the luminosity change more than 3$\sigma$ from the average luminosity. x233
and x193 are omitted because of the strong diffuse contamination in the \xmm\ luminosity measurements.}
\end{deluxetable}

\clearpage
\LongTables
\begin{deluxetable}{lccrrrrcrrl} 
\tabletypesize{\scriptsize}
\tablecolumns{11} 
\tablewidth{0pc} 
\tablecaption{M83 Point Sources: Soft X-ray Spectral Analysis\label{t:softxspec}} 
\tablehead{  \colhead{}     & \colhead{}   &  \colhead{}   &  \colhead{}   & \colhead{}    
& \colhead{}  & \colhead{}  & \colhead{log$L_{\rm X}^b$} &  \colhead{} &  \colhead{Norm} &  \colhead{} \\
 \colhead{Name}   & \colhead{}      & \colhead{}   & \colhead{$N_{\rm H}$}    & \colhead{$\Gamma$/$kT_{in}$}
& \colhead{Norm} & \colhead{} & \colhead{(0.5--8.0~keV)} & \colhead{$kT_e^c$}  & \colhead{(gas)}  & \colhead{}\\
 \colhead{ID}   & \colhead{Epoch}    & \colhead{Model$^a$}    & \colhead{(10$^{21}$~cm$^{-2}$)}   & \colhead{/(keV)}
& \colhead{(10$^{-4}$)} & \colhead{C-stat/dof}  & \colhead{(erg~s$^{-1}$)} & \colhead{(keV)}  & \colhead{(10$^{-4}$)}  & \colhead{$fr^d$}}
\startdata 
  x281 &1&PL & $0.4_{-0.4}^{+0.7}$ & $1.47_{-0.19}^{+0.26}$&$0.06_{-0.01}^{+0.01}$ & 294.6/347   & 38.0 & \nodata &  \nodata &  \nodata \\
    &1&DB& $0.4_{-0.4}^{+0.4}$  &$1.95_{-0.56}^{+0.93}$ & $1.42_{-1.02}^{+3.07}$ & 302.6/347 & 38.0 &  \nodata  &  \nodata  &  \nodata  \\
&2 &PL& $0.4_{-0.4}^{+0.5}$ & $1.74_{-0.24}^{+0.27}$  &$0.09_{-0.02}^{+0.02}$ & 216.1/275   & 38.0 & \nodata  & \nodata  & \nodata \\
  &2&DB   & $0.4_{-0.4}^{+0.3}$   &$1.53_{-0.40}^{+0.58}$ & $4.23_{-2.82}^{+7.86}$ & 230.6/275  & 38.0 &  \nodata &  \nodata &  \nodata \\
   \cline{1-11}    \\
x029  & 1  &PL& $3.6_{-1.3}^{+1.6}$ & $1.65_{-0.25}^{+0.28}$ &$0.18_{-0.05}^{+0.07}$ & 351.4/468   & 38.4 &  \nodata &  \nodata & \nodata  \\
           & 1  &DB & $1.5_{-0.8}^{+1.0}$ &$1.98_{-0.38}^{+0.56}$ & $3.15_{-1.82}^{+3.60}$ & 360.4/468  & 38.3&  \nodata & \nodata  & \nodata  \\
           & 3  &PL& $23.5_{-8.5}^{+14.8}$ & $5.01_{-1.44}^{+2.86}$ &$7.28_{-5.51}^{+86.77}$ & 337.3/509   & 38.1& \nodata & \nodata & \nodata\\
           & 3 &DB  & $12.1_{-5.0}^{+8.4}$ &$0.55_{-0.21}^{+0.23}$ & $1126.87_{-949.25}^{+9882.93}$ & 336.6/509 & 38.1& \nodata & \nodata & \nodata \\
             \cline{1-11}  \\
x145 &1 &PL & $5.8_{-2.3}^{+3.2}$ & $2.38_{-0.43}^{+0.53}$ &$0.20_{-0.08}^{+0.15}$ & 309.5/384  &  38.3 & \nodata &  \nodata & \nodata \\
          &1 &DB & $2.5_{-1.4}^{+2.0}$ &$1.16_{-0.25}^{+0.33}$ & $12.46_{-7.78}^{+21.58}$ & 312.1/384 &  38.0 & \nodata &  \nodata& \nodata \\
             \cline{1-11}   \\
x165 &1  &PL& $0.4_{-0.4}^{+1.4}$ & $1.31_{-0.23}^{+0.33}$  &$0.04_{-0.01}^{+0.02}$ & 274.9/325   & 37.9 & \nodata & \nodata & \nodata  \\
    &1&DB& $0.4_{-0.4}^{+0.7}$ &$2.51_{-0.82}^{+0.92}$ & $0.49_{-0.44}^{+1.49}$ & 280.8/325 & 37.9& \nodata & \nodata & \nodata  \\
      \cline{1-11}    \\
x152 &1&PL & $2.8_{-1.9}^{+4.3}$ & $2.17_{-0.43}^{+0.57}$  &$0.13_{-0.05}^{+0.12}$ & 302.4/365   & 38.1 & \nodata & \nodata &  \nodata\\
    &1 &DB& $0.4_{-0.4}^{+2.1}$   &$1.15_{-0.26}^{+0.26}$ & $10.12_{-5.36}^{+19.58}$ & 301.8/365  & 37.9 & \nodata & \nodata & \nodata \\
           \cline{1-11}   \\
x185   &1&PL  & $2.6_{-0.7}^{+0.8}$ & $1.96_{-0.18}^{+0.19}$ & $0.38_{-0.07}^{+0.08}$ & 479.8/629   & 38.6 & \nodata &  \nodata& \nodata\\
      &1 &DB& $0.4_{-0.4}^{+0.5}$ &  $1.49_{-0.16}^{+0.15}$ & $14.31_{-3.93}^{+7.57}$ & 490.5/629  & 38.5 &  \nodata&  \nodata& \nodata \\
&3 &PL & $3.4_{-2.0}^{+2.3}$ & $1.88_{-0.36}^{+0.39}$  &$0.40_{-0.13}^{+0.22}$ & 371.9/509  & 38.5 &  \nodata&  \nodata& \nodata \\
 &3&DB    & $1.3_{-1.3}^{+1.6}$ &$1.33_{-0.23}^{+0.34}$ & $23.16_{-13.32}^{+28.05}$ & 359.9/509  & 38.5 &  \nodata& \nodata & \nodata \\
             \cline{1-11}    \\
 x193 &  1&PL & $0.7_{-0.2}^{+0.2}$ & $2.58_{-0.14}^{+0.14}$ &$0.98_{-0.11}^{+0.12}$ & 649.9/685   & 39.2  & $0.78_{-0.02}^{+0.02}$ & $0.81_{-0.06}^{+0.07}$ & 0.43 \\
         &1 &DB&  0.4 &$0.41_{-0.02}^{+0.03}$ & $5252_{-1424}^{+1718}$ & 742.1/685  & 39.0 & 0.75& $0.68_{-0.43}^{+0.43}$ & 0.49 \\
&2 &PL& $0.9_{-0.3}^{+0.3}$ & $2.75_{-0.21}^{+0.23}$  &$1.17_{-0.18}^{+0.21}$ & 503.7/588  & 39.2 & $0.76_{-0.03}^{+0.03}$ & $1.01_{-0.10}^{+0.12}$ & 0.45 \\
&2 &DB&$0.4_{-0.4}^{+0.1}$ &$0.42_{-0.03}^{+0.04}$ & $6292.79_{-2037.69}^{+2687.55}$ & 574.3/588  & 39.1 & $0.75_{-0.03}^{+0.03}$ & $0.80_{-0.07}^{+0.07}$ & 0.47 \\
&3 &PL& 0.4 & $1.95_{-0.36}^{+0.34}$ & $0.14_{-0.05}^{+0.05}$ & 280.1/507  & 38.2 & 0.75 & $0.02_{-0.02}^{+0.02}$ & 0.08 \\
&3 &DB&0.4 &  $1.28_{-0.25}^{+0.42}$ & $10.1_{-6.6}^{+14.3}$ & 282.9/507  & 38.2 & 0.75 & $0.05_{-0.02}^{+0.02}$ & 0.21 \\
              \cline{1-11}   \\
x246 &1 &PL& $1.2_{-0.7}^{+0.8}$ & $2.34_{-0.31}^{+0.36}$  &$0.23_{-0.06}^{+0.08}$ & 491.3/517   & 38.3 & \nodata &  \nodata& \nodata\\
    &1 &DB& $0.4_{-0.4}^{+0.2}$ &$0.61_{-0.11}^{+0.15}$ & $231.28_{-136.28}^{+289.56}$ & 522.7/517  & 38.1 & \nodata & \nodata & \nodata\\
&2 &PL& $1.4_{-0.6}^{+0.6}$ & $2.33_{-0.23}^{+0.25}$  &$0.49_{-0.09}^{+0.12}$ & 511.4/498  & 38.6 & \nodata & \nodata &  \nodata\\
 &2 &DB   & $0.4_{-0.4}^{+0.1}$  &$0.87_{-0.02}^{+0.05}$ & $123.56_{-27.92}^{+34.54}$ & 537.0/498  & 38.5 & \nodata &  \nodata& \nodata\\
            \cline{1-11}   \\
x233 &1  &PL & $0.8_{-0.2}^{+0.2}$ & $2.39_{-0.09}^{+0.10}$ &$1.63_{-0.13}^{+0.14}$ & 815.8/884  & 39.4  & $0.78_{-0.02}^{+0.02}$ & $1.24_{-0.07}^{+0.08}$ & 0.39 \\
& 1&DB& $0.4_{-0.4}^{+0.0}$ &$0.78_{-0.07}^{+0.07}$ & $595.47_{-173.39}^{+283.45}$ & 1105.9/884 & 39.3 & $0.76_{-0.01}^{+0.01}$ & $1.25_{-0.06}^{+0.06}$ & 0.48 \\
& 2 &PL&  $0.7_{-0.2}^{+0.2}$ & $2.44_{-0.13}^{+0.14}$  &$1.68_{-0.19}^{+0.21}$ & 663.0/741  & 39.4 & $0.77_{-0.03}^{+0.03}$ & $1.24_{-0.10}^{+0.11}$ & 0.39 \\
&2 & DB& $0.4_{-0.0}^{+0.0}$ &$0.57_{-0.07}^{+0.11}$ & $2340_{-778}^{+110}$ & 859.3/742 & 39.3 & 0.75& $1.17_{-0.08}^{+0.08}$ & 0.54 \\
& 3& PL& $0.9_{-0.8}^{+1.0}$ & $1.87_{-0.26}^{+0.23}$  &$0.43_{-0.12}^{+0.12}$ & 441.49/508  & 38.2 & 0.75 & $0.46_{-0.15}^{+0.31}$ & 0.60 \\
& 3& DB& $2.3_{-1.6}^{+1.2}$ &  $1.97_{-0.53}^{+1.54}$ & $5.53_{-4.7}^{+13.7}$ & 451.5/508  & 39.1 &0.75  & $5.53_{-0.73}^{+19.2}$ & 0.30 \\
               \cline{1-11}    \\
x248 &1&PL  & $7.3_{-1.3}^{+1.5}$ & $2.87_{-0.24}^{+0.25}$  &$1.15_{-0.26}^{+0.36}$ & 518.5/600  & 38.9 &  \nodata& \nodata &  \nodata \\
    &1 &DB& $3.0_{-0.7}^{+0.8}$   &$0.91_{-0.09}^{+0.09}$ & $122.06_{-41.50}^{+68.67}$ & 508.3/600  & 38.6 &  \nodata& \nodata & \nodata \\
&2&PL& $4.9_{-1.3}^{+1.7}$ & $2.63_{-0.32}^{+0.37}$  &$0.65_{-0.18}^{+0.29}$ & 342.0/451   & 38.6& \nodata & \nodata & \nodata \\
 &2 &DB   & $1.6_{-0.8}^{+1.0}$  &$0.97_{-0.15}^{+0.18}$ & $67.33_{-33.70}^{+73.68}$ & 342.7/451  & 38.4& \nodata & \nodata & \nodata \\
&3&PL& $9.9_{-2.9}^{+3.6}$ & $3.43_{-0.61}^{+0.76}$  &$1.71_{-0.76}^{+1.70}$ & 333.4/509   & 38.4 & \nodata &  \nodata& \nodata \\
    &3&DB & $4.8_{-2.0}^{+2.3}$   &$0.70_{-0.14}^{+0.17}$ & $351.93_{-221.28}^{+729.67}$ & 325.7/509  & 38.3 & \nodata & \nodata & \nodata \\
              \cline{1-11}  \\
              x258 &1&PL & $0.4_{-0.4}^{+0.4}$ & $0.94_{-0.24}^{+0.23}$  &$0.04_{-0.01}^{+0.01}$ & 309.1/367   & 38.1 & \nodata & \nodata  & \nodata \\
   &1 &DB& $0.4_{-0.4}^{+0.4}$  &$8.90_{-2.85}^{+-8.90}$ & $0.02_{-0.01}^{+0.26}$ & 312.3/367  & 38.2 & \nodata & \nodata & \nodata \\
 \cline{1-11}  \\
x286 &1 &PL  & $2.8_{-0.6}^{+0.6}$ & $2.69_{-0.21}^{+0.23}$ &$0.79_{-0.13}^{+0.16}$ & 495.1/552  & 38.8 &  \nodata&  \nodata& \nodata \\
   &1  &DB& $0.4_{-0.4}^{+0.3}$  &$0.86_{-0.07}^{+0.05}$ & $143.84_{-30.14}^{+60.64}$ & 483.1/552  & 38.6 & \nodata & \nodata & \nodata \\
&2 &PL & $2.7_{-0.5}^{+0.6}$ & $2.59_{-0.20}^{+0.22}$  &$0.82_{-0.14}^{+0.17}$ & 417.4/526  & 38.7 & \nodata & \nodata & \nodata \\
  &2  &DB   & $0.4_{-0.4}^{+0.2}$  &$0.89_{-0.07}^{+0.04}$ & $139.82_{-28.31}^{+52.10}$ & 404.3/526  & 38.6 & \nodata & \nodata & \nodata \\
&3 &PL & $5.6_{-1.8}^{+2.2}$ & $2.83_{-0.36}^{+0.42}$  &$1.32_{-0.41}^{+0.70}$ & 427.8/509  & 38.6 & \nodata & \nodata & \nodata \\
 &3  &DB    & $1.6_{-1.1}^{+1.3}$  &$0.86_{-0.12}^{+0.14}$ & $180.94_{-86.68}^{+185.19}$ & 417.1/509  & 38.6 & \nodata & \nodata & \nodata \\
      \cline{1-11}  \\
x299  &1 &PL &$1.5_{-0.2}^{+0.2}$ & $1.79_{-0.05}^{+0.05}$  &$2.55_{-0.13}^{+0.14}$ & 1182.0/1220  & 39.5 & \nodata & \nodata & \nodata \\
     &1& DB&  $0.4_{-0.4}^{+0.0}$   &$1.40_{-0.04}^{+0.04}$ & $150.84_{-13.73}^{+14.98}$ & 1138.0/1220 & 39.4 & \nodata & \nodata & \nodata \\
&2&  PL &$2.6_{-0.5}^{+0.5}$ & $2.66_{-0.19}^{+0.20}$  &$0.95_{-0.15}^{+0.18}$ & 435.5/534  & 38.8& \nodata &  \nodata&  \nodata\\
  & 2  & DB&$0.4_{-0.4}^{+0.2}$ &$0.81_{-0.06}^{+0.03}$ & $227.07_{-38.97}^{+69.33}$ & 416.6/534 & 38.6& \nodata &  \nodata&  \nodata \\
&3 &PL & $2.4_{-0.7}^{+0.7}$ & $1.80_{-0.13}^{+0.13}$  &$1.92_{-0.27}^{+0.32}$ & 512.0/509   & 39.3& \nodata & \nodata & \nodata \\
   &3 & DB& $0.4_{-0.4}^{+0.4}$   &$1.52_{-0.11}^{+0.09}$ & $75.62_{-14.13}^{+24.64}$ & 501.9/509 & 39.3&  \nodata&  \nodata& \nodata \\
              \cline{1-11} \\
x284 &1&PL& $5.2_{-0.9}^{+1.0}$ & $3.58_{-0.31}^{+0.34}$ &$1.02_{-0.23}^{+0.32}$ & 436.2/450   & 38.9 &  \nodata& \nodata & \nodata \\
    &1 &DB& $1.6_{-0.5}^{+0.6}$  &$0.60_{-0.06}^{+0.07}$ & $458.55_{-183.64}^{+322.34}$ & 416.2/450 & 38.4 & \nodata & \nodata & \nodata \\
 &3 &PL & $4.8_{-2.2}^{+2.5}$ & $3.62_{-0.67}^{+0.83}$  &$1.00_{-0.42}^{+0.85}$ & 295.9/509  & 38.3 & \nodata & \nodata & \nodata \\
&3   &DB  & $0.9_{-0.9}^{+1.6}$  &$0.59_{-0.11}^{+0.12}$ & $455.02_{-255.40}^{+907.46}$ & 284.8/509  & 38.3 & \nodata &  \nodata& \nodata \\
      \cline{1-11}  \\
x321   &1 &PL& $17.9_{-3.0}^{+3.4}$ & $2.91_{-0.28}^{+0.29}$ &$2.44_{-0.77}^{+1.19}$ & 627.3/740  & 39.3 & \nodata &  \nodata&  \nodata \\
    &1 &DB& $8.7_{-1.8}^{+2.0}$ &$1.16_{-0.11}^{+0.12}$ & $72.28_{-26.96}^{+43.67}$ & 618.0/740  & 38.8 & \nodata & \nodata & \nodata \\
&2&PL& $8.4_{-2.4}^{+3.2}$ & $2.37_{-0.29}^{+0.33}$  &$1.03_{-0.32}^{+0.57}$ & 499.6/594  & 38.9 &  \nodata & \nodata & \nodata\\
 &2 &DB   & $3.5_{-1.0}^{+1.5}$   &$1.28_{-0.14}^{+0.16}$ & $41.01_{-15.99}^{+26.82}$ & 477.0/594  & 38.6 & \nodata & \nodata & \nodata \\
&3&PL& $9.6_{-3.5}^{+4.5}$ & $1.98_{-0.43}^{+0.50}$  &$0.63_{-0.27}^{+0.56}$ & 435.4/509  & 38.6 & \nodata & \nodata & \nodata \\
  &3 &DB  & $6.1_{-2.3}^{+2.9}$   &$1.43_{-0.30}^{+0.45}$ & $24.13_{-15.68}^{+40.79}$ & 429.8/509  & 38.6 &  \nodata&  \nodata& \nodata \\
                \cline{1-11}  \\   
x403  & 1 &PL  & $2.7_{-0.2}^{+0.2}$ & $3.02_{-0.10}^{+0.11}$  &$3.01_{-0.23}^{+0.25}$ & 714.1/738   & 39.4 & \nodata &  \nodata& \nodata \\
           & 1 &DB & $0.4_{-0.4}^{+0.0}$   &$0.66_{-0.02}^{+0.02}$ & $1393.98_{-167.01}^{+192.36}$ & 762.4/738 & 39.1 &  \nodata& \nodata & \nodata \\
           & 2 &PL& $2.3_{-0.2}^{+0.2}$ & $2.58_{-0.09}^{+0.10}$ &$5.05_{-0.40}^{+0.44}$ & 706.8/756  & 39.5& \nodata & \nodata & \nodata \\
           & 2 &DB& $0.4_{-0.4}^{+0.0}$   &$0.84_{-0.03}^{+0.03}$ & $1143.48_{-136.06}^{+159.54}$ & 753.0/756 & 39.4&  \nodata& \nodata & \nodata\\
           & 3 &PL& $2.4_{-0.9}^{+1.0}$ & $2.43_{-0.22}^{+0.23}$ &$1.61_{-0.32}^{+0.42}$ & 390.0/509  & 39.0& \nodata & \nodata & \nodata \\
           & 3 &DB& $0.4_{-0.4}^{+0.2}$  &$0.94_{-0.06}^{+0.02}$ & $242.79_{-32.99}^{+82.58}$ & 392.1/509  & 38.9&  \nodata&  \nodata& \nodata \\
\enddata
\tablenotetext{a}{PL: power law; DB: disk blackbody}
\tablenotetext{b}{Total intrinsic  luminosity.}
\tablenotetext{c}{Assume a solar abundance \citep{anders89}}
\tablenotetext{d}{Apec Flux Fraction.}
\end{deluxetable}

\end{document}